%% file: main.tex
\documentclass{article}

\input{packages}
\include{commands}

\begin{document}

\title{From point processes to quantum optics and back}
\date{}

\author[1]{R\'emi Bardenet\footnote{Joint first authorship and corresponding authors.}\thanks{\texttt{remi.bardenet@univ-lille.fr}}}
\author[1]{Alexandre Feller$^*$\thanks{\texttt{alexandre.feller@univ-lille.fr}}}
\affil[1]{Univ. Lille, CNRS, Centrale Lille, UMR 9189 -- CRIStAL -- Centre de Recherche en Informatique, Signal et Automatique, F-59000 Lille, France}

\author[2]{J\'er\'emie Bouttier}
\affil[2]{Universit\'e Paris-Saclay, CNRS, CEA, Institut de Physique
    Th\'eorique, 91191, Gif-sur-Yvette, France}

\author[3]{Pascal Degiovanni}
\affil[3]{
    Univ Lyon, Ens de Lyon, Universit\'e Claude Bernard Lyon 1, CNRS, 
    Laboratoire de Physique (UMR 5672), F-69342 Lyon, France
}

\author[4]{Adrien Hardy}
\affil[4]{
    Qube Research and Technologies, 75008 Paris, France
}

\author[5]{Adam Ran\c{c}on}
\affil[5]{Univ. Lille, CNRS, UMR 8523  --  PhLAM  --  Laboratoire  de Physique  des  Lasers,  Atomes  et  Mol\'ecules,  F-59000  Lille,  France}

\author[6]{Benjamin Roussel}
\affil[6]{Department of Applied Physics, Aalto University, 00076 Aalto, Finland}

\author[7]{Gr\'egory Schehr}
\affil[7]{Sorbonne Universit\'e, Laboratoire de Physique Th\'eorique et Hautes Energies, CNRS UMR 7589, 4 Place Jussieu, Tour 13, 5\`eme \'etage, 75252 Paris 05, France}

\author[8]{Christoph I. Westbrook}
\affil[8]{Univ. Paris-Saclay, Institut d'Optique Graduate School, CNRS, UMR 8501  --  Laboratoire Charles Fabry,  F-91127  Palaiseau,  France}

\setcounter{Maxaffil}{0}
\renewcommand\Affilfont{\itshape\footnotesize}
\maketitle

\begin{abstract}
[Disclaimer: this is Part I of a cross-disciplinary survey that is work in progress; All comments are welcome.] Some fifty years ago, in her seminal PhD thesis, Odile Macchi introduced permanental and determinantal point processes.
Her initial motivation was to provide models for the set of detection times in fundamental bosonic or fermionic optical experiments, respectively.
After two rather quiet decades, these point processes have quickly become standard examples of point processes with nontrivial, yet tractable, correlation structures. 
In particular, determinantal point processes have been since the 1990s a technical workhorse in random matrix theory and combinatorics, and a standard model for repulsive point patterns in machine learning and spatial statistics since the 2010s.
Meanwhile, our ability to experimentally probe the correlations between detection events in bosonic and fermionic optics has progressed tremendously. 
In Part I of this survey, we provide a modern introduction to the concepts in Macchi's thesis and their physical motivation, under the combined eye of mathematicians, physicists, and signal processers.
Our objective is to provide a shared basis of knowledge for later cross-disciplinary work on point processes in quantum optics, and reconnect with the physical roots of permanental and determinantal point processes. 
\end{abstract}
 \tableofcontents

\section{Introduction}
\label{s:introduction}
\input{introduction}

\section{Point processes and the semiclassical picture of HBT}
\label{s:point_processes}
\input{point_processes}

\section{Elements of quantum field theory}
\label{s:qft}
\input{qft}

\section{Photodetection and bosonic coherences}
\label{s:photoncoherence}
\input{photodetection}

\section{Electrodetection and fermionic coherences}
\label{s:electroncoherence}
\input{electroncoherence}

\section{Wrapping up and open questions}
\label{s:dppfermions}
\input{dppfermions}

\appendix

\bibliographystyle{plainnat}
\bibliography{biblio,remi,fermion,gregory}

\end{document}

%% file: packages.tex
\usepackage{amssymb,amsmath,amscd,mathrsfs,mathtools,amsfonts,amsthm,bbm,latexsym,lmodern,color}
\usepackage{fontawesome} 
\usepackage[a4paper]{geometry}
\usepackage{tikz,graphicx,subfigure,overpic}
\usepackage{hyperref}
\hypersetup{
  colorlinks=true,
  linkcolor=bleu,
  urlcolor=gray,
  citecolor=bleu
}
\usepackage[capitalise,noabbrev]{cleveref}
\usepackage[utf8x]{inputenc}
\usepackage{natbib}
\usepackage{xcolor}
\usepackage{siunitx}
\usepackage{authblk}
\usepackage{tikz}
\usepackage{circuitikz}
\usetikzlibrary{%
        shapes.geometric,calc,intersections,%
        decorations.markings,decorations.text,decorations,%
        patterns,decorations.pathmorphing,fpu,%
        positioning,automata,shapes.multipart,circuits,%
        circuits.ee, circuits.ee.IEC, arrows.meta, backgrounds,
        shapes.misc,graphs,decorations.pathreplacing,math,matrix
}
\usepackage{epigraph}
\usepgflibrary{curvilinear}
\usepgfmodule{nonlineartransformations}

\usepackage{dsfont}
\usepackage{todonotes}
\usepackage{physics} 

%% file: commands.tex
\renewcommand\d{\mathrm{d}}

\def\threefig{.30\textwidth}

\def\sym{\mathsf{S}}
\def\asym{\mathsf{A}}

\def\Hp{\mathcal{H}^\infty_{\text{Bosons}}}

\def\br{\mathbf{r}}
\def\bu{\mathbf{u}}
\def\bE{\mathbf{E}}

\def\cF{\mathcal{F}}
\def\cH{\mathcal{H}}

\renewcommand\tr{\mathrm{Tr}\,}
\def\balpha{\boldsymbol{\alpha}}

\def\Hfqb{H_{\text{FQ}}}
\def\Ufqb{U_{\text{FQ}}}
\def\Hff{H_{\text{FF}}}
\definecolor{darkgreen}{rgb}{0.0, 0.5, 0.0}
\newcommand*{\rb}[1]{}
\newcommand*{\af}[1]{}
\newcommand*{\ar}[1]{}
\newcommand*{\ah}[1]{}
\newcommand*{\jb}[1]{}
\newcommand*{\cw}[1]{}
\newcommand*{\bro}[1]{}
\newcommand*{\pd}[1]{}
\newcommand*{\gs}[1]{}

\newtheorem{theorem}{Theorem}[]
\newtheorem{lemma}[theorem]{Lemma}

\newtheorem{assumption}[theorem]{Assumption}
\theoremstyle{definition}

\newtheorem{remark}[theorem]{Remark}
\newtheorem{examplex}{Example}{}
\newenvironment{example}
  {\pushQED{\qed}\examplex}
  {\popQED\endexamplex}

\newcommand{\eq}{\begin{equation}}
\newcommand{\qe}{\end{equation}}

\newcommand{\N}{\mathbb{N}}

\newcommand{\R}{\mathbb{R}}
\newcommand{\C}{\mathbb{C}}
\newcommand{\E}{\mathbb{E}}
\renewcommand{\P}{\mathbb{P}}
\newcommand{\X}{\mathbb{X}}

\newcommand{\sH}{\mathscr{H}}

\newcommand{\cO}{\mathcal{O}}
\newcommand{\cN}{\mathcal N}

\newcommand{\e}{\mathrm{e}}
\renewcommand{\i}{\mathrm{i}}
\renewcommand{\d}{ {\mathrm d}}

\def\ind{\mathds{1}}
\def\id{\ind}

\newcommand{\DPP}{\mathrm{DPP}}

\renewcommand{\epsilon}{ \varepsilon}
\renewcommand{\phi}{ \varphi}

\usepackage[]{xcolor}

\definecolor{bleu}{rgb}{0.2,0.4,0.65}

\makeatletter
\def\@xfootnote[#1]{%
  \protected@xdef\@thefnmark{#1}%
  \@footnotemark\@footnotetext}
\makeatother

%% file: introduction.tex
The \emph{photoelectric effect} is the release of individual
electrons from a metal, when light falls onto that metal. 
The empirical observation that no electrons are released if the light frequency is 
beneath a certain threshold, along with an analogy with elastic 
collisions, prompted Einstein to posit the existence of light 
quanta, also known as \emph{photons}. 
A century later, the full explanation of the photoelectric effect is considered a success of the quantum theory of light.

In parallel, experimental devices have been designed to amplify the 
current resulting from a small number of released electrons, eventually 
giving detectors of light so sensitive that they are able to detect single photons.
These detectors have led to investigations on the quantum coherence properties of light. 
Coherence here means the ability of sources of light to generate 
interference patterns, such as in Young's celebrated double slit 
experiment \cite[Sections 1-3]{MaWo65}. 
One puzzling aspect of coherence was demonstrated by \cite{HaTw58}, and is called 
\emph{photon bunching} or the HBT effect, after the initials of its discoverers.
When placing a detector of single photons in the electric field created by a thermal light source, such as incandescent matter, \cite{HaTw58} showed that the detector tends to produce clicks that are grouped in time. 
Everything happens as if the photons were tightly bunched together when arriving at the detector, hence the name of photon bunching.
A few years later, with the invention of the laser, it was realized that this bunching effect disappeared for (so-called \emph{coherent}) laser light. 

At the turn of the 60s, there was intense research in finding the right mathematical framework to describe such coherence phenomena. 
A celebrated contribution is that of \cite{Glauber-1963}, who introduced concepts like coherent states and coherence functions, which quickly became textbook material for quantum optics.
Having met Glauber at Les Houches in 1964, French signal processing pioneer Bernard Picinbono launched a team in statistical optics in Orsay, determined to explore the transition between bunching and non-bunching measurements using the formalism of stochastic processes.
Picinbono assembled a small and diverse group, including PhD student Odile Macchi, who had just obtained a degree in mathematics.
The starting point of Macchi's 1972 PhD thesis was to find the right stochastic object to describe the detection times in HBT and explain photon bunching.
Her thesis turned out to be foundational in many respects.
This justified a recent translation of Macchi's thesis (originally in French) by Hans Zessin \citep{Mac17}, along with a large appendix written by Hans Zessin and Suren Poghosyan.

Odile Macchi introduced what we now call \emph{correlation functions} to describe point processes, i.e., random configurations of points in a generic space. 
She showed how common models for physical sources and detection led to point processes with closed-form correlation functions. 
In particular, for a given model of the source used in the HBT experiment, she showed that the resulting point process of detection times is a permanental point process, for which bunching can be fully characterized and related to properties of the electric field.
Teaming up with fellow PhD student in physics Christine Bénard, they used Glauber's formalism to identify the point process describing the parallel situation of detection times of electrons. 
Unlike for photons, the latter point process naturally exhibits \emph{anti-bunching}, with detection times being very regularly spaced.
\citep{BeMa73} is one the foundational stones\footnote{Together with a 1974 conference communication, that appeared later as \citep{Mac77}, and earlier work by Jean Ginibre; see the preface to the recent re-print \citep{Mac17} of Odile Macchi's thesis for a broader history.} for what we call today \emph{determinantal point processes} (DPPs).

Outside physics, determinantal point processes (DPPs) have since then become a cornerstone of the theory of random matrices \citep*{AnGuZe10,Joh06}, with applications in combinatorics \citep{BoOkOl00} and number theory \citep{RuSa96}.
DPPs are also a popular model for repulsive point pattern data in spatial statistics \citep*{LaMoRu14} and machine learning \citep{KuTa12}.
Many DPP users in these fields\footnote{including some of us prior to this work!} have little idea of the physical origin of DPPs. 
Given the late blooming of DPPs outside their original field, and some fifty years after Macchi's first generic formalization of DPPs, we felt it a natural endeavour to reconnect DPPs with their physical roots, as models for non-interacting fermions and tools to probe generalizations of the HBT effect.
This has been the purpose of two workshops so far.

We started with a two-day event\footnote{\url{https://dpp-fermions.sciencesconf.org/}} in 2019 in Lille, France, opened by Odile Macchi and featuring both theoretical and experimental physicists exposing their view of fermionic coherence to an audience of DPP users across mathematics, computer science, and signal processing.
The workshop having been met with cross-disciplinary enthusiasm, we made two decisions. 
The first was to organize an ambitious, two-week follow-up to the workshop,\footnote{\url{https://indico.in2p3.fr/event/25182/}} which took place in 2022 in Lyon, France. 
The second decision was to unite forces and write a survey on the links between point processes and optics, to pin down a common ground for discussions.
The current document is the first part of this survey.
It is both a joint introduction to point processes and quantum optics, and organized notes from a modern reading of Odile Macchi's thesis. 
The novelty of our document relies in its cross-disciplinary target audience of mathematicians, physicists, and signal processers, with a solid undergraduate background in probability and functional analysis.
In particular, while tackling topics in modern physics, we assume little physics knowledge from the reader beyond undergraduate exposure to wave optics. 
One of our leitmotivs is to sketch the thought process behind some fundamental arguments in quantum optics.
Indeed, in our experience, arguments thought as basic by physicists can be hard to grasp by people trained in mathematics, mostly because the implicitly assumed lore differs across communities.
In a reverse movement towards physicists, and following the spirit of \cite{Mac75}, we motivate most mathematical concepts by their need as models in optics, including point processes. 

We have striven to maintain a balance between mathematical rigour, clarity, and brevity, giving references whenever we had to take shortcuts.
We expect that every reader will find some of the material basic and some more exotic, depending on their background, and we hope that all readers will eventually learn something useful.
Our objective is to make the potential barrier for crossing from one discipline to another as low as possible, so that ideas can flow more easily. 
A second part of the document is in preparation, presenting selected advanced topics from the Lille workshop, including experimental measurements of the HBT effect, non-interacting trapped fermions in statistical physics \citep{dean2016noninteracting,dean2019noninteracting}, fermions in combinatorics, and electronic quantum signals. 
The two parts are ultimately to be bound in a single manuscript.

The rest of the document is organized as follows.
In Section~\ref{s:point_processes}, we introduce key examples of point processes.
Poisson, Cox, and permanental point processes are motivated there by the so-called \emph{semi-classical} derivation of the HBT photon bunching effect, treating only the detector as a quantum object, not the electric field.
Determinantal point processes are also introduced, but their physical motivation requires to go beyond the semi-classical picture, which justifies the next three sections.
Section~\ref{s:qft} is a crash course in quantum field theory, from the basics of quantum mechanics to Wick's theorem on the average of products of ladder operators. 
Wick's theorem yields two very different results depending on the commutation rules of the operators it applies to, which in turn derive from modeling either \emph{bosonic} particles (like photons), or \emph{fermionic} particles (like electrons). 
Ultimately, this dichotomy is at the origin of the appearance of permanental and determinantal point processes. 
The section concludes with a discussion of the coherent states of \cite{Glauber-1963} and their relation to time-frequency analysis in signal processing. 
Section~\ref{s:photoncoherence} covers the modern view on photodetection, culminating in the full quantum justification of the HBT effect using permanental point processes as a model for the coincidence measurements of photons, as well as considerations on the role of the source in the bunching properties of the measurements.
Following the lines of Section~\ref{s:photoncoherence}, Section~\ref{s:electroncoherence} covers the detection of electrons, finally resulting in the appearance of determinantal point processes. 
The section concludes on the comparative difficulties of recovering non-quantum arguments from the quantum treatment in the case of fermions.
Finally, Section~\ref{s:dppfermions} wraps up this first part, commenting on the generic derivation of a permanental or determinantal point process from a model of free fermions, and discussing open questions motivated by this construction.

\paragraph{A note on the style.} 
Because of its cross-disciplinary objectives, the style of our document is hybrid. 
We mostly follow a style inspired by texts in mathematics, with definitions, theorems, assumptions, and examples, sometimes merged with the main text, sometimes fleshed out to draw the reader's attention.
Assumptions, in particular, often stand out. 
By \emph{assumption} we mean a statement for the reader to accept in order to progress in a discussion or a computation. 
It can be, e.g., a modeling choice or a mathematical approximation.

Examples are often borrowed from physics; to make the text self-contained, some of these examples are relatively long. 
To help the reader to visually isolate examples from the rest of the text, we conclude each example with a $\diamond$ symbol.
Finally, footnotes abound, and are usually meant as side remarks to one of the targeted scientific communities, e.g. to discuss notational or minor conceptual differences between different domains. 

\rb{For v2, have a small table indicating our main notations and the classical corresponding notations across fields (e.g., we use $\mu$ as a reference measure, whereas a physicist may think of the chemical potential).}

%% file: point_processes.tex

In a physical experiment where a detector clicks when hit by a particle, the observation consists of a set of real numbers, the times at which the detector clicks.
The natural probabilistic model for such a situation, where both the number and the location of the observed points are uncertain, is a \emph{point process}, i.e, a random configuration of points.
In Sections~\ref{s:correlation_functions} to \ref{s:determinantal}, we introduce some of the modern vocabulary of point process theory, along with three key families of point processes: Poisson, permanental and determinantal point processes.

While optical models that feature determinantal point processes will have to wait for the fully quantum treatment of electron detection in Section~\ref{s:electroncoherence}, we already motivate Poisson and permanental point processes in this section by the so-called \emph{semi-classical} (as opposed to \emph{fully quantum}) derivation of the photon bunching effect. 
Our running examples are based on the simple setup given here as \cref{ex:physical_setup}.

\setcounter{examplex}{-1}
\begin{example}[A simple photodetection setup]
  \label{ex:physical_setup}
  As introduced in Section~\ref{s:introduction}, photodetection is based on the photoelectric effect: when light falls onto a metal, electrons are released, generating a current that we can measure. 
  Light is an electromagnetic field, so typically has an electric and magnetic component, but the physics of the photoelectric effect is essentially dependent on the electric component of the field. 

  The electric field at a point $\mathbf{r}\in\mathbb{R}^3$ in space and time $t$ is modeled by a square-integrable function $\bE:\mathbb{R}^3\times \mathbb{R}\rightarrow\mathbb{R}^3$ of space and time.
  The three components of $\bE(\mathbf r, t)$ correspond to what is called in 
  physics the \emph{polarization} of light.
  Throughout this paper, for simplicity, we assume that 
  the field is linearly polarized. 
  In this section,\footnote{
    In later sections, the field will be modeled by a collection of operators, and linear polarization will thus correspond to a different mathematical assumption.} 
  linear polarization amounts to assuming the existence of a unit vector $\bu\in\mathbb{R}^3$ such that, for all $(\br,t)\in \mathbb{R}^3\times \mathbb{R}$, $\bE(\br,t)=E(\br,t)\bu$, with $E(\br,t)\in\mathbb{R}$.
  This simplification allows us to focus here on the scalar function $E:\mathbb{R}^3\times \mathbb{R}\rightarrow \mathbb{R}$.
  Moreover, we consider a photodetector placed at a fixed position $\br\in\mathbb{R}^3$. 
  The effect of the field on the detector is assumed to only depend on the value $E(\mathbf{r},\cdot)$ of the field at the detector position, and we thus further focus on the function $t \mapsto E(\mathbf{r}, t)$ in our examples, which we also denote\footnote{
    Overloading variable names is common in physics, and we shall follow this convention when possible without confusion.
  } 
  by $E$ in this section. 

  We are interested in the times at which our detector \emph{clicks}, i.e., detects a single photon.
  Because the measured times vary from one run of the experiment to the next, we want to model them as a random set of (distinct) real numbers.
\end{example}

Finally, note that we only describe in this section an idealized 
version of the HBT experiment, following \cite[Section 4.2]{Mac75}. 
For physics arguments, we refer the reader with little prior exposition to physics to \cite[Chapter 9]{MaWo95}, which is a recent textbook treatment of the survey \citep{MaWo65} to which \cite{Mac75} originally referred.

\subsection{The correlation functions of a point process}
\label{s:correlation_functions}

Let us fix a complete 
metric space $\X$, with $\mu$ a Borel measure on the Borel sets of $\X$. 
A majority of the examples in this paper will deal with $\X=\mathbb{R}$ equipped with the Lebesgue measure.
As mentioned in the header of Section~\ref{s:point_processes}, this choice describes the ideal detection times of physical particles.
In spatial statistics, we typically encounter $\X=\R^d$ with 
$\mu$ the Lebesgue measure \citep{BaRuTu15}. 
It is also not uncommon to take $\X$ to be a compact manifold $M$ (like a sphere; \citep{BeHa19}) or a discrete space (like a large 
dataset in machine learning \citep{KuTa12}), in which case $\mu$ is typically taken to be the volume form on $M$ or the counting measure of the discrete space, 
respectively.

A point process $\gamma$ on $\X$ is a random configuration\footnote{
  \emph{Stricto senu}, we are defining here \emph{simple} point processes, i.e., such that the samples never contain a given point of $\X$ more than once. 
  Since all point processes in this work are simple, we take this shortcut and identify a (simple) point process with a random configuration.
}
of points in $\X$. 
In other words, $\gamma$ is a random variable taking its values in the space of locally finite subsets of $\X$,
$$
\mathrm{Conf}(\X):=\Big\{\gamma\subset\X:\; \#(\gamma\cap A)<\infty\; \text{for all compact }A\subset \X \Big\}.
$$
In terms of modeling beams of physical particles, the \emph{locally finite} assumption entails that for 
any given time interval, there is a finite number of particles that were
detected in that interval.

For any $k\geq 1$, the $k$-point correlation function 
$\rho_k:\X^k\to[0,\infty]$ satisfies, when it exists,
\begin{align}
\label{kCorr}
\E\left[\, \sum_{\substack{ x_1,\ldots, x_k\in \gamma \\
 x_i\neq  x_j\text{ if } i\neq j}}f( x_{1},\ldots, x_{k})\,\right]
= \int_{\X^k}f(x_1,\ldots,x_k)\; \rho_k(x_1,\ldots,x_k)\,\prod_{i=1}^k\d\mu(x_i)
\end{align}
for any bounded (or positive) and measurable test function $f$.
Here 
$$
\X^k = \overbrace{\X\times\cdots\times \X}^k$$ and the symbol $\E$ in \eqref{kCorr}
stands for the expectation\footnote{Physicists speak of an ensemble average, and denote it using angle brackets $\langle \cdot\rangle$.
We stick with the symbol $\mathbb{E}$ for classical expectations, and we reserve
brackets only for Hilbert space inner products and quantum averages.} 
under the law of the random variable $\gamma$.
Thus, the $k$-point correlation function $\rho_k$ encodes the distribution of 
$k$-tuples of points from $\gamma$. 
Indeed, an informal rewriting of 
\eqref{kCorr} reads
\begin{align}
\rho^{(k)}(x_1,\dots,x_k) \mu(\d x_1)\dots \mu(\d x_k) =  
\mathbb{P}
\begin{pmatrix}\text{There are at least $k$ points in $\gamma$,} \\
\text{one in each of the infinitesimal}\\
\text{ balls $B(x_i, \d x_i) \text{ for } i=1,\dots,k$}
\end{pmatrix}
\,.
\label{e:correlationFunctions}
\end{align}
It is common to actually define a point process $\gamma$ by a sequence 
$(\rho_k)$ of \emph{compatible} correlation functions; see \citep[]{DaVe03}.
By \emph{compatible}, we mean that not every sequence $(\rho_k)$ actually defines a point process, and that usually a mathematical argument for existence is necessary. 

Of particular practical significance are the correlation functions for $k=1$ and $k=2$. 
The first correlation function $\rho_1$ describes the marginal distribution of particles, and is called the \emph{intensity} in probability and statistics.\footnote{Physicists might prefer to call it \emph{density of particles}, and 
reserve \emph{intensity} for other physical quantities.
To avoid confusion, we will write \emph{first correlation function} in full.}
The second correlation function $\rho_2$ describes pairwise correlations, 
and is often discussed in its normalized form
\begin{align}
g(x,y) = \frac{\rho_2(x,y)}{\rho_1(x)\rho_1(y)}
\,.
\label{e:pcf}
\end{align}
Finally, when $\X=\mathbb{R}^d$ and the distribution of $\gamma$ is
translation- and rotation-invariant, $g$ only depends on 
$r=\Vert x-y\Vert$, so that we write $g(x,y)=g_0(r)$. 
The function 
$g_0$ is called the \emph{pair correlation function} of $\gamma$. 

Figure~\ref{f:three_examples} shows samples of three translation-invariant point processes in 
$\mathbb{R}$, along with their pair correlation functions.
The three point processes share the same (constant) first-order correlation function. 
Note how the pair correlation function in Figure~\ref{f:three_examples:determinantal_pcf} is lower than $1$ close to $0$, indicating fewer small pairwise distances than the reference point process in Figure~\ref{f:three_examples:poisson_pcf}: this is a sign of a very regular, more grid-like distribution of the points, as seen in Figure~\ref{f:three_examples:determinantal_samples}.
On the contrary, the pair correlation function of Figure~\ref{f:three_examples:permanental_pcf} shows more small pairwise distances than the reference: points are lumped together, as confirmed by Figure~\ref{f:three_examples:permanental_samples}.
The rest of this section describes these three examples in more detail.

\subsection{Poisson and Cox point processes}
\label{s:poisson}

\begin{figure}
  \subfigure[Poisson samples]{
  \includegraphics[width=\threefig]{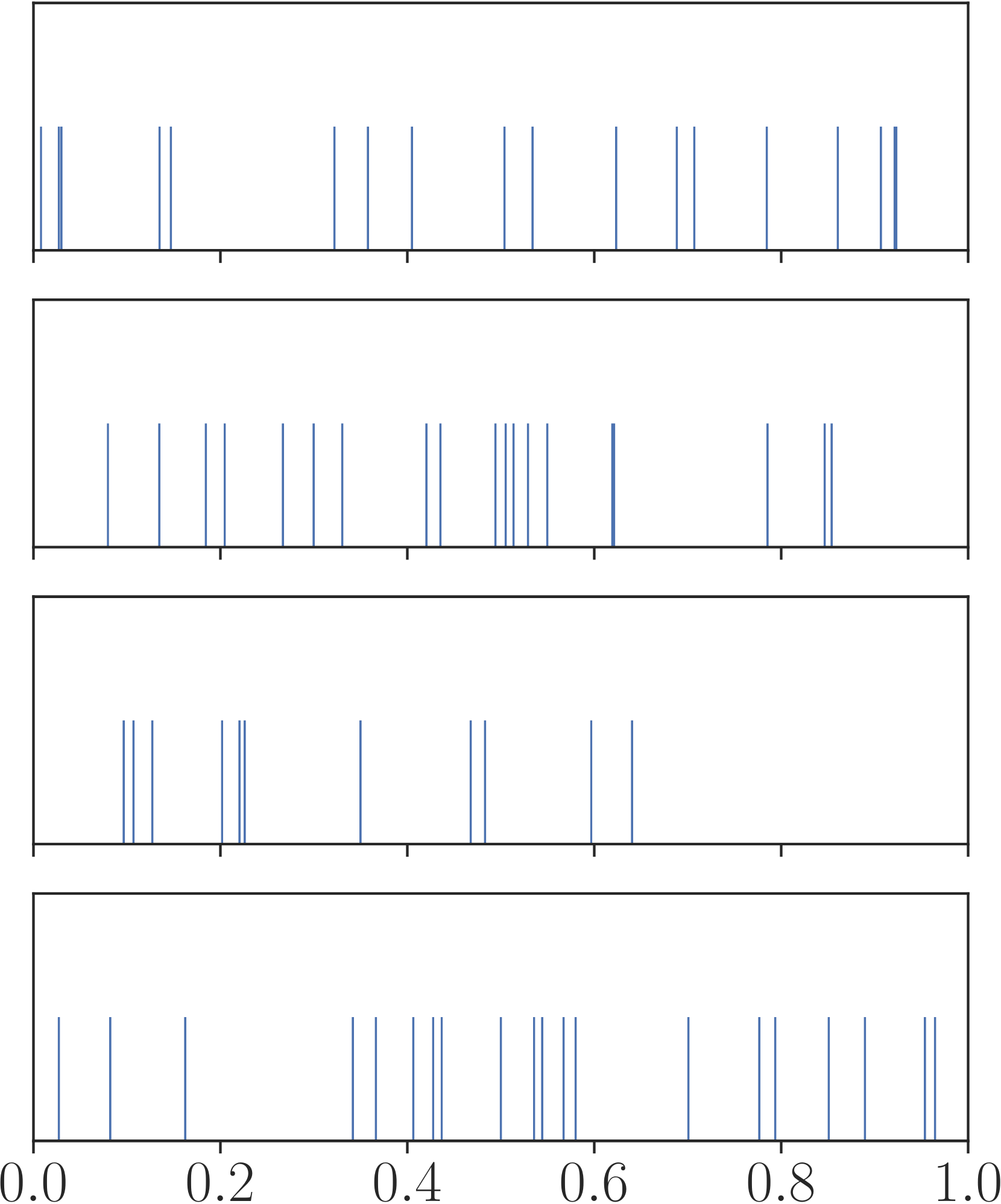}
  \label{f:three_examples:poisson_samples}
  }
  \hfill
  \subfigure[Permanental samples]{
  \includegraphics[width=\threefig]{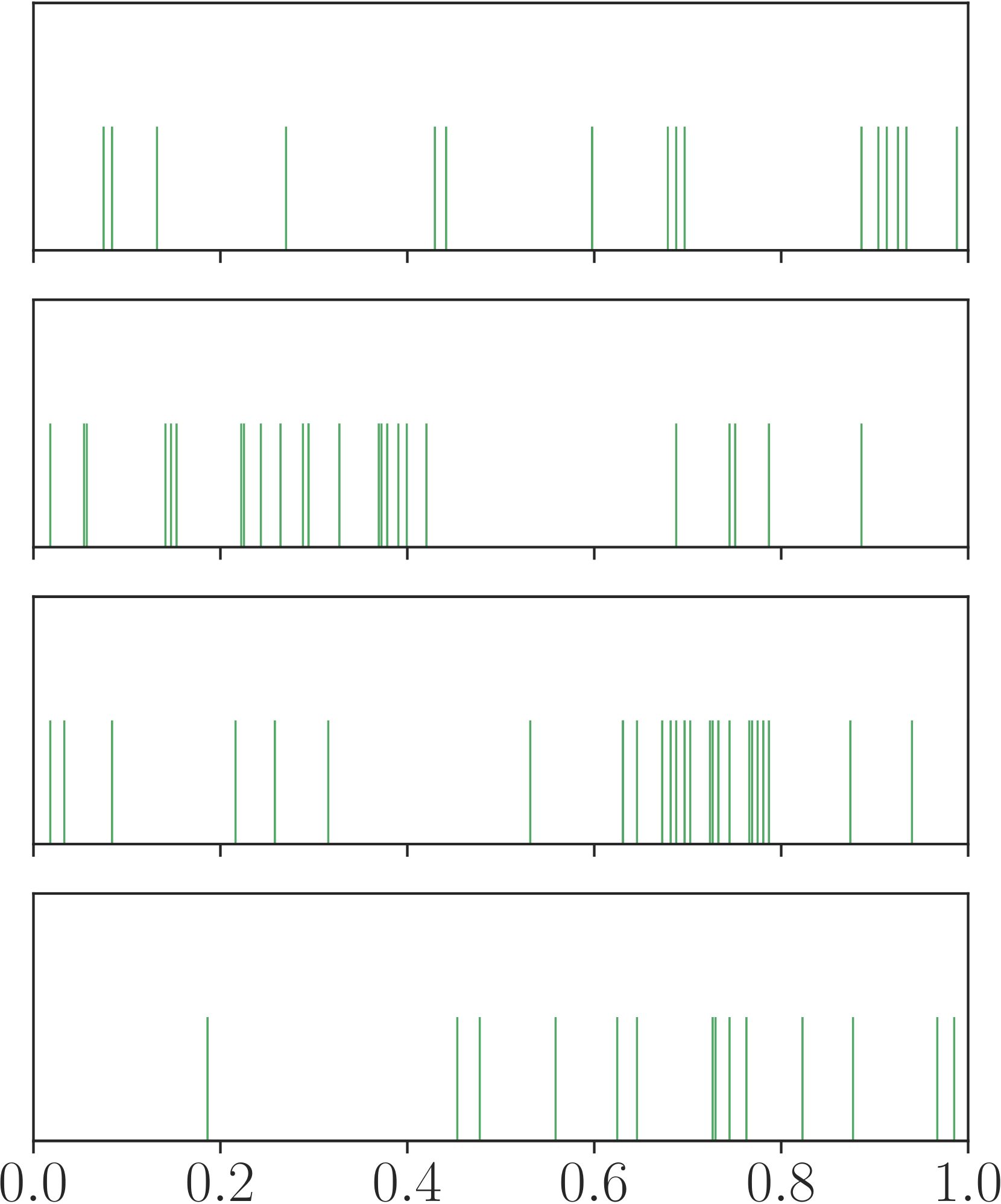}
  \label{f:three_examples:permanental_samples}
  }
  \hfill
  \subfigure[Determinantal samples]{
  \includegraphics[width=\threefig]{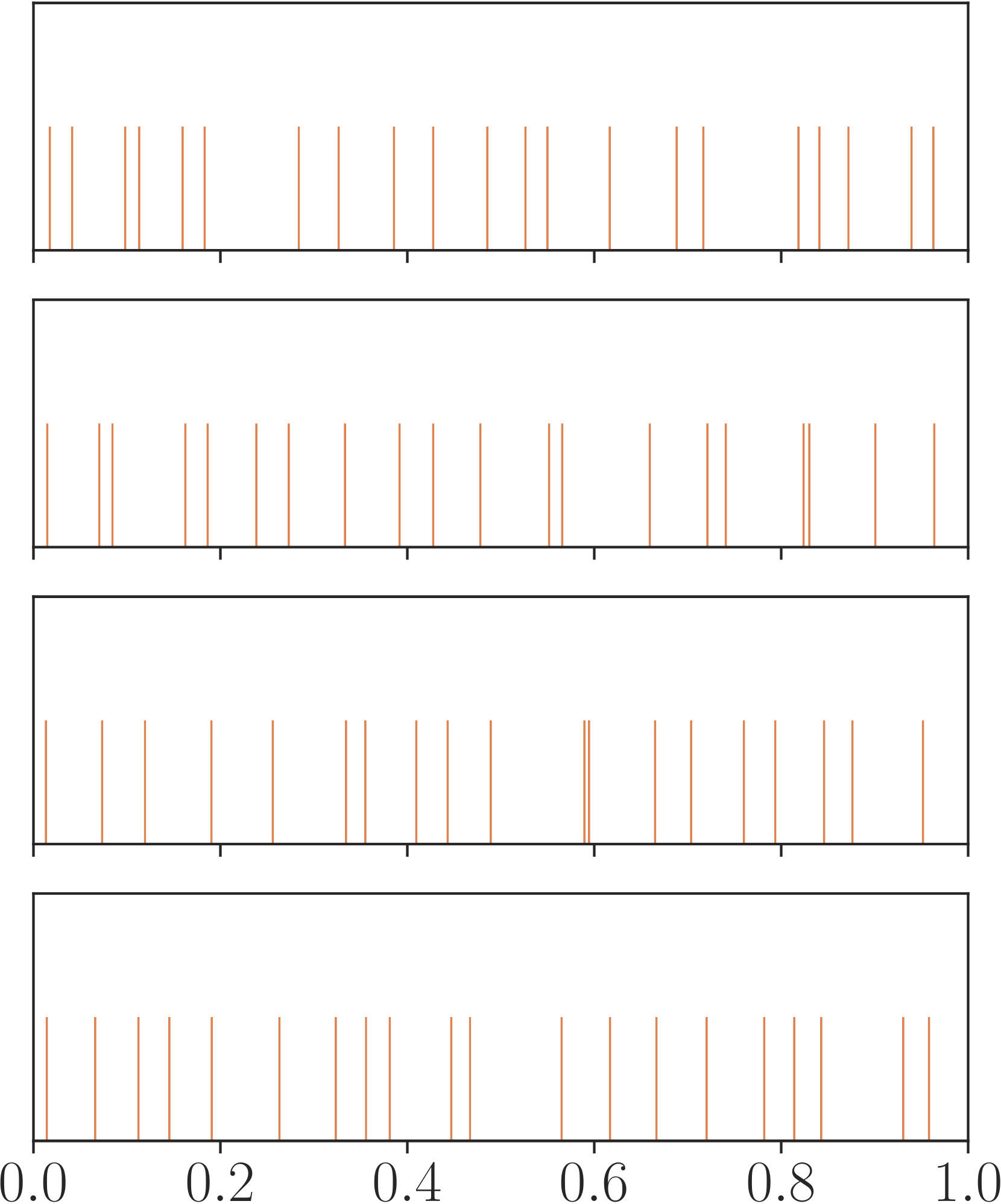}
  \label{f:three_examples:determinantal_samples}
  }\\
  \subfigure[Poisson pcf]{
  \includegraphics[width=\threefig]{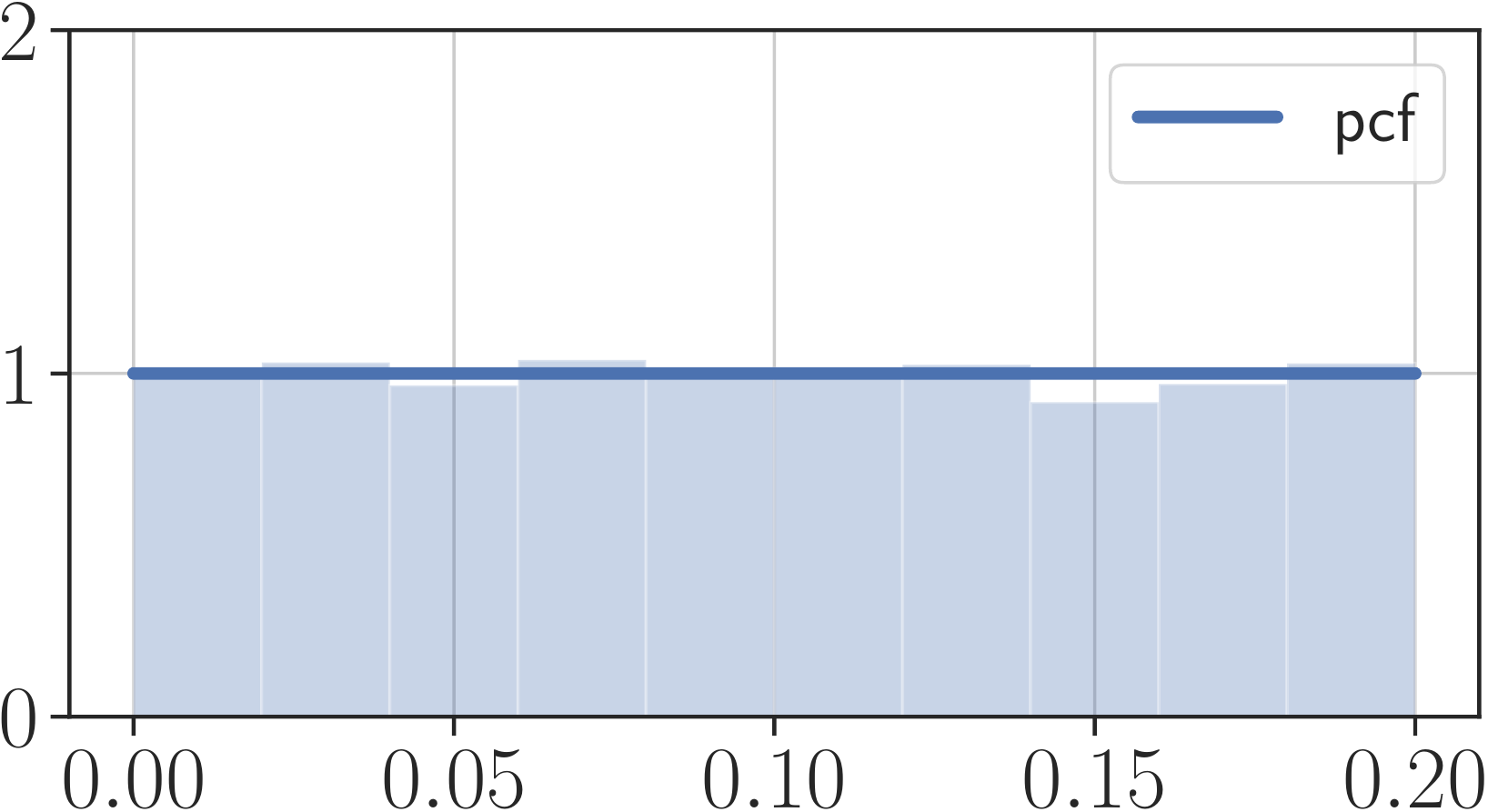}
  \label{f:three_examples:poisson_pcf}
  }
  \hfill
  \subfigure[Permanental pcf]{
  \includegraphics[width=\threefig]{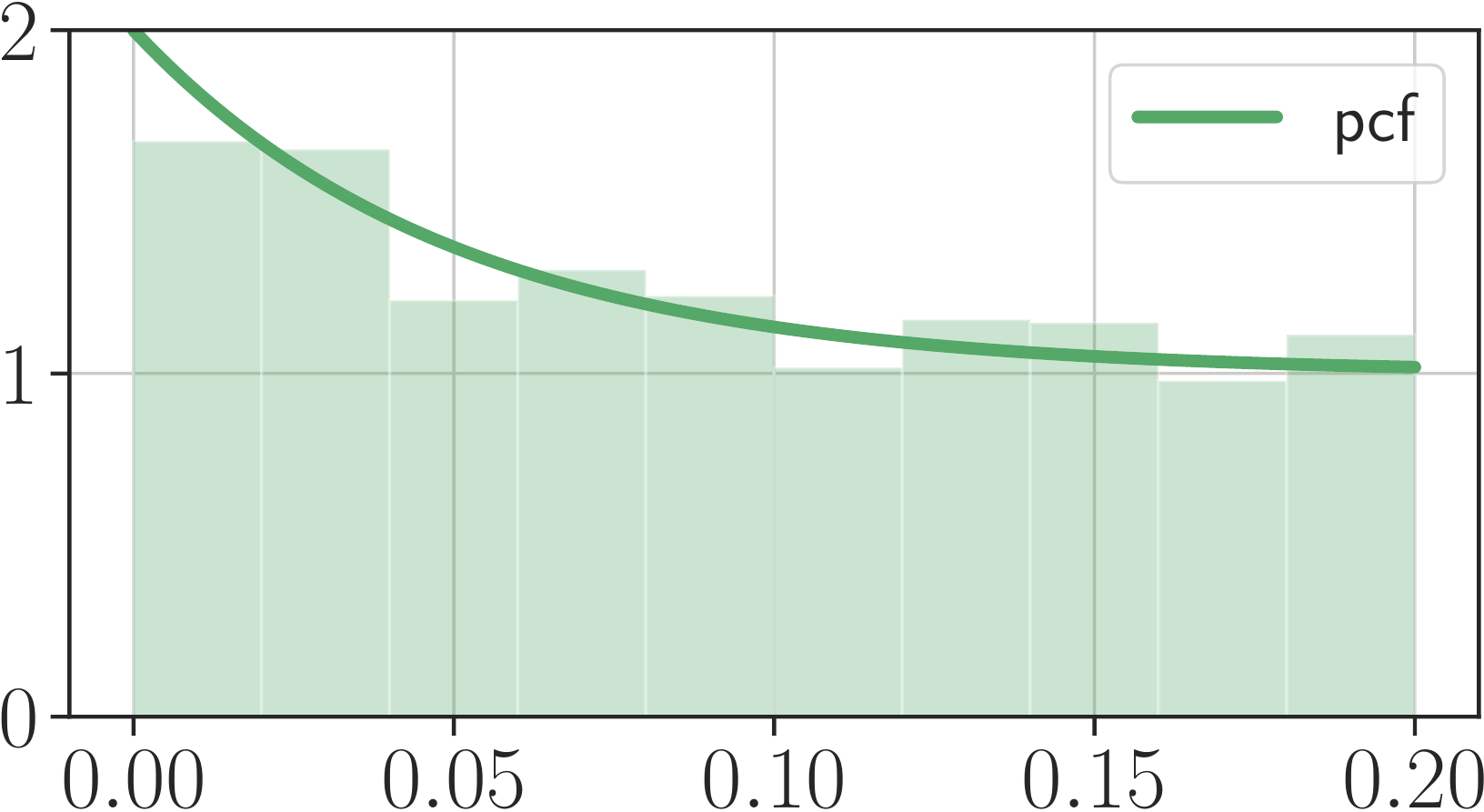}
  \label{f:three_examples:permanental_pcf}
  }
  \hfill
  \subfigure[Determinantal pcf]{
  \includegraphics[width=\threefig]{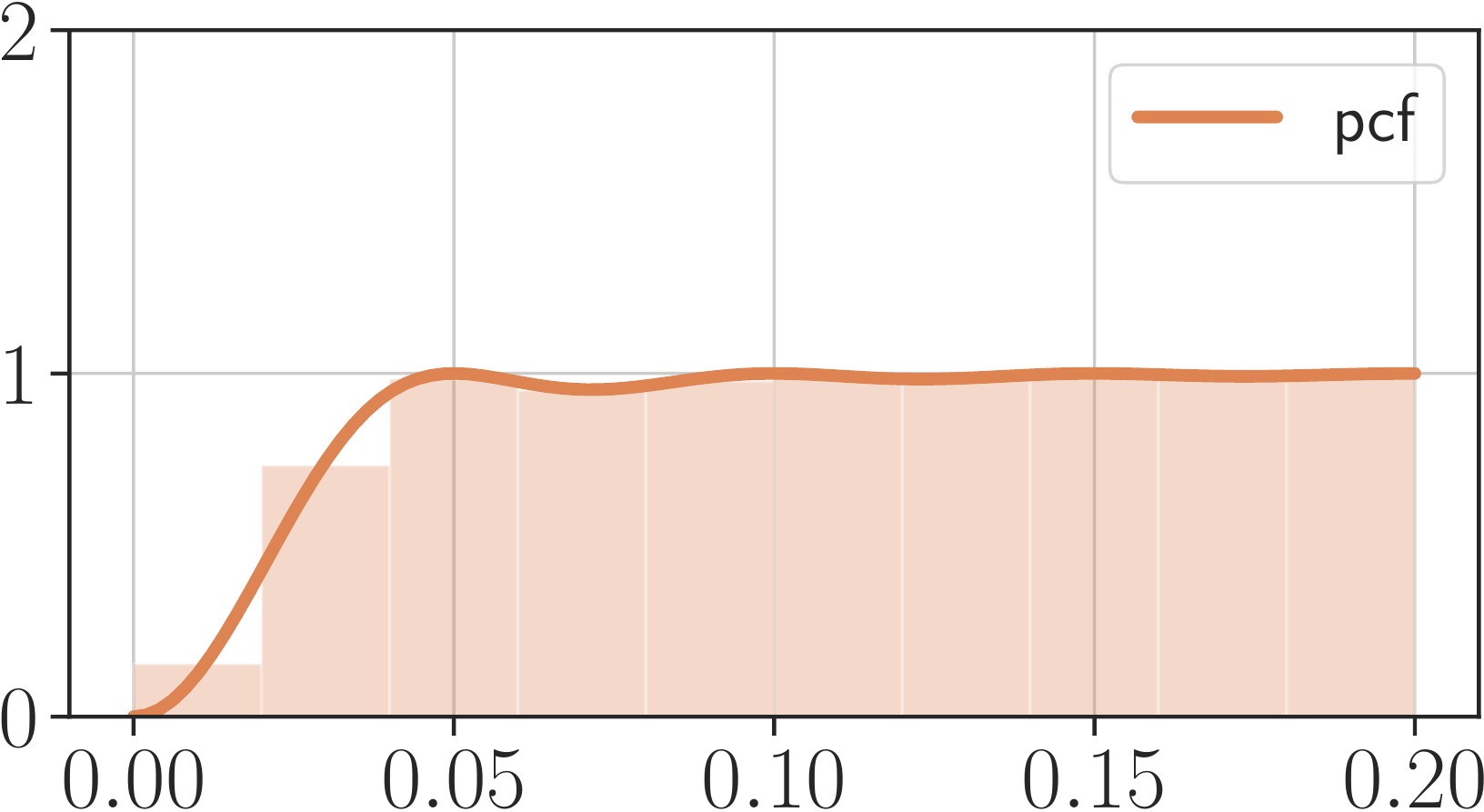}
  \label{f:three_examples:determinantal_pcf}
  }
  \caption{(a, b, c) The intersection with $W=[0,1]$ of four 
samples of each of the three types of point processes introduced 
in Section~\ref{s:point_processes}. All three point processes are 
translation-invariant and are scaled to have the same expected 
number of points falling in $W$. (d, e, f) The pair correlation function
\eqref{e:pcf} of the three point processes, and histograms of 
pairwise distances from 100 independent samples.}
  \label{f:three_examples}
\end{figure}

Let $\lambda:\X\rightarrow \mathbb{R}_+$ be locally integrable, that is,
$\int_B \lambda(x)\d\mu(x)<\infty$ for every bounded $B\subset \X$. 
Further assume for simplicity that the measure $\lambda\d\mu$ has no atom.\footnote{
  Without this assumption, a Poisson point process would not necessarily be a \emph{simple} point process. 
}
The point process with correlation functions
\begin{equation}
\rho_k(x_1,\dots,x_k) = \lambda(x_1)\dots \lambda(x_k), \quad k\geq 1,
\label{e:correlationFunctionsPoisson}
\end{equation}
always exists and is called the \emph{Poisson point process} with parameter function\footnote{
  Statisticians call this parameter function the \emph{intensity} of the process. 
  We refrain from using this naming convention, to avoid confusion with any physical intensity.
} $\lambda$.
The first correlation function of the process is thus $\rho_1=\lambda$. 
Moreover, the separable form of \eqref{e:correlationFunctionsPoisson}
implies the lowest level of correlation among the points of the process.
In particular, the pairwise correlation function $g\equiv 1$ is constant, 
so that no pairwise distances are preferred. 
Figure~\ref{f:three_examples:poisson_samples} shows a few Poisson 
samples where $\mu$ is the Lebesgue measure on $\X=\mathbb{R}$ and 
$\lambda$ is constant.

\begin{example}[Deterministic electric fields yield Poisson point processes]
  \label{e:poisson}
  Consider the photodetection setup of Example~\ref{ex:physical_setup}.
  Let us further assume that the field is \emph{quasi-monochromatic}, i.e., that the modulus $\vert\mathcal{F}E\vert$ of the Fourier transform of 
  the function $E: \mathbb{R}\rightarrow \mathbb{R}$ concentrates around $\omega$ and 
  $-\omega$ for a single value of the frequency $\omega>0$. Note that 
  the symmetry of $\vert\mathcal{F}E\vert$ is a consequence of $E$ being real-valued.
  Denote by $E^+$ the \emph{analytic signal} of the electromagnetic field $E$, i.e., 
  \begin{equation}
  E^+ = 2\mathcal{F}^{-1}  \left(\mathcal{F} (E) \times 1_{(0,\infty)}\right)\,.
  \label{e:analyticSignal}
  \end{equation}
  Taking the analytic signal \cite[Section 3.1]{MaWo95} is a partial isometry of
  $L^2(\mathbb{R})$ commonly used in signal processing, which has several
  natural properties. 
  For instance, $E^+$ removes the symmetry 
  in $\vert\mathcal{F}E\vert$, so that each frequency is represented in an unambiguous, non-redundant manner. 
  To a mathematician, the analytic signal is the boundary value of a particular analytic function of the upper-half plane, defined as the Cauchy transform of $f$; see e.g. \cite[Section 2.1]{Pug82}.
  A more physical property is that when $E$ is quasi-monochromatic, $\vert E^+\vert$ is a good approximation of the envelope of the signal $E$; see Figure~\ref{f:analytic_signal}. 
  This idea of an envelope, insensitive to rapid oscillations, along with intuition from classical electromagnetism relating a physical intensity to the square of the amplitude of a wave, is at the origin of the following modeling assumption, which we flesh our for future reference. 
  \begin{assumption}
    \label{a:modellingCoxProcess}
  For a deterministic field $E$, the detection times follow a 
  nonstationary Poisson process, with parameter function proportional to $t\mapsto \vert E^+(t)\vert^2$, where $E^+$ is given 
  by \eqref{e:analyticSignal}.
  \end{assumption}
  The full justification for the assumption can 
  be obtained by a more precise model of light and matter interaction
  using quantum theory; see \cite[Chapter 9]{MaWo95} or our \cref{s:photoncoherence}.
  Meanwhile, we see that Poisson processes naturally result from deterministic fields. 
\end{example}

Poisson point processes are useful both as a reference point process
 and as a building block in statistical modeling. 
 For instance, a Poisson 
point process with a random parameter function $\lambda$ is known as a 
\emph{Cox process} \citep[Section 6.2]{DaVe03}.
A Cox process has correlation functions 
$$
\rho_k(x_1,\dots,x_k) = \mathbb{E}\lambda(x_1)\dots \lambda(x_k), \quad  k\geq 1, 
$$
where the expectation is over $\lambda$. 
In particular, by Jensen's inequality,
\begin{equation}
  \label{e:rho2_diagonal_cox}
  \rho_2(x,x) = \mathbb{E}\lambda(x)^2 \geq \left(\mathbb{E}\lambda(x)\right)^2 = \rho_1(x)^2.
\end{equation}
If inequality is strict and $\rho_2$ is continuous, one can thus expect
that for $x, y$ close to each other, the probability that there is a 
point near $x$ and a point near $y$ in the same realization is larger
for a Cox process than for a Poisson process with the same first correlation function.
In other words, samples from a Cox process exhibit clusters, or
\emph{bunching}. 

\begin{example}[Random classical sources imply photon bunching]
  \label{ex:cox}
  When the physical field $E$ comes from a thermal source (i.e., incandescent matter or a gas discharge), experimentally observed detection events do not seem to form a Poisson process, but rather exhibit some form of \emph{bunching}, or clustering. 
  This is the HBT effect \citep{HaTw58}.
  Mathematically, it can be seen as a consequence of 
  Assumption~\ref{a:modellingCoxProcess}, as soon as one represents
  the field resulting from the thermal source by a stochastic 
  process making $x,y\mapsto\rho_2(x,y)$ smooth near the diagonal $x=y$. 
  The point process of detection times is then a Cox process that favors clusters of points, compared to a Poisson process.
\end{example}

The archetypal Cox processes are the so-called permanental point processes, which we now introduce.

\subsection{Permanental point processes}
\label{s:permanental}

A point process $\gamma$ is said to be \emph{permanental} when there 
exists a so-called \emph{correlation kernel} $K:\X\times \X\to \C$ such that 
the correlation functions of $\gamma$ read
\begin{align}
\label{e:correlationFunctionsPermanental}
\rho_k(x_1,\ldots,x_k)= 
	\text{per} \Big[K(x_i,x_j)\Big]_{i,j=1}^k, \quad  k\geq 1
\,.
\end{align}
The permanent $\text{per}\mathbf A$ in \eqref{e:correlationFunctionsPermanental} 
of a matrix $\mathbf A\in\mathbb{R}^{k\times k}$ is defined, analogously to the 
determinant, by a sum over permutations
$$
\text{per} \mathbf A = \sum_{\sigma\in\frak{S}_k} \prod_{i=1}^k a_{i\sigma(i)},
$$
where $\frak{S}_k$ denotes the symmetric group. 
Assuming existence for a moment, the point process described by 
\eqref{e:correlationFunctionsPermanental} has nontrivial correlation functions.
Taking $k=1$ and $k=2$ in \eqref{kCorr}, we obtain for instance
\begin{align}
\label{1corrPermanental}
\E\left[\sum_{ x\in \gamma} f( x)\right]
&= \int f(x) K(x,x)\mu(\d x),\\
\label{2corrPermanental}
 \E\left[\sum_{\substack{ x, y\in\gamma\\  x\neq  y}} f( x, y)\right]
&= \int f(x,y) \Big[K(x,x)K(y,y) + K(x,y)K(y,x)\Big]\mu(\d x)\mu(\d y)
\,.
\end{align}
When the kernel $K$ is Hermitian, that is, $K(x,y) = \overline{K(y,x)}$, 
the second correlation function in \eqref{2corrPermanental}, once 
normalized, becomes
\begin{align}
  g(x,y) &= \frac{\rho_2(x,y)}{\rho_1(x)\rho_1(y)} = 
	1 + \frac{\vert K(x,y)\vert^2}{K(x,x)K(y,y)}\geq 1.
  \label{e:bunching}
\end{align}
In particular, \eqref{e:bunching} shows that permanental point processes
are attractive: the larger $\vert K(x,y)\vert^2$, the more likely two 
particles at $x$ and $y$ are to co-occur. Moreover, they are always more 
likely to co-occur than if $\gamma$ were a Poisson process with the same 
first correlation function. 

The existence of a point process satisfying 
\eqref{e:correlationFunctionsPermanental} requires conditions on $K$ 
\citep{ShTa03}. 
A standard set of conditions comes from a representation 
of $\gamma$ as a Cox process. 
This representation derives from the semiclassical treatment of the HBT effect for photons by \cite{Mac75}, which we present below as \cref{e:permanental}.
To isolate the mathematical statement, let us simply mention that a Poisson point process with random first correlation function $\lambda$ taken to be the squared modulus of a Gaussian process is a permanental point process; see e.g. \citep[Proposition 35]{HKPV06}.
Making sure that the underlying Gaussian process exists in turn guarantees existence of the attached permanental point process. 

\begin{example}[Gaussian classical fields yield permanental point processes]
  \label{e:permanental}
  Continuing \cref{ex:cox} on thermal sources, quasimonochromatic thermal sources are actually represented 
  by zero-mean stationary Gaussian processes, following the intuition that
  they result from the superposition of many zero-mean, independent random
  contributions from the source at roughly the same frequency \cite[Section 4.2]{Mac75}.
  Mathematically, a random function $f$ is said to have for distribution a 
  zero-mean Gaussian process with kernel $K$ if, for any number $n$ 
  of observations and for any times $t_1,\dots,t_n\in\mathbb{R}$, we have
  \begin{equation}
    \left (f(t_1),\dots,f(t_n) \right )^T \sim \cN\left((0,\dots, 0)^T, ((K(t_i,t_j))_{1\leq i,j\leq n}\right),
    \label{e:mvn}
  \end{equation}
  where $\cN(\mu, \mathbf\Sigma)$ stands for the multivariate Gaussian with mean $\mu$ and covariance matrix $\mathbf\Sigma$.
  Assuming $E$ is such a zero-mean Gaussian
  process, and that its distribution is invariant under translations along the time axis, the linear transform $E^+$ in \eqref{e:analyticSignal} is also a zero-mean, translation-invariant Gaussian process,\footnote{However, $E^+$ is complex-valued, i.e., the vector in \eqref{e:mvn} is a complex multivariate Gaussian vector.} with 
  the specific property that $\mathbb E E^+(t)E^+(s) = 0$.

  As an example, the reader can think of $E$ as having kernel
  \begin{equation}
  K_{\text{Lorentz}}(t,s) = \exp(-\frac{\vert t-s\vert}{\sigma})\cos(\omega (t-s)) 
  = K_{\text{Lorentz}, 0}(t-s),
  \label{e:lorentz_kernel}
  \end{equation}
  for some $\omega\gg 1/\sigma$. 
  $K_{\text{Lorentz}}$ is the product of a slowly varying envelope, which we arbitrarily take to be exponential for concreteness, and a fast oscillating function.
  In particular, the Fourier spectrum of 
  $K_{\text{Lorentz}, 0}$ is concentrated around $\omega$, making 
  $E$ quasi-monochromatic in the sense of \citep{Mac75}. In that case, 
  $E^+$ has kernel
  \begin{equation}
  C_{\text{Lorentz}}(t,s) = 2 K_{\text{Lorentz}, 0}^+ (t-s) \approx 2\exp(-\frac{\vert t-s\vert}{\sigma}) \e^{\i\omega (t-s)}.
  \label{e:analytic_covariance}
  \end{equation}
  The first equality is a general property of analytic transforms of 
  second-order stationary processes, and can be proved by direct 
  computation. The approximation in \eqref{e:analytic_covariance} is a 
  consequence of Bedrosian's theorem, stating that the slowly varying envelope is preserved when taking the analytic signal; see e.g. \citep{Pic97}. 
  Finally, we note that the covariance \eqref{e:analytic_covariance} is a slight modification of the  Lorentz kernel example given by \citep{Mac75}, where we have introduced 
  a phase factor since, \emph{stricto sensu}, analytic covariance kernels 
  cannot take only real values. A sample $E$ with kernel $K_{\text{Lorentz}}$
  is shown in Figure~\ref{f:analytic_signal}. The modulus of the analytic 
  signal $E^+$ visibly plays the role of an envelope for $E$, averaging over 
  local oscillations of $E$.

  \begin{figure}
  \centering
  \includegraphics[width=0.5\textwidth]{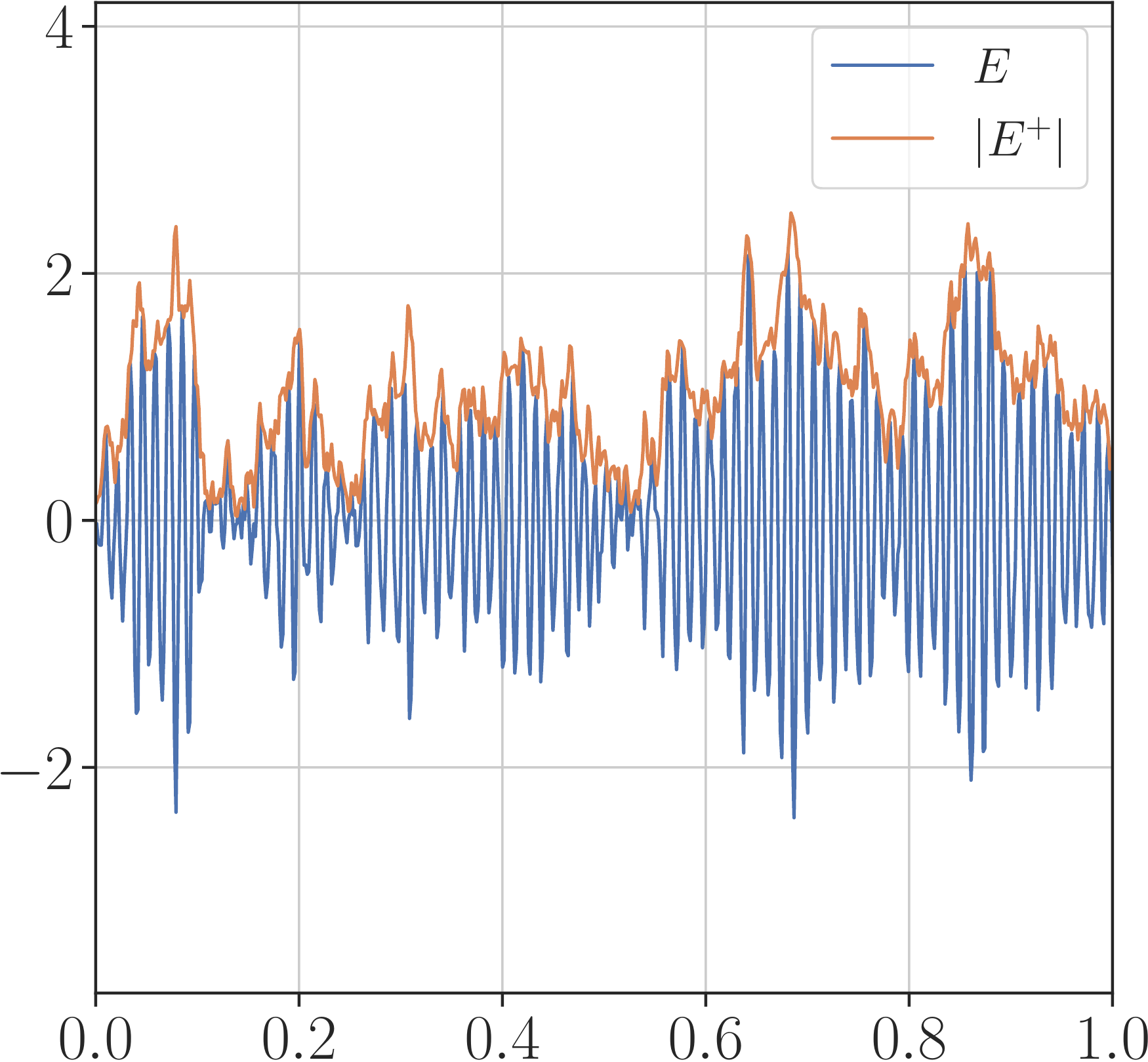}
  \caption{A sample $E$ of the zero-mean Gaussian process with Lorentz 
  kernel \eqref{e:lorentz_kernel}, and the modulus $\vert E^+\vert$ of 
  the corresponding analytic signal.}
  \label{f:analytic_signal}
  \end{figure}

  We now go back to a generic $E$ following a zero-mean Gaussian process 
  with kernel $K$. Following Assumption~\ref{a:modellingCoxProcess}, the 
  detection events now form a Poisson process with random parameter function, a.k.a.
  a Cox process; see Section~\ref{s:poisson}. 
  Its correlation functions read
  $$
  \rho_k(t_1,\dots, t_k) = \mathbb{E} \vert E^+(t_1)\vert^2 \dots \vert E^+(t_k)\vert^2 
  = \mathbb{E} E^+(t_1) \dots E^+(t_k)\\ \overline{E^+(t_1)} \dots \overline{E^+(t_k)}.
  $$
  Now the expectation of a product of Gaussians can be expressed in terms
  of pairwise expectations using a theorem by Isserlis.\footnote{This theorem is 
  also known to physicists as Wick's theorem, by analogy with a similar
  theorem for the quantum average of products of certain operators; 
  see later in \cref{s:wicktheo}.} 
  Further recalling that
  $\mathbb{E}E^+(t)E^+(s)=0$ for all $t\neq s$, we obtain
  \begin{align}
  \rho_k(t_1,\dots, t_k) = \sum_{\sigma\in\frak{S}_k} 
    \prod_{i=1}^k \mathbb{E}E^+(t_i)\overline{E^+(t_{\sigma(i)})}.
  \label{e:isserlis}
  \end{align}
  Recognizing a permanent
  in \eqref{e:isserlis}, we conclude that
  $$
  \rho_k(t_1,\dots, t_k) = \text{per} (( C(t_i,t_j) )),\quad k\geq 1,
  $$
  where $C(t,s) = \mathbb E E^+(t)\overline{E^+(s)}$ is the covariance
  kernel of the Gaussian process $E^+$. 
  The photon detection process is 
  thus a permanental point process, see \cref{s:permanental}, with kernel 
  the covariance kernel of the analytic signal of the electromagnetic field.
  In particular, for a Hermitian and translation-invariant kernel 
  $C(t,s)=C_0(t-s)=\overline{C_0(s-t)}$, we obtain, for all $s,t$ such that $t-s=r$, 
  \begin{equation}
  g(r) = \frac{\rho_2(t,s)}{\rho(t)\rho(s)} 
  = 1 + \vert C_0(r) \vert ^2.
  \label{e:pcf_permanental}
  \end{equation}
  The probability of coincidence of a pair of detection times is thus 
  larger than under a Poisson process with the same first correlation function.
  The pair correlation function $g$ in \cref{f:three_examples:permanental_pcf} 
  is actually \eqref{e:pcf_permanental} with the Lorentz kernel in \eqref{e:analytic_covariance}.
  The corresponding samples exhibit bunching when compared to the Poisson 
  samples of \cref{f:three_examples:poisson_samples}.
\end{example}

\Cref{e:pcf_permanental} provides a remarkably simple mathematical
derivation of photon bunching in terms of correlation functions, due to 
\cite[Chapter 4]{Mac75}. 
It is noteworthy that the representation of 
permanental point processes as Cox processes, 
which has become more of a
side result for probabilists \cite[Proposition 35]{HKPV06}, 
was actually the physical motivation for the introduction of permanental point processes 
by \cite{Mac75}.


\subsection{Determinantal point processes}
\label{s:determinantal}

A point process $\gamma$ is said to be \emph{determinantal} when there exists 
a so-called \emph{correlation kernel} $K:\X\times \X\to \C$ such that the 
correlation functions of $\gamma$ read, for any $k\geq 1$,
\begin{align}
\label{rhoDet}
\rho_k(x_1,\ldots,x_k)=\det \Big[K(x_i,x_j)\Big]_{i,j=1}^k
\,.
\end{align}
We write $\DPP(K,\mu)$ for the point process with correlation functions 
\eqref{rhoDet} with respect to the reference measure $\mu$. Note that, 
by definition of $\rho_k$, the kernel $K(x,y)$ has to be chosen so that the right hand  side of \eqref{rhoDet} is nonnegative for any 
$k\geq 1$ and $x_1,\ldots,x_k\in\X$; like for permanental point processes, 
not every kernel yields a well-defined DPP.

Assuming existence for a moment, we look at the first and second
correlation functions. 
Taking $k=1$ and $k=2$ in \eqref{kCorr}, we 
obtain for instance
\begin{align}
\label{1corr}
\E\left[\sum_{x\in \gamma} f(x)\right]
& =\int f(x) K(x,x)\mu(\d x),\\
\label{2corr}
 \E\left[\sum_{\substack{x,y\in\gamma\\ x\neq y}} f(x, y)\right]
& =\int f(x,y) \Big[K(x,x)K(y,y)-K(x,y)K(y,x)\Big]\mu(\d x)\mu(\d y).
\end{align}
When the kernel $K$ is Hermitian, that is, $K(x,y) = \overline{K(y,x)}$, 
the second correlation function in \eqref{2corr}, once normalized,
becomes:
\begin{align}
  g(x,y) &= \frac{\rho_2(x,y)}{\rho_1(x)\rho_1(y)} 
	= 1 - \frac{\vert K(x,y)\vert^2}{K(x,x)K(y,y)}\leq 1.
  \label{e:antibunching}
\end{align}
In particular, \eqref{e:antibunching} shows that DPPs with Hermitian 
kernels are repulsive: the larger $\vert K(x,y)\vert^2$, the less likely 
the two particles at $x$ and $y$ are to co-occur. Moreover, they are 
always less likely to co-occur than if $\gamma$ were a Poisson process 
with the same first correlation function. 
In particular, if $\gamma$ were to model the clicks 
of a detector as particles arrive, the arrival times would exhibit 
\emph{antibunching}.
We shall see examples of antibunching arrival times in \cref{s:electroncoherence}.
Unlike Poisson and permanental point processes, whose appearance results from \cref{a:modellingCoxProcess}, to understand why DPPs appear in quantum optics, we shall need first to introduce elements of quantum field theory.

Necessary and sufficient conditions for existence of a DPP are known 
when the operator $\mathcal{K}:f\mapsto \int K(\cdot,y)f(y)\d\mu(y)$ on $L^2(\mu)$ is Hermitian and locally trace-class. 
In particular, letting $(\lambda_k)$ denote the eigenvalues of $\mathcal{K}$,
$\DPP(K,\mu)$ is well-defined if and only if 
$0\leq \lambda_k\leq 1$ for any $k\in\N$, namely when $K$ is a 
contraction operator. 
This is now known as the Macchi-Soshnikov 
theorem.

\rb{
  Clarify the discussion on projection DPPs in v2 and the mixture argument.
}
One constructive proof of this existence theorem relies on the following decomposition lemma \citep{HKPV06}. 
Informally, a DPP with kernel
\begin{equation}
  K(x,y)=\sum_{k=0}^\infty \lambda_k \,\phi_k(x)\overline{\phi_k(y)},
  \label{e:spectral_decomposition}
\end{equation}
is a statistical mixture of \emph{projection DPPs}, i.e., DPPs with projection kernels.
More precisely, if $(b_k)_{k\in\N}$ are independent Bernoulli random 
variables with $\P(b_k=1)=1-\P(b_k=0)=\lambda_k$ and 
$K^b(x,y):=\sum_{k}b_k\phi_k(x)\overline{\phi_k(y)}$, then 
$\DPP(K^b,\mu)=\DPP(K,\mu)$ in law; see again \citep{HKPV06}.
In particular, if we restrict ourselves to projection kernels, 
i.e., if $\lambda_k\in\{0,1\}$ for every $k$, the cardinality of 
$\gamma\sim\DPP(K,\mu)$ is thus $\Tr K$ almost surely.
If we further assume $\Tr K=N<\infty$, the corresponding projection 
DPP generates exactly $N$ particles, with joint probability distribution on 
$\X^N$ given by
\begin{equation}
  \frac1{N!}\left|\det\Big[ \phi_{k-1}(x_j)\Big]_{j,k=1}^N\right|^2 \prod_{j=1}^N\mu(\d x_j).
\label{e:density_projection_DPP}
\end{equation}


%% file: qft.tex

To deepen our understanding of the relationship between point processes
and quantum physics, we now introduce a few elements of what is known as many-body quantum physics, quantum statistical physics or quantum field theory.
We first introduce the basic quantum formalism to describe a
single particle, its evolution in time and what information we can gather
on it, mostly following \citep{Fol08}. 
Then, we introduce how to deal with
systems made of many or an indefinite number of particles. 

\subsection{The mathematical framework of quantum theory}
\label{s:quantum_framework}

\paragraph{States.}
The state $\psi$ of a quantum system is represented by a non-zero element of a 
complex Hilbert space~$(\sH,\langle\cdot\vert\cdot\rangle)$, the space 
of all possible states of the quantum system of interest. 
Two elements of $\sH$ that are equal up to multiplication by a (complex) scalar are 
understood to represent the same physical states,\footnote{
	For brevity, 
we will stay informal on that point and ignore the subtleties coming 
from considering a projective Hilbert space rather than $\sH$; 
see \citep[Chapter 3]{Fol08}.
} 
and we henceforth always assume that 
our states are normalized,
 i.e. $\Vert \psi\Vert=\langle \psi\vert\psi\rangle = 1$.
Finally, we will often use the so-called bra-ket notation, denoting a vector 
by $\ket\psi = \psi\in\sH$ and the associated linear form on
 $\sH$ by $\bra\psi:h\mapsto \langle \psi\vert h\rangle$.

\paragraph{Observables.}
Any observable quantity is represented by a self-adjoint operator $A:\sH\to\sH$.
To understand in what sense, take a self-adjoint operator $A:\sH\to\sH$ 
and a vector $\ket\psi\in\sH$, and define an ordinary probability measure
 $\mathbb{P}_{A,\psi}$ on the spectrum $\sigma(A)\subset \mathbb{R}$ of $A$ by
\begin{equation}
\mathbb{P}_{A,\psi}:E\mapsto \expval{\ind_E(A)}{\psi}, \quad E\subset \sigma(A).
\label{e:spectral_calculus}
\end{equation}
The operator $\ind_E(A)$ in \eqref{e:spectral_calculus} is defined by 
the \emph{spectral functional calculus}; see \citep[Chapter VII]{ReSi80}. 
Informally, $\ind_E(A)$ is the 
projection operator with the same eigenvectors as $A$, but replacing 
each eigenvalue $\lambda$ by $1$ if $\lambda\in E$, and $0$ otherwise.
\rb{
	For the v2, we should maybe give a finite-dimensional example first. Maybe actually introduce the qubit here as a running example to illustrate state, observable, etc.
}
One easy setting to understand the definition of $\mathbb{P}_{A,\psi}$ is that of a compact (self-adjoint) operator $A$, for in that case the support $\sigma(A) = \{\lambda_n\}$ of $\mathbb{P}_{A,\psi}$ is discrete, with no other accumulation point than $0$, and there is an orthonomal basis $(\ket{\phi_n})$ of $\sH$ with $A\ket{\phi_n} = \lambda_n\ket{\phi_n}$; see e.g. \citep[Theorem VI.16]{ReSi80}. 
In particular, for $\lambda_n\neq 0$ with multiplicity one, $\mathbb{P}_{A,\psi}(\{\lambda_n\}) = \vert \braket{\psi}{\phi_n}\vert^2$, which indeed sums to $\Vert \psi \Vert^2 = 1$ in $n$.
Without entering into details, we finally note that the definition \eqref{e:spectral_calculus} extends to unbounded operators by representing them as multiplication operators \citep[Theorems VIII.4 and VIII.5]{ReSi80}.
Unbounded operators are necessary to represent physical observables that have an unbounded support, such as the position of a particle on the real line.

Interpreting a self-adjoint operator $A$ as an observable means that the 
experimentally accessible information about the observable $A$ is in 
$\mathbb{P}_{A,\psi}$. 
In particular, if we had a machine that could repeatedly 
prepare unrelated copies of the physical system in question in a 
state $\ket{\psi}$, we would model the measurements of observable 
$A$ on such a sequence of states as independent 
draws from $\mathbb{P}_{A,\psi}$. 
Unless $\ket{\psi}$ is an eigenvector of $A$ with 
eigenvalue $\lambda$, in which case $\mathbb{P}_{A,\psi} = \delta_\lambda$, measuring 
the physical quantity encoded by $A$ thus becomes inherently probabilistic: 
one can only talk about the probability that a measured observable 
will lie in a given $E\subset\mathbb R$. 
Finally, we note that by the law of 
large numbers, the average of a large number of measurements of $A$ 
is expected to be close to the expectation of a random variable with law $\mathbb{P}_{A,\psi}$, which is
\begin{equation}
	\int \lambda\,\d \mathbb{P}_{A,\psi}(\lambda) = \expval{A}{\psi}.
	\label{e:expectation_value}
\end{equation}
Again, the derivation of \eqref{e:expectation_value} is easier for a compact operator $A = \sum_n \lambda_n \ketbra{\phi_n}$
, say even Hilbert-Schmidt, so that $\sum \lambda_n^2 <\infty$. 
In that case, both sides of \eqref{e:expectation_value} are equal to $\sum_i \lambda_n \vert\braket{\psi}{\phi_n}\vert^2<\infty$.
The general treatment of self-adjoint operators requires to carefully define the spectral measure of an operator, and we refer to \citep[Chapters VII \& VIII]{ReSi80}.

\paragraph{Mixed states and Born's rule.}
In practice, most experimental devices are not able to repeatedly 
produce a given state $\ket{\psi}\in\sH$, but rather a noisy version of it.
To model this noise, first associate to each state $\ket{\psi}\in\sH$ the projector $\ketbra{\psi}:\ket{h}\mapsto \braket{\psi}{h} \ket{\psi}$, which we also abusively call a state. 
Now consider a linear operator on $\sH$ defined as
\begin{equation}
	\rho = \mathbb{E}_{\ket\psi\sim p} \ketbra{\psi},
	\label{e:mixed_states}
\end{equation}
with $p$ a probability measure on $\sH$. 
The linear combination \Cref{e:mixed_states} is typically interpreted as being the output of a noisy preparation device, which outputs the state $\ketbra{\psi}$ with $\ket{\psi}$ drawn from $p$. 
When the agent runs its machine many times, they will describe the output by \cref{e:mixed_states}.\footnote{
In other words, a mixed state can be used to describe epistemic uncertainty on the preparation process, while draws from $\mathbb{P}_{A,\psi}$ in \eqref{e:spectral_calculus} represent aleatoric uncertainty. 
Unlike epistemic uncertainty, aleatoric uncertainty cannot be reduced by better knowledge of the system.
}
More abstractly, a \emph{density matrix} $\rho$
is a trace-class positive operator on $\sH$ with unit trace.\footnote{The use of the word \emph{matrix} for an operator can be confusing at first glance, but it is standard here.} 
A density matrix is often called a \emph{mixed state}, as opposed to a \emph{pure state}, which corresponds to a single element of $\sH$.
In other words, pure states correspond to density matrices that are also projectors.

Generally speaking, a density matrix $\rho$ exhausts all the statistical content
that an observer can predict about a system. 
For instance, for $\rho$ defined by \eqref{e:mixed_states},
$$
\tr \left[\rho \ind_E(A)\right] = \mathbb{E} \tr \left[\ketbra{\psi_s} \ind_E(A)\right] 
= \mathbb{E} \bra{\psi_s} \ind_E(A) \ket{\psi_s}, \quad E\subset \mathbb{R}
.
$$
In particular, the map $\mathbb{P}_{A,\rho}:E\mapsto \tr (\rho \ind_E(A))$ is a probability distribution on
$\mathbb R$ that corresponds to a statistical mixture of pure states, and 
simplifies to \eqref{e:spectral_calculus} for pure states. It gives the probability
that a measurement of $A$ will belong to $E$. 
By linearity, the expectation of the random variable\footnote{Physicists talk of the \emph{expectation value} of $A$ in state $\rho$.} describing the measurement of observable $A$ when the system is in the state $\rho$, denoted 
by convention $\langle{A}\rangle_\rho$, is given by \emph{Born's rule}
\begin{equation}
	\langle{A}\rangle_\rho = \tr (\rho A)
\label{e:bornrule}
\,.
\end{equation}
This formula is a cornerstone of the quantum formalism. 
It contains \cref{e:spectral_calculus} as a special case and thus describes
both noisy and noiseless situations.

\paragraph{Conditioning on measurements.}
To define the joint distribution of the measurement of two observables $A$ and $B$ when the physical system is in a (possibly mixed) state $\rho$, one would like to use the operator $AB$ in a spectral formula like \eqref{e:spectral_calculus}. 
It turns out to be very natural to define a joint distribution using the spectral calculus when $A$ and $B$ commute, i.e. $AB=BA$.
In particular, one can then talk of the conditional distribution of the measurement of $B$ on $\rho$ given that we observed $A$ on $\rho$.
Without entering into details, Bayes' formula yields that evaluating the conditional amounts to evaluating $\mathbb{P}_{B, \tilde\rho}$, where 
\begin{equation}
	\tilde\rho = C\rho C^\dagger,
	\label{e:sandwich}
\end{equation} 
and $C$ is related to the measurement of $A$; see \citep{BoVaJa07} for a precise statement that includes monitoring a state across time, i.e., filtering. 
The ``sandwiched" updated state $\tilde\rho$ in \eqref{e:sandwich} is interpreted as the state of the system immediately after the measurement of $A$. 

\paragraph{Incompatible observables and Heisenberg's uncertainty principle.}
When $AB\neq BA$, it is not even guaranteed that $AB$ is self-adjoint, let alone that $A$ and $B$ have common eigenspaces.
Physicists associate this mathematical difficulty to the fact that non-commuting observables are \emph{incompatible}: it is not possible to obtain a joint measurement of both observables using a \emph{single} prepared copy of a state $\rho$.

Another hint that non-commuting observables are peculiar is Heisenberg's celebrated \emph{uncertainty principle}; see \citep[Section 3.3]{Fol08}.
In a nutshell, for two non-commuting observables $A$ and $B$ and a state $\rho$, the product of the standard deviations $\sigma_{A,\rho}$ of $\mathbb{P}_{A,\rho}$ and $\sigma_{B,\rho}$ of $\mathbb{P}_{B,\rho}$ is lower-bounded,
\begin{equation}
	\sigma_{A,\rho}\sigma_{B,\rho} \geq \frac12 \left\vert\tr\left[(AB-BA)\rho\right]\right\vert.
	\label{e:heisenberg}
\end{equation}
Since $A$ and $B$ cannot be measured simultaneously, this does not imply anything on measuring $A$ and $B$ on the same copy of $\rho$, but it is rather a property of the model of the overall experiment, i.e., of the physical system and the probes corresponding to $A$ and $B$.
The concrete consequences of the uncertainty relation \eqref{e:heisenberg} can be seen by repeatedly preparing a copy of the state $\rho$ and measuring either $A$ or $B$. 
If the measure represented by $A$ yields measurements with a small empirical variance, then measuring $B$ on similar repeated copies of $\rho$ will lead to a comparatively large empirical variance.

\paragraph{Schrödinger's equation and Hamiltonians.}
To complete our description of a physical system, we need to model its
evolution in time. In quantum physics, the evolution of the state of the 
system is given by a one-parameter group $\{U(t), t\in\mathbb{R}\}$ 
of unitary operators on $\sH$, that is, $U(t)U(s)=U(t+s)$ and 
$U(t)^{-1} = U(t)^\dagger = U(-t)$. 
In particular $U(0)=\id$ is the identity operator.
After time $t$, the state of a system that was in state $\ket {\psi(0)}$ at 
time $t=0$ is considered to be
\begin{equation}
\ket{\psi(t)}  = U(t)\ket{\psi(0)}
\,.
\label{e:schroedinger}
\end{equation}
Under weak assumptions on $\sH$ and $U(t)$, one can show that 
$U(t)=\exp(-\frac{\i{} t}{\hbar} H)$ in the sense of the spectral 
functional calculus again, for some (possibly unbounded) self-adjoint 
operator $H$ on $\sH$ \citep[Section 3.1]{Fol08}. 
Conversely, any 
choice of self-adjoint $H$ gives a one-parameter unitary group.
The operator $H$, thought as an observable, is called the 
\emph{Hamiltonian} of the system. 
As the generator of the dynamics,
 $H$ is often thought as the energy of the system.
In particular, the group identity of $U$ yields by 
differentiation that $U(t)^\dagger H U(t) = H$ for all $t$, so that the average 
energy $\expval{H}{\psi(t)}$ of the state $\psi(t)$ in 
\eqref{e:schroedinger} is constant over time.

Building a model in quantum physics usually boils down to choosing 
a state space $\sH$, a set of self-adjoint operators as observables, 
and a Hamiltonian to describe the evolution of the system in time.
In particular, when $s\ll 1$, \eqref{e:schroedinger} informally yields
$$
\ket{\psi(t+s)} = \left(\id -\frac{\i s H}{\hbar} + \cO(s^2)\right) \ket{\psi(t)},
$$
so that
\begin{equation}
	\i \hbar\frac{\d}{\d t} \ket{\psi(t)} = H \ket{\psi(t)},
	\label{e:schroedinger_usual}
\end{equation}
which the reader may recognize as the celebrated Schrödinger equation.
This equation of motion can be generalized to mixed states \eqref{e:mixed_states} by linearity.

To finish, note that instead of letting $\ket {\psi(t)}$ depend on time 
through \eqref{e:schroedinger} and keeping observables constant
(the so-called \emph{Schrödinger picture}), we can obtain the same 
measurement probabilities \eqref{e:spectral_calculus} by keeping 
states constant and letting observables vary as $A(t) = U(t)^\dagger A(0) U(t)$ 
(the so-called \emph{Heisenberg picture}). 
The two pictures are thus 
equivalent as to what observations they predict. 
There is a third 
equivalent convention that is commonly used, the 
\emph{interaction picture}. 
The latter is tailored to problems where 
the Hamiltonian 
$$
H = H_0 + H_I
$$
is the sum of a well-studied Hamiltonian $H_0$, typically describing standard 
prior information on a subpart of the physical system
 (say, a \emph{free field}), and an additional term $H_I$, called \emph{interaction Hamiltonian}, typically encoding the interaction between parts of the system.
In the interaction picture, one lets both states and observables evolve
in time. 
Observables evolve according to $A(t) =  U_0(t)^\dagger A(0) U_0(t)$, 
using the one-parameter unitary group $U_0$ associated to Hamiltonian 
$H_0$. 
On the other hand, states evolve as 
$\ket{\psi(t)} = V(t)\ket{\psi(0)}$, where $V(t) = U_0(t)^\dagger U(t)$ and 
$U(t)$ is the group associated to the full Hamiltonian $H$.
While the state evolution now differs from \eqref{e:schroedinger_usual}, the interaction picture
yields again the same probabilities \eqref{e:spectral_calculus}.
Unless otherwise specified, we use the interaction picture in this work.

\paragraph{Perturbation theory.}
What is convenient with the interaction picture is that it singles out
the role of $H_I$ as follows. 
To compute $U(t)$, it is enough to 
compute $V$ and then apply $U(t) = U_0(t)V(t)$, since $U_0(t)$ 
is assumed to be known. 
Now, by definition, $V$ satisfies
\begin{equation}
	\frac{\d}{\d t} V(t) 
= \frac{\i}{\hbar} U_0(t)^\dagger H_0 U(t) -  
		\frac{\i}{\hbar}U_0(t)^\dagger H U(t) 
= \frac{1}{\i \hbar} H_{I}(t)V(t).
\label{e:perturbation}
\end{equation}
Even in seemingly simple physical situations, as when modeling the 
detection of photons in \cref{s:photodetection}, solutions to 
\eqref{e:perturbation} have to be approximated. Writing 
\eqref{e:perturbation} in integral form, and noting that 
$V(0) = \id$, we obtain
\begin{equation}
\label{e:perturbation2}
V(t) = V(0) + \int_{0}^t 	\frac{\d}{\d t} V(\tau)\d \tau 
= \id + \frac{1}{\i \hbar}\int_{0}^t  H_{I}(\tau)V(\tau)\d \tau
\,.
\end{equation}
Plugging \eqref{e:perturbation2} into itself yields
\begin{equation}
\label{e:perturbation3}
V(t) = \id + \frac{1}{\i \hbar}\int_{0}^t  H_{I}(\tau)\d\tau -
	\frac{1}{\hbar^2} \int_{0}^t\d \tau\int_0^\tau\d\tau'  H_{I}(\tau)H_{I}(\tau')V(\tau').
\end{equation}
Iteratively plugging \eqref{e:perturbation2} into \eqref{e:perturbation3}
yields a series representation for $V$, the study of which is called 
\emph{perturbation theory}, and is at the heart of quantum field theory
and its current mathematical difficulties \citep[Chapter 6]{Fol08}.
In practice, physicists will often assume that stopping this iterative 
process early on yields a good approximation to $V$. For instance, 
we shall see in \cref{s:photodetection} a physical assumption that is mathematically interpreted as keeping only the first two terms in the right-hand side of \eqref{e:perturbation3}.

\subsection{Two fundamental one-particle systems}

We saw in Section~\ref{s:quantum_framework} that a quantum model consists in a Hilbert space $\sH$, a Hamiltonian $H$ specifying the dynamics, and self-adjoint operators that describe observables. 
To make things more concrete, we now describe two simple systems: the qubit and the harmonic oscillator.

\subsubsection{The qubit}
\label{s:qubit}

Consider a physical system whose state is
described by one of two labels, say $\{e,g\}$,
like whether a two-level atom is excited or in its ground state.
One would then use $\sH=\mathbb{C}^2$ with the
usual inner product. 
Let now $(\ket{e},\ket{g})$ be
any orthonormal basis of $\sH$, and
$\ket{\psi} = \alpha\ket{e} + \beta\ket{g}$ be a state.
Normalization implies $|\alpha|^2+|\beta|^2=1$.

Observables are described by self-adjoint operators
of $\sH$. For instance, the observable corresponding
to checking whether the system is in state $\ket{e}$
is the projector $\ketbra{e}$.
Following \eqref{e:spectral_calculus}, the probability to obtain
the result $e$ in the state $\ket{\psi}$ is given by
$$
\mathbb{P}_{\ketbra{e},\psi} (\{e\}) = \bra e \left(\ketbra{\psi}\right) \ket e = \vert \braket{e}{\psi}\vert^2 
		= |\alpha|^2 = 1-\mathbb{P}_{\ketbra{e},\psi}(\{g\}) = 1-|\beta|^2.
$$
More generally, every self-adjoint operator on $\mathbb{C}^2$ can be expressed
as a linear combination of four matrices. 
Expressed in the basis $(\ket{e},\ket{g})$, the four so-called \emph{Pauli matrices}
are
\begin{equation}
	\id = \begin{pmatrix} 1 & 0\\ 0 & 1\end{pmatrix},\quad
	\sigma_x = \begin{pmatrix} 0 & 1\\ 1 & 0\end{pmatrix}, \quad
	\sigma_y = \begin{pmatrix} 0 & -\i\\ \i & 0\end{pmatrix}, \quad
	\sigma_z = \begin{pmatrix} 1 & 0\\ 0 & -1\end{pmatrix}.
	\label{e:pauli_matrices}
\end{equation}
All four matrices squared are the identity, so they are all
diagonalizable with spectrum $\{\pm 1\}$. 
For instance,
an orthonormal basis of eigenvectors of $\sigma_x$ is
given by
$$
\ket {\pm_x} = 2^{-1/2}(\ket e \pm \ket g),
$$
which is simply a rotation of the basis $(\ket{e},\ket{g})$.
We can compute for instance
the probability \eqref{e:spectral_calculus}
of observing the outcome $+_x$, when the state is a generic $\ket\psi = \alpha \ket e + \beta \ket g$, namely
\begin{align*}
	\mathbb{P}_{\ketbra{+_x},\psi}(\{+_x\}) 
	 &= \bra\psi \big( \ketbra{+_x} \big) \ket\psi \\
	 &= \vert \braket{\psi}{+_x}\vert^2 \\
	 &=\frac{\vert \alpha+\beta\vert^2}{2} \leq 1.
\end{align*}
Similarly, the average value of the observable $\sigma_x$ is given
by
$$
\bra \psi \sigma_x \ket \psi = \frac{\vert \alpha +\beta\vert^2}{2} -
	\frac{\vert\alpha-\beta\vert^2}{2} = 2\frak{R}(\alpha\beta^*).
$$

Finally, we give the example of a rather common and simple Hamiltonian for the qubit, called \emph{free evolution}.
In the basis we have chosen, it reads
\begin{align}
\Hfqb = \frac{1}{2}\hbar\omega_{eg}\sigma_z \;,
\label{e:free_hamiltonian_qubit}
\end{align}
where $\omega_{eg}$ is the transition frequency between the
excited ($\ket{e}$) and ground ($\ket{g}$) states, and the label $\mathrm{FQ}$ stands for \emph{free qubit}. 
The resulting evolution in time is easy to understand. 
The states $\ket{g}$ and $\ket{e}$ are eigenstates of $\Hfqb$. 
Calling $U_0$ the one-parameter group corresponding to $\Hfqb$, and starting e.g. at state $\ket{g}$ at $t=0$, the probability of finding the evolved state $U_0(t)\ket{g}$ in state $\ket{g}$ is $\expval{U_0(t)}{g} = 1$.
The basis vectors $\ket{e}$ and $\ket{g}$ are thus stationary states. 
Finally, note that the form \eqref{e:free_hamiltonian_qubit} of the free
dynamics is mostly conventional; we could for instance add a term
proportional to the identity to shift the energy reference without
changing the physics. What matters is that the energy difference between
the two states is $\hbar\omega_{eg}$.

Richer physical behaviors are modeled by adding interaction terms to the free Hamiltonian. 
There is no general prescription as to how to write interaction Hamiltonians and they mostly depend on the physical context we wish to describe. 
One simple but useful model is to consider a qubit interacting with an external field $\mathbf{B}(t)$.
Note that we mean here a classical field, in the sense that $\mathbf{B}(t)$ is a vector $(B_x(t), B_y(t), B_z(t))^T$ of functions from $\mathbb{R}$ to $\mathbb{R}$.
Concretely, the model describes a small magnet or the spin of a particle
in an external magnetic field. 
The basic ideas behind nuclear magnetic resonance, and magnetic resonance imaging, come from this choice of interaction Hamiltonian \citep[Chapter IV, Appendix F]{CohenBookI}.
Adding the free part, we obtain the total Hamiltonian
\begin{align}
H = \frac{1}{2}\omega_{eg}\sigma_z + g \boldsymbol{\sigma}\cdot \mathbf{B}(t),
\label{e:ext_hamiltonian_qubit}
\end{align}
where $g \in \mathbb{R}$ is a coupling constant and
$\boldsymbol{\sigma}\cdot \mathbf{B}(t)$ is a shorthand
notation for the linear combination of matrices
$\sigma_x B_x(t) + \sigma_y B_y(t) +\sigma_z B_z(t)$.
When the field is time-independent, the dynamical problem
can be interpreted geometrically. 
Indeed, the Hamiltonian
is of the general form $H = \hbar\omega\mathbf{n}\cdot\boldsymbol{\sigma}$
with $\mathbf{n}$ a unitary vector.
The evolution operator is thus
$$
U(t) = \e^{-\i\omega t \mathbf{n}\cdot\boldsymbol{\sigma}},
$$
and we recognize one way of parametrizing an element of the
rotation group $\text{SU}(2)$, where $\omega t$ is the angle of the
rotation around the axis $\mathbf{n}$. 
The dynamics of a qubit in
this model can be represented as a vector, representing the state
$\ket{\psi}$, which precesses around an axis fixed by the external field; see \citep{PIC} for details.

\subsubsection{The harmonic oscillator}
\label{s:harmonic_oscillator}

To model the position of a particle living in $\mathbb{R}^d$,
the state space is commonly taken to be $\sH=L^2(\mathbb{R}^d)$.
A natural observable is the position of the particle.
We associate\footnote{Choosing what operator to associate to
what classical quantity is a process known as quantization and
is a whole research area; see \citep{Fol08} and references therein.}
the measurement of the $j$th coordinate to the multiplication operator
$X_jf:x\mapsto x_j f(x)$, defined for all functions $f$ of
$\sH=L^2(\mathbb{R}^d)$ such that $x\mapsto x_j f(x)\in L^2(\mathbb{R}^d)$.
Note that the only candidates to be common eigenvectors of
all $X_j$ are delta functions, and are thus not in $L^2$.
When performing computations, however, it is convenient to
consider the tempered distributions
$$
	\bra x := \delta_x, \quad x\in\mathbb{R}^d
$$
as generalized states, and interpret
$\int \delta_a(x)\delta_b(x)\d x = \delta(a-b)$ and
$f = \int f(x)\delta_x(\cdot) \d x$ as making $\ket x$
a ``generalized orthogonal basis".
In particular, for a smooth state $\ket{\psi}\in\sH$, $\braket{x}{\psi} =
\psi(x)$, and the probability distribution \eqref{e:spectral_calculus}
corresponding to $X_j$ in a smooth state $\ket{\psi}$ becomes
$$
	\mathbb{P}_{X_j,\psi}(E_j) = \int_{\mathbb{R}\times \dots \times \mathbb{R} \times E_j \times \mathbb{R} \times \dots\mathbb{R}} \vert\psi (x)\vert^2\d x, 
\quad 
E_j\subset\mathbb{R}
\,.
$$
The reader may recognize here the common interpretation
of the squared modulus $\vert\psi(x)\vert^2$ of the wave
function as the probability density function for the position of
the particle. 

Similar considerations allow for considering
the \emph{momentum} operator $P_j$, defined for $1\leq j\leq d$
through its Fourier transform
$$
	\cF (P_jf)(k) = \hbar k_j \cF (f)(k),
$$
for all $f\in\sH$ such that $k_j \cF( f)(k)\in L^2(\mathbb{R}^d)$.
The ``generalized basis" of tempered distributions $\bra k : f\mapsto \cF( f)(k)$
allows talking about the \emph{momentum representation}
$\braket{k}{\psi}$ of a smooth state $\ket{\psi}$. Furthermore, by the
inverse Fourier transform, we can express the momentum operator
$P_j$ in the position basis as a derivative operator,
$$
P_j = -\i \hbar \partial_j
\,.
$$

A natural Hamiltonian in this setting is the so-called quantum harmonic
oscillator, modelling the movement of a single particle in a quadratic potential.
It writes
\begin{align}
H = \sum_{j=1}^d \frac{1}{2m}P_j^2 + \frac12 m \omega^2 Q_j^2
\,,
\label{e:freeHamiltonianoscillator}
\end{align}
where $m>0$ is called the mass and $\omega\in\mathbb{R}$ the angular
frequency of the oscillator.
For $\psi$ a smooth (Schwartz) function, we get
$$
\bra{x} H \ket \psi =  - \frac{\hbar^2}{2m} \Delta \psi (x)
+ \frac12 m \omega^2 \Vert x\Vert_2^2 \psi(x).
$$
In particular, the eigenequation $\bra{x} H \ket \psi = \lambda \psi(x)$ leads
to a well-known differential equation. 
The solutions are tensor products of Hermite functions \citep[Section 3.4]{Fol08}, and form an orthonormal
basis of $L^2(\mathbb{R}^d)$ of (smooth) eigenfunctions of $H$.

The harmonic oscillator can also be diagonalized algebraically. 
Define the so-called \emph{ladder operators} 
$$
a_j := \sqrt{\frac{m\omega}{2\hbar}}\left( Q_j +\frac{\i}{m\omega}P_j \right) \quad\text{ and }\quad a^\dag_j := \sqrt{\frac{m\omega}{2\hbar}}\left( Q_j -\frac{\i}{m\omega}P_j \right).
$$
They satisfy the commutation relation $a_i a_j^\dagger-a_j^\dagger a_i = \delta_{ij}$. The Hamiltonian can now be rewritten as
\begin{equation}
	H=\hbar\omega \sum_{j=1}^d \left ( a_j^\dagger a_j + \frac12 \right ),
	\label{e:free_hamiltonian_with_Ps_and_Qs}
\end{equation}
and its spectrum is given by that of the number operators $N_j = a_j^\dagger a_j $. Using that the number operator is positive, one can easily show that its eigenvalues are $n_j\in \mathbb N$, with corresponding eigenstates $|n_j\rangle$, i.e. $N_j|n_j\rangle=n_j|n_j\rangle$, called the Fock states \citep[Chapter V]{CohenBookI}. 
The corresponding wavefunctions $\langle x|n_j\rangle$ are the Hermite functions discussed above.

The ladder operators were introduced first by Dirac to solve the harmonic oscillator of a single quantum particle. 
But these operators have taken a life of their own, to describe many-particles quantum states, as we discuss now.

\subsection{Modelling a finite number of particles}

Section~\ref{s:quantum_framework} focused on describing systems made 
of one particle. In this section, we show how to combine such systems 
to describe one made of a (known) finite number of particles.

\subsubsection{Subsystems, bosons and fermions}

Let us now consider a system of $N<\infty$ identical particles.
Concatenating several physical systems is represented by tensor products,
so that one is tempted to consider the Hilbert space
$\mathcal{H}^{\otimes N} = \bigotimes_{i=1}^{N}\mathcal{H}$,
where $\mathcal H$ is the Hilbert space corresponding to a one-particle system.
A generic state would thus be a linear combination of states of the
form
\begin{align}
|\psi\rangle=\bigotimes_{i=1}^{N}|\psi_{i}\rangle
	    =: |\psi_{1}\cdots\psi_{N}\rangle
\,.
\end{align}
However, experiments rather suggest to use a state that encapsulates the property
that the particles are \emph{indistinguishable}, i.e., that permuting
particles leaves the state invariant. Arguments from (projective)
representation theory of the symmetric group \citep[Chapter 1.2]{PIC} lead to two
types of state spaces, at least when $d > 2$,
bosons and fermions.
\emph{Bosons} have symmetric states, i.e., they are represented
by states $\ket{\psi}\in \mathcal{H}^{\otimes N}$ such that
for all permutations  $\sigma\in\frak{S}_N$
$$
\ket{\psi_1,\dots,\psi_N} = U_\sigma\ket{\psi_1,\dots,\psi_N} :=
	\ket{\psi_{\sigma(1)},\dots,\psi_{\sigma(N)}}
\,,
$$
where we have implicitly defined the unitary representation $U_\sigma$ of the permutation group on the Hilbert space $\mathcal{H}^{\otimes N}$.
In other words, the Hilbert space for $N$ bosons is the symmetric
subspace of $\mathcal{H}^{\otimes N}$, namely the range of
the orthogonal projector
$$
\sym = \frac{1}{N!}\sum_{\sigma \in \mathfrak{S}_{N}} U_{\sigma}.
$$
The other type of particles is called \emph{fermion}, and corresponds
to antisymmetric states
$$
\ket{\psi_1,\dots,\psi_N} = \epsilon(\sigma)
	U_\sigma\ket{\psi_1,\dots,\psi_N}, \quad\sigma\in \mathfrak{S}_{N},
$$
where $\epsilon(\sigma)$ is the signature of $\sigma$.
Systems of $N$ fermions are thus represented by states in the
antisymmetric subspace of $\mathcal{H}^{\otimes N}$, namely
in the range of the orthogonal projector
$$
\asym = \frac{1}{N!}\sum_{\sigma \in \mathfrak{S}_{N}} \epsilon
	\left(\sigma\right)U_{\sigma}.
$$

Note that observables that are permutation-invariant leave both the ranges
of $\sym$ and $\asym$ invariant. If the Hamiltonian is permutation-invariant,
bosons thus stay bosons across time, and fermions stay fermions.
One particular case is that of $N$ \emph{free} (a.k.a. \emph{non-interacting}) particles.
Letting $H$ be the Hamiltonian for one particle, the dynamics
of $N$ free particles is given by the Hamiltonian
\begin{equation}
	\label{e:free_hamiltonian_N_particles}
	H^{(N)} = \sum_{n=1}^N \id\otimes\cdots\otimes \id\otimes
	\underbrace{H}_{n\text{-th term}}
	\otimes\id\otimes\cdots\otimes\id.
\end{equation}

\subsubsection{Example $N$-particle states}
\label{s:example_N_particle_states}
Consider the harmonic oscillator setting of \cref{s:harmonic_oscillator}, say with $d=1$ for simplicity.
Since the Hamiltonian is interpreted as measuring the energy of the system, and since the eigenstates of the Hamiltonian $H$ in \eqref{e:freeHamiltonianoscillator} form a basis of $L^2(\mathbb{R})$, the state with minimal energy is the eigenfunction of $H$ with minimal eigenvalue. 
In position representation, it is the first Hermite function
$\phi_0$, that is, a normalized Gaussian function.

Now, if we consider $N$ free bosons, each coming with the same $1$-dimensional harmonic oscillator Hamiltonian, 
the ground state of $H^{(N)}$ in \eqref{e:free_hamiltonian_N_particles} is easily
seen to be, in position representation,
$$
\psi_0^{\rm{Sym}}(x_1,\dots,x_N) = \phi_0(x_1)\dots\phi_0(x_N),
$$
so that, in this ground state, the probability density of having a configuration of $N$ free bosons
at $(x_1,\ldots,x_N)\in\R^{N}$ is given by
$$
|\psi_0^{\rm{Sym}}(x_1,\ldots,x_N)|^2=|\phi_0(x_1)|^2\cdots |\phi_0(x_N)|^2.
$$
This is the law of $N$ independent random variables with identical
distribution $|\phi_0(x)|^2\d x$, i.e. $N$ i.i.d. Gaussians.

Still within the harmonic oscillator setting, but this time for $N$ free
fermions, the situation is less straightforward. 
With similar eigendecomposition arguments, one can show
that the ground state of $H^{(N)}$ on
 $\asym \mathcal{H}^{\otimes N}$ is given by
\eq
\label{Slater}
\psi_0^{\rm{Asym}}(x_1,\ldots,x_N)=\frac1{\sqrt{N!}}\det\left[\begin{matrix}
\phi_0(x_1) & \cdots & \phi_0(x_N)\\
\vdots &  & \vdots\\
\phi_{N-1}(x_1) & \cdots & \phi_{N-1}(x_N)
\end{matrix}\right],
\qe
where $\phi_{k-1}\in L^2(\R^d)$ is the (unit-norm) eigenstate
associated with the $k^{\rm{th}}$ smallest eigenvalue of $H$, i.e. here, the $k$th Hermite function. 
The right hand side of \eqref{Slater}
is known as a Slater determinant. 
In particular, in this ground state, the density
probability of having a configuration at $(x_1,\ldots,x_N)\in\R^{Nd}$
is given by
\begin{align*}
|\psi_0^{\rm{Asym}}(x_1,\ldots,x_N)|^2& =\frac1{N!}\det\left|\left[\begin{matrix}
\phi_0(x_1) & \cdots & \phi_0(x_N)\\
\vdots &  & \vdots\\
\phi_{N-1}(x_1) & \cdots & \phi_{N-1}(x_N)
\end{matrix}\right]\right|^2\\
& =\frac1{N!}\det \Big[K_N(x_i,x_j)\Big]_{i,j=1}^N
\end{align*}
where
$$
K_N(x,y):=\sum_{k=0}^{N-1}\phi_k(x)\overline{\phi_k(y)}.
$$
Since the $\phi_k$'s are orthonormal,
$K_N$ is a projection kernel. 
In other words, the $N$ positions
of the free fermions form a projection DPP on $\X=\R$, as
introduced in \cref{s:determinantal}. 
In our setting where $d=1$ and the $\phi_k$ are Hermite functions, this DPP is known as the \emph{Gaussian unitary ensemble} (GUE); see e.g. \citep{dean2019noninteracting}. 
The GUE is a fundamental example in random matrix theory, as it arises as the eigenvalues of a $(A+A^*)/2$, where $A$ is filled with i.i.d. complex Gaussian random variables \citep{AnGuZe10}.

More generally, any projection DPP can be obtained as the position representation of the ground state of an (adhoc and non-necessarily physically realizable) free Hamiltonian for fermions.
This is our first encounter of DPPs as arising from a fermionic construction.\footnote{
	In Section~\ref{s:dppfermions}, we shall come back to the idealized free fermion construction of this section, introducing a temperature parameter to obtain non-projection DPPs. 
}

\subsubsection{Occupation number representation}
Since the Hilbert space for modeling $N$ bosons,
respectively $N$ fermions, is $\sym \cH^{\otimes N}$, respectively
$\asym \cH^{\otimes N}$, it is useful to have an orthonormal basis
of these spaces \citep[Section 4.5]{Fol08}. 
A convenient way of writing states of indistinguishable particles
is through the \emph{occupation number basis}.

The rationale behind this basis is simply to count the number
of particles that are in a specific basis state of the one-particle
Hilbert space, chosen to make observables simple to express.
Let $\left( e_{i} \right)$ be a Hilbert basis of $\cH$.
Then an orthonormal basis of $\sym\cH^{\otimes N}$ is given by
$$
\ket{n_1,n_2,\dots} := \frac1Z \sym \ket{\underbrace{e_{1},\cdots,e_{1}}_{n_{1}},
		\underbrace{e_{2},\cdots,e_{2}}_{n_{2}},\cdots}, \quad \text{where } \sum_{i\geq 1} n_i = N,
$$
with $Z$ the normalization constant. 
The label of the
state $\ket{n_1,n_2,\dots}$ represents occupation numbers: $n_i$ represents the number of particles in the state $\ket{e_i}$.
By construction, the inner product between two states with different occupation numbers is zero. 
The normalization constant $Z>0$ is fixed by requiring the state to be normalized, i.e., 
$$
1 = \frac{1}{Z^2} \expval{\sym^{\dagger}\sym}{n_1,n_2,\dots} = \frac{1}{Z ^{2}} \frac{1}{N!}\prod_{i=1}^{\infty}n_{i}!
\,.
$$

The same construction applies for fermions. 
An orthonormal basis of $\asym\cH^{\otimes N}$ is given by
\begin{align}
\ket{ n_{1},n_{2},\cdots} = \sqrt{N!} \asym\ket{\underbrace{e_{1},\cdots,e_{1}}_{n_{1}},
		\underbrace{e_{2},\cdots,e_{2}}_{n_{2}},\cdots},
 \quad \text{where } \sum_{i\geq 1} n_i = N
 \,,
\end{align}
with the important difference that for fermions, $n_{i}\in\lbrace 0,1 \rbrace$ for all $i$.
Indeed, if there is some $j$ such that $n_j\geq 2$, then the corresponding state of $\cH^{\otimes N}$ is in the kernel of the antisymmetrization operator $\asym$. 
Physicists identify this property with
\emph{Pauli's exclusion principle}: no two fermions of the same
multi-particle state can be \emph{in the same mode}, i.e., correspond
to the same vector of the basis $e_i$ of the single-particle space $\cH$.

\subsection{Modeling an indefinite number of particles}
\label{s:fock_spaces}

In order to form a state space that can accommodate any
number of particles, and in particular to model uncertainty
in that number, physicists introduce so-called \emph{Fock}
spaces.

\subsubsection{Fock spaces for bosons and fermions}
\label{s:fock}

Denote by $\mathcal{H}^{\otimes 0}\cong\mathbb{C}$
the Hilbert space consisting of all multiples of some unit norm state
labeled as $|0\rangle$, called the \emph{vacuum}, and
representing the absence of any particle.
Taking the convention $\sym h = \asym h = h$ for $h\in \cH^{\otimes 0}$,
we define two Hilbert spaces $\mathcal{H}^{\infty}_{\text{Bosons}}$
and $\mathcal{H}^{\infty}_{\text{Fermions}}$, respectively called 
the bosonic and fermionic Fock spaces, as orthogonal direct
sums
\begin{equation}
\mathcal{H}^{\infty}_{\text{Bosons}} := \bigoplus_{k\geq 0} \sym \cH^{\otimes k},
\quad
\mathcal{H}^{\infty}_{\text{Fermions}} := \bigoplus_{k\geq 0} \asym \cH^{\otimes k}.
\label{e:fock}
\end{equation}
Several remarks are in order.
Note first that in \eqref{e:fock}, by an abuse of notation, we write the 
symmetrization and antisymmetrization operators in any dimension as
$\sym$ and $\asym$, respectively. 
Second, the orthogonal sum in \eqref{e:fock} makes two vectors coming from different summands have inner product zero by definition. 
Third, we underline that the absence of particles is modeled by the vacuum state $\ket 0$, not by the null element of either Hilbert space
$\mathcal{H}^{\infty}_{\text{Bosons}}$ or $\mathcal{H}^{\infty}_{\text{Fermions}}$.
In particular, unlike any null element, the vacuum state represents a 
physical state, and thus has norm $1$ in the Fock space.

The Fock spaces in \eqref{e:fock} are defined as orthogonal sums of 
Hilbert spaces, and are thus Hilbert spaces themselves.
For instance, $\mathcal{H}^{\infty}_{\text{Bosons}}$ is the set of collections $h=(h_k)_{\geq 0}$, where $h_k\in \sym \cH^{\otimes k}$ and $\Vert h\Vert:= \sum_k \Vert h_k\Vert^2< \infty$.
It is sometimes convenient to work instead with the \emph{algebraic sum} 
of the same vector spaces, meaning that all but a finite set of coefficients  are constrained to be zero. This is the case, for instance, when defining the number operator through
\begin{equation}
N \ket{n_1,n_2,\dots} := \left(\sum_i n_i\right)\ket{n_1,n_2,\dots},
\label{e:number_operator_photons}
\end{equation}
which counts the particles in a state, and is defined by the same formula 
for bosons and fermions. $N$ is then uniquely extended to an 
unbounded operator on the Fock space \citep[Section 4.5]{Fol08}.
In this text, we will assume that this extension step can always be 
done unequivocally, and work directly with the (Hilbert) Fock spaces \eqref{e:fock}.
Finally, a state of the form $\ket{n_1,n_2,\dots}$ with $\sum_{i\geq 1} n_i<\infty$
is called a \emph{Fock state}, and together they form an orthonormal
basis of the corresponding Fock space. Any state of a Fock space can thus
be represented as a linear combination of Fock states.

\subsubsection{Creation and annihilation operators}

Besides the number operator, we now introduce two types of operators on
Fock spaces that are respectively thought of as creating and destroying
a particle in a given one-particle state. The choice of the symmetrization 
operator $\sym{}$ or $\asym{}$ has important consequences on the 
commutation relations between these operators, and we thus separate 
the treatment of bosons and fermions.
We paraphrase here \cite[Chapter 4.5]{Fol08}.

\paragraph{Bosons.}
For $v\in \sH$, define the operator $b(v)$ on the algebraic sum corresponding to \eqref{e:fock}, i.e., finite  linear combinations of Fock states, by
$$
b(v) \sym(u_1\otimes \dots \otimes u_k) = \frac1k \sum_{j=1}^k \braket{v}{u_j} 
	\sym (u_1\otimes \dots \otimes u_{j-1} \otimes u_{j+1}\otimes \dots \otimes u_k),
$$
for any $k\geq 1$ and $u_1, \dots, u_k\in\sH$.
Now, for later ease of writing, we renormalize $b$ and define the operator
$a(v)$ on the algebraic sum corresponding to \eqref{e:fock}, i.e., finite 
linear combinations of Fock states, by $ a(v) w = \sqrt{k} b(v) w$ 
whenever $w\in\sym \cH^{\otimes k}$. Alternately, using the number 
operator introduced in \cref{s:fock}, $a(v) = \sqrt{N+I}\, b(v)$.
The operator $a(v)$ is called the annihilation operator in state $v$.
Its adjoint $a^\dagger(v) := a(v)^\dagger$ on the same algebraic sum 
is defined as $a(v)^\dagger = \sym \sqrt{N} b(v)^\dagger$, and is called
the creation operator in state $v$. Together with $a(v)$, they are called 
\emph{ladder operators}.\footnote{
	Note that like the number operator later on, the ladder operators 
are defined only on the dense subset of finite linear combinations of Fock states; see \cite{Fol08} for discussions on their extension.
} 
\footnote{We also note that these ladder operators were inspired by those introduced by Dirac to solve the harmonic oscillator; see Section~\ref{s:harmonic_oscillator}. However, one should keep in mind that they do not act on the same spaces. In the rest of the paper, we only use the ladder operators acting on the Fock space.}

Writing $[u,v] = uv-vu$ for the commutator of two operators, we can
check that the ladder operators satisfy the so-called 
\emph{canonical commutation relations} (CCR),
\begin{align}
	\left[ a(u),a(v) \right] = \left[ a^\dagger(u),a^\dagger(v) \right] = 0,
	  \qquad
	\left[a(u), a^{\dagger}(v) \right] = \braket{u}{v}I.
	\label{e:CCR}
\end{align}	

To see why the creation and annihilation operators bear such names, 
consider again a basis $(e_i)$ of $\sH$, and let $a_i = a(e_i)$ for $i\geq 1$. 
It can be checked that the ladder operators act on Fock states as
\begin{subequations}
\begin{align}
a_{i}^{\dagger}| n_{1},\cdots,n_{i},\cdots\rangle
  &= \sqrt{n_{i}+1}| n_{1},\cdots,n_{i}+1,\cdots\rangle, \\
a_{i}| n_{1},\cdots,n_{i},\cdots\rangle
  &= \sqrt{n_{i}}| n_{1},\cdots,n_{i}-1,\cdots\rangle
\,,
\label{e:a_i_second}
\end{align}
\end{subequations}
where, by convention, the right-hand side of \eqref{e:a_i_second} is the null element of the Fock space whenever $n_i=0$. 
Physically, $a_{i}^{\dagger}$ thus models the
creation of a particle in the state $\ket{e_i}$, while $a_i$
removes a particle from the same state. 
Note also that $a_{i}\ket{0} = 0$ is the null element of the Fock space; we say that the annihilation operator annihilates the vacuum.


In terms of ladder operators, the number operator
\eqref{e:number_operator_photons} can be rewritten
\begin{align}
N = \sum_{i=1}^{\infty}a^{\dagger}_{i}a_{i},
\label{e:number_operator_using_a_and_a_dagger}
\end{align}
with $N\ket{0} = 0\ket{0} = 0$, the null element of the Fock space, 
as expected from an operator that counts particles. 
Note also that the CCRs imply
$
[ N,a^{\dagger}_{i}] = a^{\dagger}_{i}
$ and
$
[ N,a_{i}] = -a_{i}.
$
Finally, any Fock state can be rewritten as the action of creation
operators on the vacuum as
\begin{align}
| n_{1},\cdots,n_{i},\cdots\rangle =
  \frac{1}{\sqrt{\prod_{i=1}^{\infty}n_{i}!}}
  \prod_{i=1}^{\infty}\left(a_{i}^{\dagger} \right)^{n_{i}}|0\rangle
\,.
\label{e:fock_state_using_creation}
\end{align}
Note that by definition of Fock states, the products in 
\eqref{e:fock_state_using_creation} consist of a finite number of terms.

\paragraph{Fermions.}

Following the same lines, we can now define the creation
and annihilation operators for fermions on the Fock space
$\mathcal{H}^{\infty}_{\text{Fermions}}$.
Formally, $a(v)$ is defined on finite linear combinations of 
Fock states\footnote{and then extended to the whole Fock space.} by
$$
a(v) \asym (u_1\otimes \dots \otimes u_k) = \frac{1}{\sqrt{k}} 
	\sum_{j=1}^k (-1)^j \braket{v}{u_j} 
		\asym (u_1\otimes u_{j-1} \otimes u_j \otimes \dots \otimes u_k),
$$
while one can check that its adjoint $a^\dagger(v) := a(v)^\dagger$ satisfies
$$
a^\dagger(v) \asym (u_1\otimes \dots \otimes u_k) = 
	\sqrt{k+1} \asym u_1\otimes \dots \otimes u_k.
$$
As for bosons, these definitions lead to particular commutation relations.
Writing $\{u,v\}= uv+vu$ for the so-called \emph{anti-commutator} of two 
operators, the fermionic annihilation and creation operators satisfy
\begin{align}
\lbrace a(u),a(v) \rbrace = \lbrace a^\dagger(u),a^\dagger(v) \rbrace  = 0
\qquad
\lbrace a(u),a^{\dagger}(v) \rbrace = \braket{u}{v}I,
\label{e:CAR}
\end{align}
known in the literature as the \emph{canonical anti-commutation
relations} (CARs). Here again, the presence of an anti-commutator is
interpreted by physicists as a manifestation of Pauli's exclusion
principle. For instance, the identity 
$\lbrace a^{\dagger}(u),a^{\dagger}(u) \rbrace = 2a^{\dagger}(u) a^{\dagger}(u)=0$ 
translates the fact that there cannot be two fermionic particles in the same
quantum state. 

Again, the ladder operators $a_i = a(e_i)$ in a given basis apply 
straightforwardly to the corresponding Fock states. For $i \geq 1$, and 
remembering that $n_{i} \in \{0,1 \}$, we obtain
\begin{subequations}
\begin{align}
a_{i}^{\dagger}| n_{1},\cdots,n_{i},\cdots\rangle
  &= \delta_{n_{i}0}\epsilon(\sigma)|n_{1},\cdots,n_{i}+1,\cdots\rangle
\\
a_{i}| n_{1},\cdots,n_{i},\cdots\rangle
  &= \delta_{n_{i}1} \epsilon(\sigma)|n_{1},\cdots,n_{i}-1,\cdots\rangle,
\end{align}
\end{subequations}
where $n=\sum n_i<\infty$, $\sigma$ is the permutation
\begin{align}
\sigma =
\left(
\begin{array}{ l l l l l l }
     1  & \cdots & S_{i}+1 & S_{i}+2 & \cdots & n+1  \\
     S_{i}+1 & \cdots  & S_{i} & S_{i}+2 & \cdots & n+1
\end{array}
\right),
\end{align}
where $S_{i}=\sum_{j=1}^{i-1}n_{j}$ 
are the partial sums of the sequence of occupation numbers.

These fermionic ladder operators call for comments again. 
First, the presence of the delta symbols ensures that in order 
to create a fermion in a given 
mode, i.e. in a state described by one of the basis vectors, this 
mode must be empty; similarly, destroying a fermion in a mode 
requires that mode to be initially occupied by one particle.
This is a natural implementation of the exclusion principle which 
requires that a mode can only be either empty or occupied by one 
and only one particle.

Finally, we note that the fermionic number operator can
be written as $N = \sum_{i=1}^\infty a_i^{\dagger}a_i$, and that the
following commutators are identical to their bosonic counterparts, 
\begin{align}
\label{e:commutation_N_a}
\left[ N,a^{\dagger}_{i} \right] =
  a^{\dagger}_{i} \qquad \left[ N,a_{i} \right] = -a_{i}.
\end{align}

\subsubsection{Modes}
\label{s:modes}
As seen in Section~\ref{s:fock_spaces}, a Fock space $\cH^\infty$ is built starting from a single-particle Hilbert state $\cH$ of arbitrary dimension, of which we single out a basis $(e_k)$.  
Basis vectors of the single-particle Hilbert space $\cH$ are typically called \emph{modes} in physics. 
Fock states $\ket{n_1, n_2, \dots}$ form a basis of the Fock space, where the notation stands for the (anti-)symmetrization of the tensor product of $n_1$ times the basis vector (mode) $e_1$, $n_2$ times the basis vector $e_2$, etc.

There is a natural isomorphism of Hilbert spaces between the Fock space $\cH^\infty$ built on $\cH$ and the tensor product $\bigotimes_k (\mathbb{C}e_k)^\infty$ of the Fock spaces built on each of the one-dimensional (or ``single-mode") Hilbert spaces $\mathbb{C}e_k$. 
Indeed, taking bosons as an example, one simply needs
to map the Fock state $\ket{n_1, n_2, \dots}\in\cH^\infty$ to 
$$
\bigotimes_k \sym (\underbrace{e_k \otimes \dots \otimes e_k}_{n_k\text{ times}}),
$$
where $\sym$ is the symmetrization operator, and should be replaced by $\asym$ for fermions.

Thinking of the Fock space as a product of Fock spaces across modes is often implicit in physics texts.\footnote{
	Arguably, one already has this isomorphism in mind when writing $\ket{n_1, n_2, \dots}$ for a Fock state.
} 
It is, for instance, customary to introduce sophisticated states assuming a single mode, i.e., that $\cH$ is one-dimensional, and then to write tensor products of such states across modes to cover the case of an $\cH$ of arbitrary dimension.
A concrete example will be bosonic coherent states in \cref{s:coherent_states}.

\subsubsection{Field operators}

There are often several natural bases for a physical situation, for 
instance a natural basis to describe a source of radiation, and a 
natural basis to describe a measurement. 
The linearity of the definition
of ladder operators yields easy ``change of basis formulas'' connecting
ladder operators in two different bases.
Formally, let $\left( |u_{i}\rangle \right)_{i\in I}$ 
and $\left( |v_{i}\rangle \right)_{i\in I}$ be two bases of the single-particle Hilbert space $\mathcal{H}$.
The creation and annihilation operators transform exactly 
as a regular change of basis, i.e.,
\begin{equation}
a^{\dagger}(v_{k})
  = \sum_{i}a^{\dagger}(u_{i})\langle u_{i}|v_{k} \rangle \quad\text{ and }\quad
a(v_{k})
  = \sum_{i}a(u_{i})\langle v_{k}|u_{i} \rangle.
\label{e:change_of_basis}
\end{equation}
Note that we implicitly assumed that the two bases were actual Hilbert bases(i.e., countable), but the formula naturally extends in a weak sense to 
``generalized orthogonal bases", such as the position basis for the 
harmonic oscillator in \cref{s:harmonic_oscillator}.
The creation and annihilation \emph{position field operators}
are precisely the creation operator and the annihilation operator 
corresponding to the position basis, and are usually denoted 
by $\psi^{\dagger}\left( x \right)$ and $\psi\left( x \right)$, respectively.
Their physical interpretation is that the operator $\psi^{\dagger}\left( x \right)$
creates a particle of a certain type at the position $x$.
They are called (position) field operators 
because they are the building blocks for position-dependent operators,
which we shall see are the mathematical description of quantum physical fields.
As ladder operators, the field operators also satisfy the commutation 
relations corresponding to the particles being described, say for bosons
\begin{align}
\left[
  \psi( x ),\psi( y )
\right]=0
\,,
\qquad
\left[
  \psi^{\dagger}( x ),
  \psi^{\dagger}( y )
\right]=0
\,,
\qquad
\left[
  \psi( x ),
  \psi^{\dagger}( y )
\right]=\delta(x-y)
\,.
\end{align}
To illustrate a generalized case of the change of basis in \eqref{e:change_of_basis}, consider
the other fundamental basis for the harmonic oscillator, the momentum
basis $\ket{k}$. Since, in an appropriate sense 
$
\braket{k}{x} =
  \frac{\e^{-\i k \cdot x}}{\left( 2\pi \right)^{d}}
$,
\cref{e:change_of_basis} then yields the Fourier transform relationship 
between position and momentum field operators,
\begin{equation}
\label{Op_champs_impul}
\psi^{\dagger}( x )
  = \int a^{\dagger}(k)
     \e^{-\i k\cdot x}
	\frac{\d k}{\left( 2\pi \right)^{d}}
\quad\text{ and }\quad
\psi( x )=
   \int a(k) \e^{\i k\cdot x}
     \frac{\d k}{\left( 2\pi \right)^{d}}
\,,
\end{equation}
with $a^{\dagger}(k)$ the creation operator corresponding to the
momentum basis, creating a particle of momentum $k$.
Once again, the important algebraic relations are the
commutation relations written in the momentum basis, say for bosons,
\begin{align}
\left[ a(k),a(\ell) \right]=0
\,,
\qquad
\left[ a^{\dagger}(k),a^{\dagger}(\ell) \right]=0
\,,
\qquad
\left[ a(k),a^{\dagger}(\ell) \right]=
\left( 2\pi \right)^{3}\delta_{k\ell}
\,.
\end{align}
Note that the whole section translates to fermions, with the 
anticommutation relations \eqref{e:CAR} replacing the commutation 
relations of bosons.

Finally, we make a point on notation. 
While it is customary in physics to use $\psi(x)$ for the field operator of bosons and fermions,
to avoid confusion, we write $a(x)$ for a position \emph{bosonic} field operator, and reserve $\psi(x)$ for a position \emph{fermionic} field operator.

\subsubsection{Observables}

Ladder operators allow one to define observables that
model clusters of interacting particles. 
Depending on the number of 
particles interacting, we call the corresponding operators \emph{one-particle} 
observables, \emph{two-particle} observables, etc.

\paragraph{One-particle observables.}
An example of one-particle observable is the number operator $N$ in 
\eqref{e:number_operator_using_a_and_a_dagger}. 

A generic one-particle operator $O^{(1)}$ is an operator that takes as in input a 
one-particle state $\ket{u_i}$ and gives as an output $\ket{u_j}$, in a
given basis $(u_i)$ of the one-particle Hilbert space $\cH$.
This means that we destroy the particle in the state $\ket{u_i}$ and 
create one in the state $\ket{u_j}$. The operator $O^{(1)}$ can actually 
be rewritten \citep[Section XV.B]{CohenBookIII} as
\begin{equation}
O^{(1)} = \sum_{ij} O_{ij}^{(1)} a^\dagger(u_j) a(u_i)
\,.
\label{e:one_particle_observable}
\end{equation}
The number operator $N$ is diagonal, in the sense that the only nonzero
terms in the sum \eqref{e:one_particle_observable} are those for which 
$i=j$. A typical non-diagonal one-particle observable correspond to a hopping between states.
For example, if $i,j$ are labels for sites on a lattice, then $\sum_{ij} J_{ij} a^\dagger_i a_j$ with $J_{ij}$ a symmetric matrix implements the hopping of particles from $i$ to $j$ with amplitude $J_{ij}$.

\paragraph{Two-particle observables.}
Two-particle observables are used to describe the interaction between
two particles. The intuition is that an interaction takes two particles in a
state $\ket{u_k} \otimes \ket{u_l}$ and yields a new two-particle state 
$\ket{u_i}\otimes \ket{u_j}$. 
Acting on the Fock space, a generic two-body operator $O^{(2)}$ can be written using ladder operators as 
\begin{align}
O^{(2)} = \sum_{ijkl}
		O_{ijkl}^{(2)}
		a^{\dagger}(u_i) a^{\dagger}(u_j)a(u_l) a(u_k)
\,;
\end{align}
see \citep[Chapter XV, Section C]{CohenBookIII} for manipulations using multi-particle observables.
A simple Hamiltonian featuring one- and two-particle observables is the Bose-Hubbard model \citep{Fisher1989},
\begin{equation}
	H = \sum_i \hbar\omega_i \, N_i + \sum_{i,j} J_{ij} \, a^\dagger_ia_j + \sum_i U_i \, N_i(N_i + 1),
	\label{e:example_H}
\end{equation}
where $(\omega_i,J_{ij}, U_i)$ are real numbers that parametrize the model, and $N_i = a_i^\dagger a_i$ counts the particles in mode $i$.
The Hamiltonian \eqref{e:example_H} is composed of both one-particle and two-particle contributions. 
The two-particle term in $N_i (N_{i}+1)$ models the extra energetic cost to pay to add a particle on a site $i$ when some are already present. 

\subsection{Wick's theorem}
\label{s:wicktheo}

Wick's theorem is a cornerstone of quantum field theory and the backbone
of perturbation theory. It gives rise to the famous Feynman diagrams, 
and importantly for us, it yields the permanents and determinants in 
the correlation functions of fields, later to be turned into correlation 
functions of point processes when we introduce physical detection in 
\cref{s:photoncoherence} and \cref{s:electroncoherence}.

In essence, Wick's theorem is a generalization of the calculation of 
the moments of multivariate Gaussian distributions to the case of 
Gaussian density matrices.  We follow the derivation from 
\citep[Appendix C XVI]{CohenBookIII}.

\subsubsection{Gaussian density matrices}

A commonly used mixed state on the Fock space $\sH^\infty$ of either 
fermions or bosons corresponds to the so-called 
\emph{Gaussian density matrices}, defined, when it exists, as the unit-trace, self-adjoint operator
\begin{equation}
\rho = \frac{1}{Z} e^{-\sum_{i,j}
	a_i^\dagger M_{ij} a_j},
\label{e:gaussian_density}
\end{equation}
with $M_{ij} = \overline{M_{ji}}$.
As in \cref{s:quantum_framework}, the definition \eqref{e:gaussian_density} relies on the spectral calculus: $\rho$ is the operator with the same eigenvectors as the argument of the exponential, but with the exponential applied to the corresponding eigenvalues, followed by division by $Z$.
The normalization constant
$$
Z=\tr \e^{-\sum_{i,j}
a_i^\dagger M_{ij}a_j}
$$ 
is called the
\emph{partition function} in statistical physics.
The use of of the word \emph{Gaussian} in the name is sometimes confusing to non-physicists. 
It is motivated by the quadratic expression in the exponential, and the fact that, as we shall see below, the moments of products of creation and annihilation operator under these density matrices behave similarly to the moments of multidimensional Gaussian distributions. 
Finally, we note that there are conditions on the operator appearing in the exponential in \eqref{e:gaussian_density} for $\rho$ to be a proper mixed state, in particular to guarantee that $Z<\infty$. 
These conditions are best discussed for any particular $M$.


\begin{example}[Grand canonical ensemble]
	Consider a system described by a Hamiltonian $H$ acting on a Fock space of either bosons or fermions, with the Hamiltonian being quadratic in the ladder operators. 
	In other words, we require that the Hamiltonian is a one-particle observable \eqref{e:one_particle_observable}, like the Bose-Hubbard Hamiltonian \eqref{e:example_H} with $U_i=0$. 
	Physicists usually consider additional fluctuations in energy and number of particles, and describe the system at thermal equilibrium by the so-called \emph{grand canonical ensemble}.
	Formally, this amounts to considering the Gaussian density matrix 
	\begin{equation}
	\rho_\text{GC}=\frac{\e^{-\beta ( H -\zeta  N)}}{Z_\text{GC}}
	\,
	\label{e:grandcanonical}
	\end{equation}
	where $ N=\sum_i  a^\dagger_i  a_i$ is the number operator, $\beta>0$ is the inverse temperature and $\zeta>0$ is the chemical potential, which can be adjusted to change the average number of particles in the system. 
	The normalization 
	$Z_\text{GC} = \tr \e^{-\beta ( H -\zeta  N)}$ is the so-called \emph{grand-canonical
	partition function}.
\end{example}


As with any state $\rho$, a natural question is to compute the 
expectation $\langle A\rangle_\rho = \tr (\rho A)$ of an observable $A$; see 
\cref{s:quantum_framework}. 
It turns out that for Gaussian density 
matrices, this is a simple mechanical computation using canonical (anti)commutation relations, as soon as $A$ is 
a product of linear combinations of annihilation and creation operators.
The result of this computation is precisely Wick's theorem, a close parent to Isserlis' theorem in classical statistics on the computation of moments of a Gaussian distribution.
Up to a change of basis, and because we only care about linear combinations of ladder operators, it is enough to treat the case of 
\begin{equation}
	\rho = \frac{1}{Z} e^{-\sum_{i}
		\nu_i a_i^\dagger a_i}.
	\label{e:gaussian_density_diagonal}
\end{equation}
In the rest of this section, we henceforth assume $\rho$ to be in diagonal form \eqref{e:gaussian_density_diagonal}.
	
\subsubsection{Where permanents and determinants appear}

For brevity, every average in this section is implicitly meant as under the Gaussian density $\rho$ in \eqref{e:gaussian_density_diagonal}, and we thus write $\langle\cdot\rangle$ instead of $\langle\cdot\rangle_\rho$. 
Additionally, to treat both bosons and fermions in a single theorem, we define
the generalized commutator $[ u, v]_\eta := uv - \eta vu$, where $\eta=\pm1$. 
We note that both the CCR \eqref{e:CCR} and the CAR \eqref{e:CAR} can be rewritten as
\begin{align}
	\label{e:generalized_CRs}
\left[ a_{i},a_{j} \right]_\eta = 0,
  \quad
\left[ a^{\dagger}_{i},a^{\dagger}_{j} \right]_\eta = 0,
  \quad \text{and } \quad
\left[a_{i},a^{\dagger}_{j} \right]_\eta = \delta_{ij},
\end{align}
with $\eta=1$ for CCR and $\eta=-1$ for CAR. 

We note right away that the density matrix \eqref{e:gaussian_density_diagonal} conserves the number of particles, in the sense that $[\rho,N]=0$, with $N=\sum_i a^\dagger_i a_i$. 
This leads to the following important lemma.

\begin{lemma}[Creation and annihilation numbers must match]
	\label{lemma:wick}
	Consider the Gaussian density matrix $\rho$ given by \eqref{e:gaussian_density_diagonal}, acting on either the bosonic or fermionic Fock space, and an operator $O_{n_c,n_a}$ composed of the product, in any order, of $n_c$ creation operators $a^\dagger_{i_1}, \dots,a^\dagger_{i_{n_c}}$ and $n_a$ annihilation operators $a_{i_1}, \dots,a_{i_{n_a}}$. 
	Then $\langle  O_{n_c,n_a}\rangle=0$ unless $n_c=n_a$.
\end{lemma}
Note that this implies that if the total number of creation and annihilation operator is odd, the average necessarily vanishes.

\begin{proof}
The proof is simply based on the fact, to be shown below, that $[N,O_{n_c,n_a}]=(n_c-n_a)O_{n_c,n_a}$. 
Then it follows that $(n_c-n_a)\langle  O_{n_c,n_a}\rangle=\langle  [N,O_{n_c,n_a}]\rangle=\tr (\rho N O_{n_c,n_a}-\rho O_{n_c,n_a}N )=0$ by the cyclicity of the trace and $[\rho,N]=0$.

To prove $[N,O_{n_c,n_a}]=(n_c-n_a)O_{n_c,n_a}$, we proceed by recurrence. 
It is true for $\max(n_c,n_a)\leq 1$ by the commutation relations of the creation and annihilation operators with the number operator \eqref{e:commutation_N_a}. 
Now assume that it is true up to some $n_c,n_a$, and write $O_{n_c,n_a+1}=O_{m_c,m_a}a_iO_{p_c,p_a}$ with $m_c+p_c=n_c$ and $m_a+p_a=n_a$. Then
\begin{equation}
\begin{split}
[N,O_{n_c,n_a+1}] =& [N,O_{m_c,m_a}]a_iO_{p_c,p_a}+O_{m_c,m_a}[N,a_i]O_{p_c,p_a}+O_{m_c,m_a}a_i[N,O_{p_c,p_a}],\\
=&(m_c-m_a)O_{n_c,n_a+1}-O_{n_c,n_a+1}+(p_c-p_a)O_{n_c,n_a+1},\\
=& (n_c-(n_a+1))O_{n_c,n_a+1}.
\end{split}
\end{equation}
The case $O_{n_c+1,n_a}$ is similar.
\end{proof}

Now, for an even integer $N=2k$, define a \emph{contraction} of order $N$ as a permutation $\sigma\in\frak{S}_N$ such that $\sigma(1) < \sigma(3) < ... < \sigma(2k-1)$, and $\sigma(2i-1)<\sigma(2i)$ for $i=1, \dots, k$. 
For instance, there are three contractions of order $4$, namely 
\begin{equation}
	\label{e:contractions_of_order_4}
	\left(
\begin{array}{ l l l l }
     1  & 2  & 3 & 4 \\
     1  & 2  & 3 & 4
\end{array}
\right),
\left(
	\begin{array}{ l l l l }
		 1  & 2  & 3 & 4 \\
		 1  & 3  & 2 & 4
	\end{array}
	\right), \text{ and }
	\left(
\begin{array}{ l l l l }
     1  & 2  & 3 & 4 \\
     1  & 4  & 2 & 3
\end{array}
\right).	
\end{equation}
In words, contractions are built as follows. 
Starting from $Z = \{1, \dots, N\}$ and $i_1=1$, pair $i_1$ with an arbitrary $i_2\in Z\setminus\{i_1\}$. Then select the smallest element $i_3$ of $Z\setminus \{i_1,i_2\}$ and pair it with any element of $Z\setminus\{i_1,i_2,i_3\}$. 
Repeat the procedure until all integers in $Z$ have been paired. 
The corresponding permutation $\sigma$ is the one such that $(\sigma(1),\sigma(2),\ldots,\sigma(N))=(i_1,i_2,\ldots,i_N)$.
This constructive definition shows that there are actually $(N-1)!! = 1 \times 3 \times 5 \dots \times (N-3) \times (N-1)$ contractions of order $N$.

\begin{theorem}[Wick's theorem]
	\label{th:wick}
	Consider the Gaussian density matrix $\rho$ given by \eqref{e:gaussian_density_diagonal} and $N=2k$ linear combinations $b_1, \dots, b_N$ of creation and annihilation operators, in any basis.
	Then
	\begin{equation}
	\langle  b_1\ldots  b_N\rangle =
		\sum_{\sigma \text{ contraction}} \eta^{\epsilon(\sigma)}
			\langle  b_{\sigma(1)} b_{\sigma(2)}\rangle
			\langle  b_{\sigma(3)} b_{\sigma(4)}\rangle
			\ldots
			\langle  b_{\sigma(N-1)} b_{\sigma(N)}\rangle
	,
	\label{e:main_wick}
	\end{equation}
	where $\eta=+1$ for bosons and $\eta=-1$ for fermions.
	The sum is over contractions of order $N$, with parity $\epsilon(\sigma)$. 
	
\end{theorem}

Before proving Wick's theorem for the sake of completeness, we give two illustrative examples. 
First, when $N=4$, remembering the three contractions \eqref{e:contractions_of_order_4} of order $4$, we have
\begin{equation}
\expval{b_1b_2b_3 b_4} =
	\expval{b_1b_2}
	\expval{b_3 b_4}
	+\eta \langle  b_1b_3\rangle\langle b_2 b_4\rangle+\langle  b_1b_4\rangle\langle b_2 b_3\rangle
.
\label{e:simple_wick}
\end{equation}
Second, Wick's theorem is the mathematical reason why permanents and determinants will appear when we consider point processes of detection times in Sections~\ref{s:photoncoherence}, \ref{s:electroncoherence}, and more abstractly in Section~\ref{s:dppfermions}.
This application is so fundamental to our paper that we highlight it here. 

\begin{example}[Wick's theorem for coherence functions]
	\label{ex:wick_for_correlation_functions}
	We are interested in the expected number of $k$-uplets of particles simultaneously appearing in (distinct) modes $1$ to $k$. 
	Since the operator $N_i = a_i^\dagger a_i$ counts the number of particles in mode $i$, we aim to compute
	$$
	\expval{N_1\dots N_k} = \expval{a_1^\dagger a_1 \dots a_k^\dagger a_k}.
	$$
	Note that all $N_i$s commute, so that the order is irrelevant.
	We now use the (anti-)commutation relations \eqref{e:generalized_CRs} to bring all creation operators to the front,
	$$
	\expval{N_1\dots N_k} = \eta^{1+2+\dots+(k-1)}\expval{a_1^\dagger a_2^\dagger \dots a_k^\dagger a_1 a_2 \dots a_k}.
	$$
	By convention, we also put the annihilation operators in decreasing order, which removes the sign, 
	$$
	\expval{N_1\dots N_k} = \expval{a_1^\dagger a_2^\dagger \dots a_k^\dagger a_k a_{k-1} \dots a_1}.
	$$
	Now, we are ready to apply \cref{th:wick}. 
	Upong noting, thanks to \cref{lemma:wick}, that pairing two creation or two annihilation operators results in a zero average, the only non-zero terms in \eqref{e:main_wick} result from permutations $\sigma\in\frak{S}_{2k}$ of the form 
	$$
	\sigma = \left(
	\begin{array}{ l l l l l l l l }
		1  & 2  & 3 & 4 & 5 & \cdots & 2k-1 & 2k  \\
		1 & \nu(k+1)  & 2 & \nu(k+2) & 3 & \cdots & k+1 & \nu(2k)
	\end{array}
	\right),
	$$
	where $\nu$ is a permutation of $\{k+1, \dots, 2k\}$.
	Composing $\sigma$ with $2\times (1+2+\dots+(k-1))$ transpositions, we obtain
	$$
	\left(
	\begin{array}{ l l l l l l }
		1 & \cdots & k & k+1 & \cdots & 2k  \\
		1 & \cdots & k & \nu(k+1) & \cdots & \nu(2k)
	\end{array}
	\right),
	$$
	so that $\epsilon(\sigma) = \epsilon(\nu)$.
	In particular, Wick's theorem yields 
	$$
	\expval{N_1\dots N_k} = \sum_{\nu\in\frak{S}_k} \eta^{\epsilon(\nu)} \prod_{i=1}^k \expval{a_i^\dagger a_{\nu(i)}},
	$$
	where we recognize the permanent or determinant of the matrix $(\langle a_i^\dagger a_j\rangle)_{1\leq i,j \leq k}$, depending on $\eta=\pm1$.
\end{example}

We now prove Wick's theorem for the sake of completeness.
\begin{proof}


We want to compute the expectation $\tr(O_N \rho )$, where
$ O_N= b_1 b_2\ldots  b_N$, 
$$
b_i=\sum_\alpha \left(A_{i,\alpha} a_\alpha^\dagger+B_{i,\alpha} a_\alpha\right),
\quad\text{ and }\quad
\rho =e^{-\sum_\alpha \nu_\alpha  a^\dagger_\alpha  a_\alpha}/Z,
$$
where we have expanded the $b_i$'s in the basis in which the density matrix is diagonal.

Let us start with two general remarks.
First, by Lemma \ref{lemma:wick}, we can restrict the number of $b_i$'s to be even, since otherwise the average vanishes.
We henceforth assume that $N=2k$ is even.
%
We are going to repeatedly use the generalized commutation relations. 
In particular, we note that, for any $\eta$ and
$i,j$, $[ b_i, b_j]_\eta$ is a multiple of the identity and
thus commutes with all operators.

We are now ready to compute
$\langle O_N \rangle =  
\langle b_1\dots b_N\rangle = \tr(b_1\dots b_N \rho)$. 
The general idea is $(i)$ to push $ b_1$
through all the other operators $b_i$ using the (anti)commutation relations.
Then, $(ii)$ using the cyclicity of the trace, we bring $b_1$ back in front
of the rest. 
And finally, $(iii)$ using that $\rho$ is Gaussian to recover the original average $\langle  O_N\rangle$ up to a constant.

Starting this programme, we have
\begin{align*}
\langle O_N \rangle &= [ b_1, b_2]_\eta\langle  b_3\ldots  b_N\rangle+
	\eta\langle  b_2 b_1\ldots  b_N\rangle,\\
&= [ b_1, b_2]_\eta\langle  b_3\ldots  b_N\rangle+
	\eta[ b_1, b_3]_\eta\langle  b_2 b_4\ldots  b_N\rangle+
	\eta^2\langle  b_2 b_3 b_1\ldots  b_N\rangle.
\end{align*}
Iterating the same steps, we write
\begin{align}
\langle  O_N\rangle &= 
	\eta^{N-1}\langle  b_2 b_3\ldots  b_N b_1\rangle+
	\sum_{j=2}^N[ b_1, b_j]_\eta \eta^{j-2}\langle  b_2
		\ldots b_{j-1}b_{j+1}\ldots  b_N\rangle.
\label{e:wickcommutation}
\end{align}
Now, by the cyclicity of the trace, we have
\begin{align}
\langle  b_2 b_3\ldots  b_N b_1\rangle
	&=\tr\left(\rho  b_2 b_3\ldots  b_N b_1\right),\\
	&=\tr\left( b_1\rho  b_2 b_3\ldots  b_N\right)
\,.
\end{align}

Let us for a moment assume that $ b_1$ is either a specific
creation or annihilation operator, i.e. $ b_1= a_\alpha^{(\dagger)}$
for some specific $\alpha$. 
Now, using the explicit Gaussian form
of the density matrix, we have $ b_1 \rho=\lambda \rho b_1 $, with
$\lambda=\e^{-\nu_\alpha}/Z$ if $ b_1=  a_\alpha$, and
$\lambda=\e^{\nu_\alpha}/Z$ if $ b_1=  a_\alpha^\dagger$,
as can be seen by checking the action of these operators on the
basis of Fock states. We thus come to
\begin{equation}
\langle  b_2 b_3\ldots  b_N b_1\rangle =
	\lambda \langle  b_1\ldots  b_N\rangle
\,.
\end{equation}
Since $\eta=\pm1$, \cref{e:wickcommutation} is simplified into:
\begin{equation}
\langle  b_1\ldots  b_N\rangle=
	\sum_{j=2}^N\eta^{j}\frac{[ b_1, b_j]_\eta}{1-\eta\lambda}
	\langle  b_2\ldots b_{j-1}b_{j+1}\ldots  b_N\rangle.
\label{e:wick1}
\end{equation}
In particular, note that \eqref{e:wick1} with $N=2$ becomes
\begin{equation}
\langle  b_1 b_j\rangle=\frac{[ b_1, b_j]_\eta}{1-\eta\lambda}.
\end{equation}
Plugging this into \eqref{e:wick1}, we conclude that
\begin{equation}
\langle  b_1\ldots  b_N\rangle=
	\sum_{j=2}^N\eta^{j}
	\langle  b_1 b_j\rangle \langle  b_2\ldots b_{j-1}b_{j+1}\ldots  b_N\rangle.
\label{e:tool_wick}
\end{equation}
Since this last expression is explicitly linear in $ b_1$, it is
actually valid for any linear combination of creation and
annihilation operators. 
In particular, \eqref{e:tool_wick} replaces
the calculation of an average of a product of $N$ operators by the
weighted sum of $N-1$ averages of products of $N-2$ operators.
Repeating the procedure for each product of $N-2$ operators in
\eqref{e:tool_wick}, we obtain Wick's theorem.
\end{proof}

This is the simplest form of Wick's theorem. It can be further generalized
to Hamiltonians with in addition the ``anomalous'' quadratic terms
$\sum_{nm} A_{nm} a_n a_m+\sum_{nm} \bar A_{nm} a^\dagger_n
 a^\dagger_m$, which arise in mean-field theories of
 interacting Bose gases and superconductors, and generate $\langle a_i a_j\rangle \neq 0$; or to the case of Hamiltonians
 with linear terms, which generate nonzero $\langle a_i\rangle$ \citep[Chapter 3]{Berezin1966}.
 
\begin{remark}
		The proof of Wick's theorem also holds for the vacuum, and the vacuum is also usually said to be a Gaussian state. The vacuum can also be obtained as a limiting state when the temperature $1/\beta$ is taken to zero and $\zeta=0$; see the computations in \cref{s:dppfermions}.
\end{remark}

 Finally, note that the definition of a Gaussian state is not universal and may be community-dependent. 
Quantum opticians, for instance, might define a Gaussian state as one that has a Gaussian Wigner transform; see Section~\ref{s:photoncoherence}.
 
\subsection{Bosonic coherent states model classical fields}
\label{s:coherent_states}
Many quantum systems have a classical (i.e., non-quantum) description, and physicists also
sometimes start from a classical description to build a quantum theory, a procedure known as quantization. 
But in the end, a physical system is fundamentally quantum and we should try to understand how
a classical behavior emerges from a quantum description. 
This phenomenon is known as \emph{decoherence}, and is a vast modern research programme. 
An important part of the answer is to
build quantum states that behave as closely as possible to classical ones. 
\cite{Glauber-1963} made seminal contributions in that regard, defining \emph{coherent states} for systems of photons.\footnote{The name coherent state comes from
the concept of coherence in optics.}
We follow the introduction of bosonic coherent states in \citep[Chapter 11]{MaWo95}, before pointing out the strong connections with the subfield of signal processing called time-frequency analysis. 
We defer the more subtle discussion on fermionic coherent states to \cref{s:recovering_classical_currents}.

\subsubsection{Definition and properties}

As discussed around \cref{e:sandwich}, measuring an observable is modelled by sandwiching the state operator. 
After measurement, the quantum state is thus different from the initial state. 
This is very different from classical physics, where it is possible, in principle, to perform \emph{passive} measurements, e.g. looking at a screen, or recording the intensity of the electromagnetic field.
Consider henceforth the boson Fock space built using the modes of the harmonic oscillator from Section~\ref{s:harmonic_oscillator}.
If we are to build a state of this Fock space that behaves classically, we would intuitively like to take a state that remains unchanged when we measure it. 
Since most measurements involve absorbing bosons (think photons in optics; see
Section~\ref{s:photodetection}), good candidates for classical-like states are the left eigenstates of the annihilation operator, or equivalently the right eigenstates of the creation operator.

To make things concrete, consider a single mode for simplicity, that is, a Fock space built on a one-dimensional $\cH = \mathbb{C}e_1$; see Section~\ref{s:modes}.
Denote the ladder operators by $a = a(e_1)$ and $a^\dagger = a^\dagger(e_1)$. 
Our candidate classical-like states are $\ket{\alpha}\in \cH^\infty_\text{Bosons}$ such that
\begin{align}
 \label{e:coherentstatedef}
	a\ket{\alpha} = \alpha \ket{\alpha}, \quad \alpha\in\mathbb{C}
\,.
\end{align}
Informally,\footnote{For details, see the Lindblad equation in \citep{BoVaJa07}.} the sandwiched state \eqref{e:sandwich} that appears when modeling absorption mechanisms is then $a \ketbra{\alpha} a^\dagger = \vert\alpha\vert^2 \ketbra{\alpha}$, so that coherent states are stable under absorption of a photon.

Solving the eigenstate equation \eqref{e:coherentstatedef} for $\ket \alpha=\sum_n c_n \ket n$
expressed in the Fock basis, we write
$$
\sum_{n\geq 1} c_n \sqrt{n} \ket{n-1} = \sum_{n\geq 0} c_n \ket n,
$$
so that $c_n = \alpha c_{n-1}/\sqrt{n}$ for $n\geq 1$.
This leads to
\begin{equation}
\ket \alpha = e^{-\vert \alpha\vert^2/2} \sum_{n\geq 0} \frac{\alpha^n}{\sqrt{n!}}\ket n
\,,
\label{e:coherent_state}
\end{equation}
up to a complex number of unit modulus, where we have determined
the modulus of the normalizing constant by imposing $\braket \alpha = 1$.
The states \eqref{e:coherent_state}, labeled by nonzero complex
numbers $\alpha$, are called \emph{canonical coherent states}.
By construction, they are mixed states formed as an infinite linear combination
of Fock states. 
If we were to measure the number of photons in
$\ket \alpha$, we would obtain $n$ with probability
$\vert\braket{\alpha}{n}\vert^2 \propto \vert\alpha\vert^{2n}/n!$, and we recognize the Poisson distribution with mean $\vert \alpha\vert^2$.
In particular, the Poisson distribution is in a sense the maximum entropy distribution for independent counts \citep{Har01}: for a fixed mean $\lambda>0$, the maximal entropy of a sum of $k$ independent Bernoulli variables is increasing with $k$, and converges to the entropy of the Poisson variable with mean $\lambda$. 
In that limited sense, one can think of the coherent states as states with a maximally uncertain number of particles.

An equivalent point of view on coherent states, which sometimes serves as their definition in mathematical physics \citep{AlAnGa00}, 
is that they are naturally associated to a projective representation of the translation group in phase space,\footnote{
	The name \emph{phase space} comes from the connection through \eqref{e:free_hamiltonian_with_Ps_and_Qs} to the classical phase space with position-momentum coordinates.
} 
the complex plane indexed by $\alpha$.
To see this, start from the description of Fock states $\ket{n} = (a^\dagger)^n/\sqrt{n!}\ket{0}$.
The coherent state \eqref{e:coherent_state} is thus obtained from the vacuum by the action of an operator $D(\alpha)$ on the boson Fock space called a \emph{displacement operator},
\begin{align}
 \ket{\alpha} = D(\alpha)\ket{0}, \quad D(\alpha) = \e^{\alpha a^{\dagger} - \alpha^{*} a}\,.
 \label{e:displacementop}
\end{align}
This formula is a consequence of the CCRs and the so-called Baker–Campbell–Hausdorff formula expanding the exponential of a sum of (non-commuting) operators.
The same formula yields properties of $D$ like
\begin{align}
\label{e:displacement_operator_action_on_a}
 D^{-1}(\alpha)\,a\,D(\alpha) &= a + \alpha\,
\\
D(\alpha)\, D(\beta) &=\e^{\i\Im{(\alpha\,\beta^*)}}D(\alpha+\beta)
 \,.
\end{align}
These relations are central to the theory of coherent states. 
First, they show that $D$ is a projective representation of
the translation group in phase-space. 
Second, they naturally generalize to other groups, yielding coherent states for different systems than the harmonic oscillator; see e.g. spin coherent states. 
In fact, coherent states have been defined for any locally compact Lie group \citep{Perelomov-1972,Arecchi-1972,
Zhang-1990}. 

Another way to justify that coherent states are almost classical is through Heisenberg's uncertainty principle \eqref{e:heisenberg}. 
Remember that for the harmonic oscillator, position and momentum operators $X, P$ are incompatible observables, and that both can be expressed as sums of ladder operators; see the discussion around \cref{e:free_hamiltonian_with_Ps_and_Qs}.
With the notation of Section~\ref{s:quantum_framework}, one can check that 
the product of $\sigma_{X,\ketbra{\alpha}}\sigma_{P,\ketbra{\alpha}}$ is minimal among states. 
One says that coherent states \emph{saturate} Heisenberg's uncertainty relations.

Another quasi-classical aspect of coherent states is their dynamical evolution.
Since the Fock states are eigenvectors of the Hamiltonian of a harmonic oscillator $H = \hbar\omega a^\dagger a$, $U(t) = \e^{-\i tH}$ applied to $\alpha$ decomposed on the Fock basis \eqref{e:coherent_state} yields
\begin{align}
	\ket{\alpha(t)} = \ket{\alpha\e^{-\i\omega t}}
	\,.
\end{align}
In phase space, i.e., the complex plane parametrized by $\alpha$, the coherent state parameter $\alpha(t)$ simply rotates over time, with an angular frequency $\omega$.
This is precisely the phase-space behavior of a classical harmonic oscillator. 

Finally, we mention two more properties of coherent states. 
First, they are an overcomplete family of the Fock space, with a reconstruction formula
\begin{align}
	\frac{1}{\pi}\int_{\mathbb{C}} \ket{\alpha}\!\!\bra{\alpha} 
		\,\d\alpha = \id
	\,.
\end{align}
The fact that they are not an orthonormal basis is further seen from the overlap
\begin{align}
	\langle \beta | \alpha \rangle =
	\exp\left( \alpha^*\beta - \frac{1}{2}|\alpha|^2 - \frac{1}{2}|\beta|^2\right), \quad \alpha, \beta\in\mathbb{C}.
\end{align}
A good reference for such properties of coherent states of the harmonic oscillator is \citep{Fol89}, or, actually, books on harmonic analysis applied to time-frequency signal processing \citep{Gro01}, on which we say a few words in the next section.

Finally, we have built here bosonic coherent states assuming a single-mode Fock state. 
Multi-mode coherent states are naturally obtained as tensor products of single-mode coherent states, see Section~\ref{s:modes}, and one typically writes them
$$ \ket{\balpha} = \ket{\alpha_1} \otimes \ket{\alpha_2} \otimes \dots$$

\subsubsection{The Husimi distribution and time-frequency analysis}

Single-mode bosonic coherent states of the harmonic oscillator are intimately linked
to the subfield of signal processing called time-frequency analysis; 
see e.g. \citep{Fla98, Gro01}. Consider e.g. the short-time Fourier 
transform $V_g:L^2(\mathbb{R})\rightarrow L^2(\mathbb{R}^2)$, defined as 
$$
V_g(f):t, \omega \mapsto \int f(\tau)\overline{g(\tau-t)}\e^{-2\i\pi \omega \tau}\d \tau\,,
$$
where $g(t)=2^{1/4} e^{-\pi t^2}$ is a unit-norm Gaussian window.
Intuitively, if $f$ is a signal, say representing an audio recording, then 
$\vert V_g(f)(t, \omega)\vert$ will be large whenever frequency $\omega$
is present at time $t$.  In other words, $V_g(f)$ is akin to a musical score. 
Time-frequency transforms such as $V_g$ are thoroughly used in signal
processing, for tasks such as detection or estimation of signal corrupted
with noise \citep{Fla98}. Mathematically, $V_g$ is a unitary linear 
operator, which can be inverted \citep{Gro01}. Furthermore, it is tightly 
linked to decomposing a state into coherent states. Indeed, up to a 
non-vanishing factor, 
\begin{equation}
	V_g(f)(t,-\omega) \propto \braket{\alpha}{\phi}, 
	\label{e:link_stft_husimi}
\end{equation}
where $\alpha = t+\i \omega$ indexes a coherent state 
\eqref{e:coherent_state}, and the state $\phi$ is described in the Fock 
basis by $\braket{\phi}{n} = \int f h_n^* \d t$, where $h_n$ is the 
$n$th Hermite function, a special basis of $L^2(\mathbb{R})$ that 
we already met in \cref{s:harmonic_oscillator}.
Physicists call the squared modulus of the right-hand side of \cref{e:link_stft_husimi} the 
Husimi distribution of $\phi$, and think of it as a phase space 
representation of the state $\phi$, in pretty much the same way signal 
processers think of the \emph{spectrogram} $t, \omega\mapsto \vert V_g(f)(t,\omega)\vert^2$ as a time-frequency representation of the signal $f$.

As a final note and to introduce another related point process, 
one can prove that, in a suitable sense, the STFT $V_g$ of white 
Gaussian noise is equal to the so-called \emph{planar Gaussian analytic 
function}, up to an nonvanishing term  \citep{BaFlCh18,BaHa19}. 
Similarly, if one had access to a white noise-like state $\ket\xi$ such 
that $\braket{\xi}{n}\sim \mathcal{N}_\mathbb{C}(0,1)$ are i.i.d. 
complex unit Gaussians, then its coherent-state ``decomposition" $\alpha\mapsto\braket{\xi}{\alpha}$ 
would be the so-called \emph{planar Gaussian analytic function}, 
up to a nonvanishing term again. This planar Gaussian analytic function,
along with the point process of its zeros, plays a role in the analysis 
of chaotic dynamical systems in statistical physics \citep{Non13}, 
where it is also called the \emph{chaotic analytic function} \citep{Han98}. 
The zeros of the planar Gaussian analytic function are not a DPP.
Yet, being the zeros of a random \emph{smooth} function, they have a repulsive behaviour. 
They actually share many properties with the Ginibre ensemble, itself a 
fundamental DPP \citep{HKPV09}.

%% file: photodetection.tex
\rb{For a v2, explain better the correspondence with point processes: Markov models, counting statistics, etc.? Role of modelling the detector?}
To obtain signals from a physical situation, the picture to have in mind consists in the processes of \emph{emission}, \emph{propagation},
and \emph{detection} of radiation. 
The first two processes are usually described by physicists using one or several sources, classical or quantum, and a field theory, classical or quantum. 
Questions that arise regard, for instance, the dynamics of the source(s),
how they generate radiation, and the physical properties of this
radiation. 
The third process, detection, lies at the interface of physics and signal processing. 
In this section, we discuss simple standard models for the sources, the field, and the detection for bosons, having in mind photodetection. 
Following the footsteps of \cite{Mac75}, our goal is to arrive at the description of the measurement of arrival times at a simple detector as a point process.
We shall see how some of the point processes introduced in Section~\ref{s:point_processes} naturally appear from Glauber's coherent state decompositions. 
As a guiding thread, we will comment on the HBT effect in its different guises.

In \cref{s:photodetection}, we decribe simple models for sources, fields, and detectors, and arrive at the marginal probability of detecting an event at a given time.
In \cref{s:correlation_photodetection}, we examine the correlation between more than one detection events. 
The central objects that encapsulate information about detection times are the coherence functions. 
In \cref{s:identifying_point_processes}, we examine how to turn the coherence functions of a physical detection setup into the correlation functions of a point process. 
We discuss special particular cases, and recover the permanental point processes already announced using a semi-classical treatment in \cref{s:permanental}.
Finally, in \cref{s:single_photon_sources}, we show that single-atom sources can yield anti-bunching detection events.
This is a warning that bosons should not be identified with bunching particles in general, but that the properties of the source should be mentioned.

\subsection{Modeling photodetection events}
\label{s:photodetection}

Photomultipliers are experimental devices that turn incoming radiation
into a measurable electric current. 
The theory of photodetection aims at understanding what kind of signals can be observed as the output of a photomultiplicator. 
As we shall see, these signals are directly related to the so-called coherence functions\footnote{Often called \emph{correlation functions} in physics, but we refrain from using \emph{correlation} here to avoid confusion with the concept of \emph{correlation function} of a point process; see Section~\ref{s:point_processes}.} 
of the field, quantum or classical.

\subsubsection{Modeling the radiation}

Before discussing how we introduce coherence functions, we need to
specify how we describe the radiation we want to probe, using
the objects of quantum field theory introduced in Section~\ref{s:qft}.
Note that, while the framework below is enough to describe
photodetection, a complete description would
require to go into the details of the construction of (relativistic)
quantum field theories and especially its dynamical content 
\citep{Cohen-Atom}.

The electromagnetic field is described by a set of quantum fields 
satisfying some commutation relations. 
For photodetection, the dominant contribution comes from the 
\emph{electric} field at the detection device.
As a reminder, a field is a 
quantity that depends on both position and time.
A \emph{bosonic quantum field} is a collection of operators associated to an
indefinite number of bosonic particles, called e.g. photons, on a Fock space
$\cH^\infty_\text{Bosons}$ (see \cref{s:fock_spaces}). 
Formally,
the state of the field is built by acting on the vacuum $\ket{0}$
with creation and annihilation operators 
$a^\dagger_{\mathbf{k},\boldsymbol{\epsilon}}$ and 
$a_{\mathbf{k},\boldsymbol{\epsilon}}$
\citep[Chapter 10]{MaWo95}. 
The quantities $\mathbf{k}$ and $\boldsymbol{\epsilon}$ index modes of the field (i.e., solutions to Maxwell's equations), and typically correspond to the momentum and the polarization of the photon.

For simplicity, and because it already contains the ingredients 
that relate the point processes of Section~\ref{s:point_processes} 
to photodetection, we consider a fixed linear polarization, and henceforth
drop the index $\boldsymbol{\epsilon}$. 
This corresponds to an assumption we made in the 
semi-classical treatment of Section~\ref{s:permanental}. 
The electric field is then described  by a single (instead of one per 
space coordinate) time- and space-dependent operator $E(\mathbf{r},t)$, 
acting on the Fock space $\cH^\infty_\text{Bosons}$.

To obtain the explicit form of the electric field operator, the stategy
is to consider the field as enclosed in a (large) box and perform 
the so-called \emph{canonical quantization} of Maxwell's equations. 
We refer the reader to the literature for such details, e.g. 
\citep[Chapter 10]{MaWo95}, and consider as given the resulting 
form of the field, namely
\begin{align}
E(\mathbf{r},t) = 
	\sum_{\mathbf{k}} \mathcal{N}_\mathbf{k}
	\left[
		a_{\mathbf{k}}
			\e^{\i (\mathbf{k}\cdot \mathbf{r} - \omega_{\mathbf{k}} t)} +
		a_{\mathbf{k}}^{\dagger}
			\e^{-\i (\mathbf{k}\cdot \mathbf{r} - \omega_{\mathbf{k}} t)}
	\right]
\,, 
\label{e:field_operator}
\end{align}
where $\mathcal{N}_\mathbf{k}$ is a normalization coefficient, $\mathbf{r}\in\mathbb{R}^d$ 
indexes space and $t\in\mathbb{R}$ indexes time. The time dependence in 
\eqref{e:field_operator} is a consequence of working in the interaction 
(or equivalently, so far, Heisenberg) picture. The explicit plane-wave
form corresponds to the evolution implied by the free electromagnetic Hamiltonian
\begin{align}
\label{e:free_hamiltonian_field}
\Hff = \hbar \sum_{\mathbf k} \omega_{\mathbf{k}}
	a_{\mathbf{k}}^{\dagger}a_{\mathbf{k}}
\,.
\end{align}

\subsubsection{Modeling the detector: the first-order coherence function}
\label{s:first_order_coherence}

The usual image to have in mind for a detector is a two-level atom like
the qubit described in Section~\ref{s:qubit}, described
quantum-mechanically with an eigenfrequency $\omega_{eg}$.
The atom interacts with the electric field $E(\mathbf{r},t)$.
In mathematical terms, the Hilbert space of the detector
is finite-dimensional, isomorphic to $\mathbb{C}^2$; see Section~\ref{s:qubit}.
Since the Hilbert space of the electric field is $\Hp$, the joint system
of the detector and the field is $\mathbb{C}^2\otimes \Hp$.

To describe the time evolution of the system, we use the interaction
picture and decompose the Hamiltonian as a free and an interaction 
part
\begin{equation}
H = H_0 + H_I
\,,
\label{e:hamiltonian_photodetection}
\end{equation}
see Section~\ref{s:quantum_framework}.
The free part is taken to be $H_0 = \Hfqb \otimes \ind + \ind \otimes \Hff$,
a combination of the free Hamiltonians of a qubit and the field, respectively
defined in \eqref{e:free_hamiltonian_qubit} and \eqref{e:free_hamiltonian_field}.
To describe the interaction Hamiltonian, let first $q\in\mathbb{R}$
model the modulus of the dipolar moment multiplied by the
charge of an electron. 
Once gain, Maxwell's equations hint that the interaction can be taken as
\citep{Cohen-Atom}
\begin{equation}
H_I(t) = - q\, \sigma_x(t) \otimes E(\mathbf{r}, t),
\label{eq/onedethamil}
\end{equation}
where
$$
\sigma_x(t) = \Ufqb(t)\sigma_x(t) \Ufqb(-t) =
	\Ufqb(t)\begin{pmatrix} 0 & 1 \\ 1 & 0\end{pmatrix}\Ufqb(-t),
$$
is the observable $\sigma_x$ from Section~\ref{s:qubit}, evolved through
time using the group $\Ufqb(t)$ corresponding to the free part of
\eqref{e:hamiltonian_photodetection}, as befits the interaction picture.\footnote{
Physicists will recognize here an instance of the more traditional
form $H_I(t) = -\mathbf{d}(t) \cdot \mathbf{E}(\mathbf{r},t)$, where 
we assumed the dipolar moment $\mathbf{d}(t)$ to be aligned with the
polarization $\mathbf{u}$ of the field $\mathbf{E}(\mathbf{r},t) = 
E(\mathbf r, t)\mathbf u$.
}

We want to compute the probability $p_{\omega_{eg}}(\mathbf{r},t)$
of a detector at position $\mathbf{r}$ to be in its excited state
$\ket{e}$ after a time $t$, knowing that the initial state of the detector is $\ket{g}$ and the field is any arbitrary initial state, but irrespective of the final state of the field.
We assume that the initially prepared state of the detector
and the field is $\ket{g} \otimes \ket{i} = \ket{g,i}$, where $\ket{g}$ is
the ground state of the detector and $\ket{i}$ is an arbitrary initial state of the field.
The corresponding density matrix is thus the projector $\rho(0) = \ketbra{g,i}$.
In the interaction picture described in Section~\ref{s:quantum_framework},
the state evolves until time $t$ through
the action of the evolution operator $V(t)=U_0^\dag(t)U(t)$ as
$\rho(t) = V(t) \rho(0) V^\dagger(t)$.
Measuring the detector in its excited state corresponds to the observable
\begin{equation}
	\ketbra{e}\otimes \id = 
	\ketbra{e} \otimes\sum_f \ketbra{f} = 
	\sum_f \ketbra{e,f}
\,,
\label{e:observable}
\end{equation}
where $f$ indexes Fock states, and we decomposed the identity operator
onto the Fock states, $\id = \sum_f \ketbra{f}$. Now the observable 
\eqref{e:observable}, in the interaction picture, evolves as
$$ 
U_0^\dag(t) \sum_f \ketbra{e,f} U_0(t).
$$
We are now ready to compute the probability $p_{\omega_{eg}}(\mathbf{r},t)$
of the detector being measured in its excited state at time $t$.
By Born's rule (see Section~\ref{s:quantum_framework}), this probability is the
average value of the evolved observable onto the evolved state $\rho(t)$,
namely
\begin{subequations}
\begin{align}
p_{\omega_{eg}}(\mathbf{r},t) 
&= \tr\left( U_0^\dag(t) \left[\sum_f \ketbra{e,f}\right] U_0(t) V(t)\ketbra{g,i} V^\dag(t) \right)\\
&= \sum_f \tr\left( \ketbra{e,f} U(t)\ketbra{g,i} U(-t) \right) \\
&= \sum_f \left\vert \bra{e,f} U(t) \ket{g,i} \right\vert^2 \\
&= \sum_f \left\vert \bra{e,f} V(t) \ket{g,i} \right\vert^2, 
\label{e:probability_of_detection}
\end{align}
\end{subequations}
where the last line comes from the definition $V(t)=U_0^\dag(t)U(t)$ and the fact
that $\ket{e,f}$ is an eigenvector of $H_0(t)$, so that applying $U_0(t)$
to $\ket{e,f}$ simply multiplies by a complex number of modulus $1$.
Note also that the dependence in $\mathbf{r}$ is hidden in $V(t)$.



Up until now, all the discussion is exact. We now introduce a physical 
assumption that allows for a simple expression of $V(t)$.

\begin{assumption}[Weak Coupling]
	\label{a:weak_coupling}
The interaction between the detector and the field is assumed to
be weak, allowing us to treat the dynamics using perturbation theory.
Formally, this means that we assume the existence of a series expansion
of the evolution operator in terms of the interaction Hamiltonian; 
see the discussion around \eqref{e:perturbation3}.
To first order, this amounts to
\begin{align}
	\label{e:weak_coupling}
	V(t)
	\simeq
	\id - \frac{\i}{\hbar} \int_0^t H_I(t') \d t'
\,.
\end{align}
\end{assumption}
Using \eqref{e:weak_coupling} and
$\langle \e |\sigma_x(t)|g \rangle = \e^{\i \omega_{eg} t}$, 
we find
\begin{align}
	\langle e,f|V(t)|g,i \rangle
	=
	\frac{\i q}{\hbar}
	\int_0^t
		\e^{\i \omega_{eg} t'}
		\langle f|E(\mathbf{r},t')| i \rangle
	\d t'
\,.
\end{align}
Plugging this into \eqref{e:probability_of_detection}, and remembering
that the Fock states $\ket{f}$ form a basis of the Fock space, we obtain
\begin{align}
	p_{\omega_{eg}}(\mathbf{r},t)
	&=
	\left(\frac{q}{\hbar}\right)^2
	\int_{[0,t]^2}
		\e^{\i \omega_{eg} (t'-t'')}
		\langle i | E(\mathbf{r},t'') E(\mathbf{r},t') | i \rangle
		\,
	\d t' \d t''
\,.
\end{align}
This equation is straightforwardly generalized to the case where
the initial state of the field is not a pure state but a mixed state, 
described by a density matrix $\rho= \sum_{i} p_i \ket{i}\!\!\bra{i}$
in some basis $(\ket{i})$ of initial states; see Section~\ref{s:quantum_framework}.
The probability of a detection
event between $0$ and $t$ then becomes
\begin{align}
	p_{\omega_{eg}}(\mathbf{r},t)
	&=
	\left(\frac{q}{\hbar}\right)^2
	\int_{[0,t]^2}
		\e^{\i \omega_{eg} (t'-t'')}
		\langle E(\mathbf{r},t'') E(\mathbf{r},t') \rangle_{\rho}
		\,
	\d t' \d t'' \;.
\,
\label{e:probability_of_detection}
\end{align}
Equation~\eqref{e:probability_of_detection}
underlies most of photodetection theory. It relates properties of the field, 
encoded in a \emph{coherence function} 
$\langle E(\mathbf{r},t'') E(\mathbf{r},t') \rangle_{\rho}$, to the excitation 
probability of the detector.

The second important approximation usually made in photodetection
theory is the \emph{rotating wave approximation} (RWA), which
translates the intuition that a proper photodetection event is an absorption
of an excitation. To motivate and illustrate it, we consider first a 
simple example where the field is assumed to always
be in a coherent state.

\begin{example}[Monochromatic classical wave]
\label{ex:monochromatic}
We assume here that the field always remains in a coherent state 
as time evolves and is factorized with respect to the state of the qubit; 
see Section~\ref{s:coherent_states}. Since the free part of the 
Hamiltonian preserves coherent states up to a phase, what we are 
assuming is that the evolution under the interaction 
Hamiltonian approximately  preserves coherent states. 
Physically, this assumption corresponds to, for instance, a macroscopic source 
emitting a coherent state or a statistical mixture thereof, like a common 
lightbulb: the fact that one observes the emitted light does
not change its classical character. We also assume, for simplicity, 
that we have an incoming monochromatic coherent state at
frequency $\omega$. 
Formally, we reduce the sum in 
\eqref{e:field_operator} to a single term $\mathbf{k}$ such that 
$\omega_\mathbf{k}=\omega$, and denote by $\mathcal{E}(t)
=\mathcal{E}(\mathbf{r},t) $ the average value of this field at 
the fixed position $\mathbf{r}$ of the detector.
The terms in $\mathbf{r}$ can be omitted by changing the time origin. 
More explicitly, we set
$\mathcal{E}(t) = \mathcal{E} \sin\omega t$. 
We want a coherent state $\ket{\alpha}$ such that the average
value of the electric field operator $$
\expval{E(t)}{\alpha} = \expval{a_{\mathbf{k}}
			\e^{-\i\omega t} +
		a_{\mathbf{k}}^{\dagger}
			\e^{\i\omega t}
}{\alpha} = \cN_{\mathbf{k}} (\alpha\e^{-\i\omega t} + \alpha^* \e^{\i \omega t}) 
$$ 
is equal to $\mathcal{E}(t)$.
We thus choose the coherent state parameter $\alpha$ to be $-\mathcal{E}/2\i\cN_{\mathbf{k}}$.\footnote{
	The normalization constant $\mathcal{N}_k$ is reinserted
	into the coherent state parameter in this derivation to simplify the
	notation.
}
Furthermore, assume that we initially prepare the detector in its 
ground state $\ket{g}$. In the interaction picture, our approximation that the state of the field remains the same coherent state\footnote{Strictly speaking, we should allow $\alpha$ to depend on time, but the argument below does not change.} and factorized allows to write
\begin{equation}
	V(t) \ket{g} \otimes \ket{\alpha}
	\approx (V_{\text{sc}}(t)\ket{g}) \otimes \ket{\alpha}\;.
\label{e:evolution_operator_sc}	
\end{equation}

The operator $V_{\text{sc}}(t)$ in \eqref{e:evolution_operator_sc} is
defined as the interaction picture evolution operator of the so-called 
\emph{semi-classical} Hamiltonian 
$H_{\text{sc}}(t) = -q \mathcal{E}(t) \, \sigma_x(t)$.
Using the weak coupling approximation to evaluate
the time evolution, we obtain 
\begin{subequations}
\begin{align}
p_{\omega_{eg}}(\mathbf{r},t) &=
\tr\left(  
	U_0^\dagger(t)
	\big[\ketbra{e}\otimes \id\big] U_0(t) \,
	V(t) \big[ 
	\ketbra{g} \otimes \ketbra {\alpha} \big]
	V^\dagger(t)
\right)
\\
&=
\tr\left(  
	\big[\ketbra{e}\otimes \id\big] \,
	V(t) \big[ 
	\ketbra{g} \otimes \ketbra {\alpha} \big]
	V^\dagger(t)
\right)\\
&=
\tr\big(
V_{\text{sc}}(t)\ketbra{g}V^\dagger_{\text{sc}}(t)
\ketbra{e}\big)
\\
&=
	\left| \frac{\i q}{\hbar}
	\int_0^t
		\mathcal{E}(t') \, \e^{\i \omega_{eg} t'}
	\,\d t'
	\right|^2
\\
&=
\left|
\frac{q\mathcal{E}}{2\hbar\i}
	\left(
		\frac{1-\e^{\i(\omega_{eg}-\omega) t}}{\omega_{eg}-\omega}
		-
		\frac{1-\e^{\i(\omega_{eg}+\omega) t}}{\omega_{eg}+\omega}
	\right)
\right|^2
\,,
\label{e:prerwa}
\end{align}
\end{subequations}
where the second line is obtained because the state $\ket{e}$ is an
eigenvalue of the free evolution operator $U_0(t)$, the third follows
from the semi-classical assumption that the state of the field 
always stays in a coherent state, the fourth from the weak coupling assumption and the form of $H_{\text{sc}}$, and the last from the monochromatic assumption.

At this point, we make an extra assumption called the \emph{rotating wave approximation} (RWA). 
In our special case, the RWA amounts to neglecting the second term of the
right-hand side of \eqref{e:prerwa}. 
Because both $\omega$ and $\omega_{eg}$ are strictly
positive, the denominator of the second term never vanishes, unlike the denominator of the first term. 
Around resonance $\omega \approx \omega_{eg}$, the
first term will largely dominate the second.
\end{example}

We make the RWA from Example~\ref{ex:monochromatic} a general principle that applies to arbitrary quantum fields.
It echoes the use of the analytic signal in representing a real classical field; see Section~\ref{s:point_processes}.



\begin{assumption}[Rotating wave approximation]
\label{a:rwa}
The photodetection response is a function of the positive frequency 
part of the quantum field only.
\end{assumption}
With Assumption~\ref{a:rwa}, the transition probability \eqref{e:probability_of_detection}
now reads
\begin{align}
	p_{\omega_{eg}}(\mathbf{r},t)
	=
	\left(\frac{e\, q}{\hbar}\right)^2
	\int_0^t
		\e^{\i \omega_{eg} (t'-t'')}
		G^{(1)}_{\rho}\big (\mathbf{r}\, t', \mathbf{r}\, t''\big)
		\,
	\d t' \d t''
\,,
\label{eq/probtwolevel}
\end{align}
where\footnote{
	The missing commas in the arguments are a convention that will become handy when we consider higher-order coherence functions.
} we introduced the \emph{first-order coherence function}
\begin{align}
	\label{e:first_order_coherence}
	G^{(1)}_{\rho}\big ( \mathbf{r}'\, t', \mathbf{r}\, t\big )
	=
	\langle E^-(\mathbf{r}',t') E^+(\mathbf{r},t) \rangle_\rho
\,,
\end{align}
where $E^+$ is associated to the positive part of the spectrum
(annihilation operators) and $E^-$ is associated to the negative
part (creation operators).  The coherence function 
\eqref{e:first_order_coherence} was originally introduced by \cite{Glauber-1963}.\footnote{
	As a side remark, remember that this coherence function 
	is defined after choosing a specific polarization of the field. In the
	general case, the electric field should be expanded along a basis 
	of the three-dimensional polarization space and we should consider
	all combinations of coherence functions
	$$	
	G^{(1)}_{ij}(t', t)
	=
	\langle E_j^-(t') E_i^+(t) \rangle_\rho
	$$ where
	$i,j$ label the basis elements.
}

At least informally, it is not difficult to generalize the computations leading 
to \eqref{e:probability_of_detection} to more sophisticated detectors. 
In experiments, for instance, photodetection is often performed through 
ionization: an electron of an atom is extracted by the incident light and 
this event is amplified and detected. This detector is best described by a 
system having, on top of its ground $\ket{g}$ state, a countable set of 
excited states $\ket{e_i}$, and a continuum of so-called \emph{diffusive} 
states $\ket{p},p\in\mathbb{R}$. This leads to non-compact Hamiltonian
operators, having a spectrum that is partly discrete, partly continuous. 
Nonetheless, the computation of the transition probability $p(\mathbf{r}, t)$
follows from the same lines as above, with the new observation corresponding 
to projecting on all states that are not the ground state, that is, 
$(\sum_n\ketbra{e_n}+\int \ketbra{p} \d p) \otimes \id$ 
instead of $\ketbra{e}\otimes \id$. Because we are working 
with a first-order approximation in Assumption~\ref{a:weak_coupling}, this 
leads to a probability of transition that is an expectation of the single-state 
probability of transition \eqref{e:probability_of_detection}, namely
\begin{align}
	p(\mathbf{r}, t)
	= \int p_{\omega_{eg}}(\mathbf{r}, t) \d\xi(\omega_{eg})
	=
	\int_{[0,t]^2} \kappa(t'-t'') G^{(1)}_{\rho}\big (\mathbf{r}\, t', \mathbf{r}\, t''\big) \, \d t' \d t''
\,,
\label{eq/generalprob}
\end{align}
where $\xi$ has a support that is the union of a discrete set, corresponding 
to the states $\ket{e_n}$, and an uncountable set, corresponding to $\ket{p},
 p\in\mathbb{R}$, and the structure function is defined as
 $$
 \kappa(\tau) := \left( \frac{e}{\hbar} \right)^2 \int \e^{-\i \omega \tau} \d\xi(\omega).
 $$
We can think of either $\xi$ or $\kappa$ as being a characteristic
of the detector, describing its time or frequency efficiency. 
For instance, if $\xi$ is supported on a single point or closely 
around a single point, we only collect photons at a very specific energy, 
and call the detector \emph{narrow-band}. If $\xi$ is supported on a 
large interval, we collect all photons equally, and call the detector 
\emph{broad-band}.

For completeness, we close this section by recovering the semi-classical
 treatment of Section~\ref{s:permanental} through our quantum treatment, 
using the coherent state formalism of Section~\ref{s:coherent_states}.
\begin{example}[General rotating wave]
By linearity arguments, the monochromatic light case of Example~\ref{ex:monochromatic} can be used to derive the excitation
probability in the case of a more general field.
Thinking of the field being enclosed in a large box, so as to be able to work with Fourier series instead of Fourier transforms for simplicity, let us decompose the electric field  $\mathcal{E}_{\balpha}(\mathbf{r},t) $ on a set of  monochromatic plane waves
\begin{align}
\mathcal{E}_{\balpha}(\mathbf{r},t) = \sum_\mathbf{k} \mathcal{N}_\mathbf{k}
	\left[
		\alpha_\mathbf{k} \e^{\i (\mathbf{k}\cdot \mathbf{r} - \omega_{\mathbf{k}} t)} +
		\alpha^*_\mathbf{k} \e^{-\i (\mathbf{k}\cdot \mathbf{r} - \omega_{\mathbf{k}} t)}
	\right]
\,,
\label{e:classical_field_in_a_box}
\end{align}
with $\mathcal{N}_\mathbf{k}$ a normalization coefficient and 
$\alpha_\mathbf{k}\in\mathbb{C}$ the amplitude of the mode $\mathbf{k}$. 
Note that $\mathcal{E}$ is a (real-valued) function, not an operator: we
are still treating the field as a classical field. The first part of the
sum corresponds to the positive frequency part of the field (the ``analytic" field), while the second part corresponds to the negative frequency part $\mathcal{E}^-$.

An important observable of the electromagnetic field is its
intensity. 
A general result of Maxwell's equations \citep{MaWo95} 
is that the energy carried by the field is given by the squared modulus
of the field. 
Under our rotating wave Assumption~\ref{a:rwa},
the intensity $I(\mathbf{r},t)$ of the field reads
\begin{align}
I(\mathbf{r},t) = \mathcal{E}_{\balpha}^-(\mathbf{r},t) \mathcal{E}_{\balpha}^+(\mathbf{r},t)
	= |\mathcal{E}_{\balpha}^+(\mathbf{r},t)|^2
\,.
\label{eq/intensity}
\end{align}

Investigating this intensity, and especially its correlation
structure in the case of random sources, has led to the major
discoveries of photon bunching and in the end of the general
theory of optical coherence in quantum theory \citep{MaWo95}. 
More precisely, say the field \eqref{e:classical_field_in_a_box} 
is not perfectly known; it is then reasonable to model it as a 
stochastic process, i.e., a random function. 
A very common assumption is to assume the process to be a stationary 
Gaussian process.
One partial justification is that the field at a given spacetime
point is the sum of the contributions of many independent sources, 
and is well approximated by a Gaussian. 
Formally, the field state becomes 
a statistical mixture of coherent states, a \emph{mixed state} in the
vocabulary of \cref{s:quantum_framework}. The probability of excitation
remains of the form \eqref{e:probability_of_detection}, with first-order coherence function 
\begin{align}
	G^{(1)}_{\rho}\big ( \mathbf{r}\, t', \mathbf{r}\, t'\big )
	&= \mathbb{E}_{\balpha} \expval{E^-(\mathbf{r}',t')E^+(\mathbf{r},t)}{\balpha}\\
	&= \mathbb{E}_{\balpha} \mathcal{E}_{\balpha}^-(\mathbf{r}',t')\mathcal{E}_{\balpha}^+(\mathbf{r},t)\\
	&= \int p_\text{classical}(\balpha) \mathcal{E}_{\balpha}^-(\mathbf{r}',t')\mathcal{E}_{\balpha}^+(\mathbf{r},t)
\, \d\balpha
\,.
\end{align}
The expectation is over the stochastic process $\mathcal{E}$. 
Physicists prefer to denote it informally using a probability density 
$p_\text{classical}$ over the Fourier coefficients 
$\balpha = (\alpha_\mathbf{k})$ of the field in
\eqref{e:classical_field_in_a_box}, and we use that notation here for future notational compatibility with Glauber-Sudarshan decompositions.
In  Section~\ref{s:correlation_photodetection}, we shall use Wick's 
theorem to derive higher-order coherence functions, and recognize the set of correlation functions of a permanental point process.
%

\end{example}

\subsubsection{Role of the detector structure function}
\label{s:structure_functions}

In this section, we consider the influence of the detector response 
$\kappa$ on the probability of excitation \eqref{eq/generalprob}. 

\paragraph{Broad-band detector.}
Consider first a broad-band detector, i.e. a detector with structure 
function that is sharply peaked in time or, equivalently, does not
depend much on the frequency. Ideally, this corresponds to 
$\kappa (t-t') = \kappa_0 \, \delta(t-t')$ with $\kappa_0$ a real constant. 
The probability of excitation \eqref{eq/generalprob} becomes
\begin{align}
	p(\mathbf{r}, t)
	=
	\kappa_0
	\int_0^t G^{(1)}_{\rho}\big ( \mathbf{r}\, t', \mathbf{r}\, t'\big ) \, \d t'
\,.
\label{e:detection_probability_broadband}
\end{align}
Anticipating over Section~\ref{s:identifying_point_processes}, we would like to map this probability to a point process on the positive half-line $\mathbb{R}_+$. 
One way to associate the two notions is to imagine $t>0$ infinitely small in \eqref{e:detection_probability_broadband}, and ``re-setting" the detector in its ground state at each time multiple of $t$, keeping track of whether the detector was found in its excited state, say in the middle of each small time interval. 
Loosely speaking, in the limit $t\rightarrow 0$, these detection times would form a point process on the positive half of the real line, with first
correlation function $G^{(1)}_{\rho}\big ( \mathbf{r}\, t', \mathbf{r}\, t'\big )$.
To obtain a more formal correspondence, one needs to further model the detection process, for instance including de-excitation of the detector, see e.g. \cite{BoVaJa07}.

\paragraph{Narrow-band detector.}
A narrow-band detector is a detector that is well-resolved in frequency, 
like the two-level atom we used in our derivation of the photodetection 
signal. At the extreme, we think of the structure function as a delta
distribution in Fourier space at a frequency $\omega_0$, so that
$\cF \kappa(\omega) = \kappa_0 \delta(\omega - \omega_0)$, and thus
$\kappa(t) = \kappa_0/\sqrt{2\pi} \e^{\i \omega_0 t}$.
The excitation probability \eqref{eq/generalprob} now reads
\begin{align}
	p(\mathbf{r}, t)
	&=
	\frac{\kappa_0}{\sqrt{2\pi}}
	\int_{[0,t]^2} G^{(1)}_{\rho}\big ( \mathbf{r}\, t', \mathbf{r}\, t''\big )
		\e^{\i \omega_0 (t'-t'')} \, \d t' \d t'' \;.
\, ,
\end{align}
The meaning of this equation is even more transparent when
we switch to the frequency representation of the coherence
function. Indeed, let us define the Fourier representation of the first-order coherence function through the Fourier transform(-like) equation
\begin{align}
	G^{(1)}_{\rho}(\mathbf{r}'\, t', \mathbf{r}''\, t'') =
	\int_{\mathbb{R}^2}
		G^{(1)}_{\rho}(\mathbf{r}'\, \omega', \mathbf{r}'' \, \omega'')
		\e^{-\i\omega' t'} \e^{\i\omega'' t''}
	\, \frac{\d \omega' \d \omega''}{2\pi}
	\,,
	\end{align}
	where we follow the physicists' convention to denote both the function and its Fourier transform by the same notation and index the function by the name of its variable. 
Putting everything together, we arrive at 
\begin{align}
p(\mathbf{r}, t) = \frac{\kappa_0}{\sqrt{2\pi}}
\int_{\mathbb{R}^2}
	G^{(1)}_{\rho}(\mathbf{r}\, \omega, \mathbf{r}' \, \omega')
	\, 2 t \frac{\sin(t(\omega-\omega_0))}{(\omega-\omega_0)}
\, \frac{\d \omega \d \omega'}{2\pi} \;.
\end{align}
In the long time limit\footnote{
	usually the coherence function varies over some timescale, so large time means $t$ sufficiently large compared to that timescale
}, the integration window is approximately
equal to $t \delta(\omega-\omega_0)$. We end up with
the simple expression
\begin{align}
	p(\mathbf{r}, t)
	\approx 
	\frac{\kappa_0 t}{\sqrt{2\pi}}
	G^{(1)}_{\rho}(\mathbf{r}\, \omega_0,  \mathbf{r} \, \omega_0)
\,.
\end{align}
The interpretation of this expression is easier now. 
For a narrow-band detector in frequency space centered around a
frequency $\omega_0$, the excitation probability
is proportional to the diagonal $\omega_0$ part of the
first-order coherence function which is none other than
the average intensity of the mode $\omega_0$.

\subsection{Correlation between photodetection events}
\label{s:correlation_photodetection}
We now examine the joint probability of detecting two, and then more, photons.

\subsubsection{Second-order coherence}

Consider two detectors labelled
$1$ and $2$ and placed at two distinct positions $\mathbf{r}_1$ and $\mathbf{r}_2$. 
They are excited by an incident radiation, and we ask for the probability that detector $1$ gets excited at time $t_1$ \emph{and} detector $2$ gets excited at time $t_2$.
Actually, the derivation of the joint probability of detection is quite similar to the derivation for the one photodection event in Section~\ref{s:first_order_coherence},
and we will only give the general principles of the method here again.

The dynamics are once again modeled through a Hamiltonian, whose
interaction term is given by the following generalization of \eqref{eq/onedethamil},
\begin{align}
H_I(t) = q_1 \sigma^{(1)}_x(t) \otimes \id \otimes E(\mathbf{r}_1,t) + 
	q_2 \id \otimes \sigma^{(2)}_x(t) \otimes E(\mathbf{r}_2,t)
\,.
\label{e:interaction-2det}
\end{align}
Such a form is natural to physicists, in the sense that it models the local interaction of each of two detectors with the field at their respective position.

Computing the joint probability of excitation requires to compute first 
a matrix element of the evolution operator in the interaction picture,
which is again done assuming weak coupling. 
We will see that in order
to have a non-trivial result, we need to push the expansion in \cref{a:weak_coupling} up to second order. 
\begin{assumption}[Weak Coupling; second order]
	\label{a:weak_coupling-2}
The interaction between the detectors and the field is assumed to
be weak, allowing to treat the dynamics using perturbation theory.
To second order, this amounts to
\begin{align}
	\label{e:weak-coupling-2}
	V(t)
	\simeq
	\id - \frac{\i}{\hbar} \int_0^t H_I(t_1) \d t_1
	-\frac{1}{\hbar^2} \int_0^t \int_0^{t_1}  H_I(t_1) H_I(t_2) 
	\, \d t_2 \d t_1
\,.
\end{align}
\end{assumption}
Being interested in the joint probability of excitation after a time $t$, 
we have to compute the matrix element
$\langle e_1,e_2,f|V(t)|g_1, g_2, i \rangle$ where $(\ket{g_1}, \ket{g_2})$
and $(\ket{e_1},\ket{e_2})$ are the ground and excited states
of both detectors. 
Since they are all pairwise orthogonal,
the $\id$ term in \eqref{e:weak-coupling-2} will give a zero contribution. 
As for the linear term, and given the form of the interaction \eqref{e:interaction-2det}
as a sum of two contributions from each detector, it will split into
two terms, each acting trivially on one of the detectors,
again yielding a zero result. 
For instance 
\begin{align*}
	\langle e_1,e_2,f|\sigma^{(1)}_x(t) \otimes \id \otimes E(\mathbf{r}_1,t)|g_1, g_2, i \rangle
=
\matrixel{e_1}{\sigma^{(1)}_x(t)}{g_1} \braket{e_2}{g_2} \matrixel{f}{E(\mathbf{r}_1,t)}{i}
=0
\end{align*}
by orthogonality of $\ket{e_2}$ and $\ket{g_2}$. 
This is the reason why we have to go to the second order in perturbation
theory to compute the joint probability distribution.
In general, to compute the joint probability distribution of $N$ detectors clicking, we will need to push the expansion of \cref{a:weak_coupling-2} to the order $N$, since none of the lower-order terms contributes. 
Going back to $N=2$, the quadratic term also simplifies with only
the cross terms $E(\mathbf{r}_2,t_2) E(\mathbf{r}_1,t_1)$ giving
a non-zero contribution. 
Using again the RWA of \cref{a:rwa}, we end up with
\begin{align}
	\langle e_1,e_2,f|& V(t)|g_1, g_2, i \rangle\\
	&=
	-\frac{1}{\hbar^2}
	\int_0^t \int_{t_1}^t 
		\e^{\i \omega_{eg_1} t_1} \e^{\i \omega_{eg_2} t_2}
		\langle f|E^+(\mathbf{r}_2,t_2) E^+(\mathbf{r}_1,t_1)| i \rangle
	\, \d t_2 \d t_1
+
1\leftrightarrow 2 \nonumber \\
	&=
	-\frac{1}{\hbar^2}
	\int_0^t \int_0^t 
		\e^{\i \omega_{eg_1} t_1} \e^{\i \omega_{eg_2} t_2}
		\langle f|E^+(\mathbf{r}_2,t_2) E^+(\mathbf{r}_1,t_1)| i \rangle
	\, \d t_2 \d t_1
\,.
\end{align}
The notation $1\leftrightarrow 2$ is a shorthand for the same term with the two indices transposed.
The second line is obtained by grouping the two terms into one
and changing the integration range, noting that the involved operators commute.
Just like we did in \cref{s:first_order_coherence}, we use Born's rule and sum over the final state of the field, since we 
are only interested in the state of the detectors, to obtain the joint excitation probability.
Once again, we express it as the Fourier transform of a second-order 
coherence function. 
Assuming a fixed polarization for 
the field for simplicity, we have 
\begin{align}
	G^{(2)}_{\rho}(\mathbf{r}_1' \,t_1', \mathbf{r}_2' \, t'_2, \mathbf{r}_2 \,t_2, \mathbf{r}_1 \, t_1)
	:=
	\langle E^-(\mathbf{r}_1',t'_1) E^-(\mathbf{r}_2',t'_2)
			E^+(\mathbf{r}_2, t_2) E^+(\mathbf{r}_1, t_1)
	\rangle_\rho
\,.
\label{eq/G2}
\end{align}

By once again introducing structure functions $\kappa_1$ and
$\kappa_2$ for both detectors, we obtain our final expression for the joint excitation probability
\begin{align}
	p(\mathbf{r}_2\, t, \mathbf{r}_1 \, t)
	=
	\int_0^t \kappa_1(t'_1-t''_1) \kappa_2(t'_2-t''_2)
		G^{(2)}_{\rho}(\mathbf{r}_1\, t''_1, \mathbf{r}_2\, t_2'', \mathbf{r}_2\, t_2', \mathbf{r}_1\, t_1'))
		\, \d t_1' \d t_1'' \d t_2' \d t_2''
\,,
\label{eq/generaljointprob}
\end{align}

\paragraph{Intensity correlations.}
Of particular experimental relevance is the diagonal of the second-order coherence function, namely 
\begin{align}
	G^{(2)}_{\rho}(\mathbf{r}_1\,t_1, \mathbf{r}_2\, t_2, \mathbf{r}_2 \,t_2, \mathbf{r}_1 \, t_1)
	=
	\langle E^-(\mathbf{r}_1,t_1) E^-(\mathbf{r}_2,t_2)
			E^+(\mathbf{r}_2, t_2) E^+(\mathbf{r}_1, t_1)
	\rangle_\rho
\,.
\label{eq/G2intensity}
\end{align}
For reference, if we were treating a classical field (i.e., modeled as a real- or complex-valued stochastic process rather than a collection of noncommuting operators), all the fields would commute and from the
definition of the field intensity, we would obtain
\begin{align}
	G^{(2)}_{\rho}(\mathbf{r}_1\,t_1, \mathbf{r}_2\, t_2, \mathbf{r}_2 \,t_2, \mathbf{r}_1 \, t_1)
	=
	\mathbb{E} I( \mathbf{r}_2,t_2) I(\mathbf{r}_1,t_1)
\,.
\label{eq/G2intensityclass}
\end{align}
\cref{eq/G2intensity} is the non-commuting counterpart to \eqref{eq/G2intensityclass}, which gives us information about the intensity correlations of the field.
The order of the operators in \eqref{eq/G2intensity}, called \emph{normal order}, is crucial for the
interpretation of the photodetection signal as the absorption of excitations
by the detector, but also to take properly into account the fundamental
quantum fluctuations that affect the detection signal.

Intensity correlations can be probed experimentally. 
For instance, assume a perfect temporal resolution for each detector, and that they are functioning in the time intervals $[0,t_i]$, $i=1,2$.
This is modeled by choosing for structure functions 
$$
\kappa (t_i-t_i') = \kappa_i \, \Theta_{[0,t_i]} \delta(t_i-t_i'), i=1,2,
$$
with $\kappa_1, \kappa_2>0$.
The probability of joint excitation is then given by
\begin{align}
	p(\mathbf{r}_2\, t_2, \mathbf{r}_1 \, t_1)
	= \kappa_1\kappa_2
	\int_0^{t_2} \int_0^{t_1}
		G^{(2)}_{\rho}(\mathbf{r}_1\, t', \mathbf{r}_2\, t'', \mathbf{r}_2\, t'', \mathbf{r}_1\, t'))
		\, \d t' \d t''
\,.
\end{align}
In particular, taking the derivative in $t_1$ and $t_2$ yields the so-called transition rate, 
\begin{align}
\Gamma(\mathbf{r}_2\, t_2, \mathbf{r}_1 \, t_1)
	= \kappa_1\kappa_2
		G^{(2)}_{\rho}(\mathbf{r}_1\, t_1, \mathbf{r}_2\, t_2, \mathbf{r}_2\, t_2, \mathbf{r}_1\, t_1)
\,,
\end{align}
which is proportional to \eqref{eq/G2intensity}, and reduces to the intensity correlation function in the classical setting.

\subsubsection{Higher-order coherences}

The previous derivation can be generalized straightforwardly
to obtain the joint probability of $n$ detection events. 
Denoting
the spacetime coordinates by $x_i = (\mathbf{r}_i,t_i)$, we define the $n$th order
coherence function by
\begin{align}
G^{(n)}_{\rho}(x_1, \cdots, x_{2n}) =
	\langle
		E^-(x_1) \cdots E^-(x_n)
		E^+(x_{n+1}) \cdots E^+(x_{2n})
	\rangle_\rho
\,,\quad n\geq 1.
\label{e:nth_coherence_function}
\end{align}
This definition generalizes the first- and second-order coherence functions. 
Upon noting that $\tr(\rho O^\dagger O) \geq 0$ for any operator $O$, we can derive general inequalities for the coherence functions \eqref{e:nth_coherence_function}.
For instance, consider the operator
$O = \sum_{k=1}^m \lambda_k E^+(x_k)$, for arbitrary $\lambda_k$'s. 
The nonnegativity of the quadratic form $\tr(\rho O^\dagger O)$ implies a contraint in the underlying determinant. 
For $m=2$, we obtain
$$
G^{(1)}_{\rho}(x,x)G^{(1)}_{\rho}(y,y) \geq |G^{(1)}_{\rho}(x,y)|^2,
$$
which we interpret as a sign of bunching.

Finally, we note that for general states, higher-order coherence functions are typically hard to compute. 
One special case is for a Gaussian state \eqref{e:gaussian_density}, where Wick's \cref{th:wick} applies.
The $n$th-order coherence function can then be expanded as
\begin{align}
G^{(n)}_{\rho}(x_1, \cdots, x_{2n})  =
\sum_{\sigma\in\frak{S}_n} \prod_{i=1}^n
	\langle
		E^-(x_i) E^+(x_{n+\sigma(i)})
	\rangle_\rho
\,.
\end{align}
We now have all the tools to identify the point processes behind the detection of photons. 

\subsection{Turning coherence functions into correlation functions}
\label{s:identifying_point_processes}

There is a specific set of states for which the coherence functions are particularly simple, namely the coherent states of Section~\ref{s:coherent_states}. 
Specifically, take a multi-mode coherent state $\ket{\balpha} = \otimes_{\mathbf k} \ket{\alpha_{\mathbf k}}$. 
These states are
eigenstates of the electric field \eqref{e:field_operator} itself, with eigenvalue the classical field from \eqref{e:classical_field_in_a_box}
, i.e., 
\begin{align}
E^+({\mathbf r}_1, t_1) \ket{\balpha} = \mathcal{E}^+_{\balpha}({\mathbf r}_1, t_1) \ket{\balpha} = \left(\sum_{\mathbf k} \mathcal{E}^+_{\alpha_{\mathbf k}}({\mathbf r}_1, t_1)\right) \otimes_{\mathbf k} \ket{\alpha_{\mathbf k}}
\,.
\end{align}
Physically, coherent states correspond to ideal lasers.
Now, by construction, the quantum coherence functions \eqref{e:nth_coherence_function} are reduced to the classical
coherence functions when the field is prepared in such a state, i.e., writing $x_i = (\mathbf{r}_i, t_i)$, 
\begin{align}
	G^{(n)}_{\ketbra{\boldsymbol{\alpha}}}(x_1, \cdots, x_{2n}) &=
	\expval{
		 E^-(x_1) \cdots  E^-(x_n)
		E^+(x_{n+1}) \cdots E^+(x_{2n})}{\boldsymbol{\alpha}}\nonumber\\
		&=  \prod_{i=1}^n \overline{\mathcal E^+_{\boldsymbol{\alpha}}(x_i)} 
			\mathcal E^+_{\boldsymbol{\alpha}}(x_{n+i}).
\label{e:coherent_correlation_functions}
\end{align}
This factorization across coherent states is what allows us recovering the correlation functions of a Poisson point process. 
\begin{example}[Ideal lasers yield Poisson processes]
	Take $x_i=x_{n+i}=(\mathbf{r}, t_i)$ for some fixed position $\mathbf{r}\in\mathbb{R}^3$ in \eqref{e:coherent_correlation_functions}, and assume a broad-band detector placed at $\mathbf{r}$, see \cref{s:structure_functions}. 
	The joint detection probabilities yield measurement times that follow a point process with correlation functions
	$$
	\rho_n(t_1,\dots,t_n) = \vert \mathcal E^+_{\boldsymbol{\alpha}}(t_1) \cdots \mathcal E^+_{\boldsymbol{\alpha}}(t_n) \vert^2.
	$$
	Such a point process indeed exists: it is the Poisson point process of first correlation function $\vert \mathcal E_{\boldsymbol{\alpha}}^+\vert^2$; see Section \ref{s:poisson}. 
\end{example}

Now, consider a slightly more general state $\rho$ that can be written 
as a statistical mixture 
$
\rho = \mathbb{E}_{\boldsymbol{\alpha}\sim p_\text{classical}} \ketbra{\boldsymbol{\alpha}}
$
of coherent states. 
Concretely, this state describes non-ideal
laser light, e.g. where there is uncertainty on which light amplitude is emitted. 
For instance, a single-mode randomly phased laser falls
into this category, where $p_\text{classical}$ is uniform over a circle in the complex plane: the mixture is over all possible phases
with a fixed intensity $|\alpha|^2$. 
As for the coherence functions,
by linearity, we obtain
\begin{align*}
	G^{(n)}_{\rho}(x_1, \cdots, x_{2n})
		&=  \mathbb{E}_{\boldsymbol{\alpha}\sim p_\text{classical}} 
		\prod_{i=1}^n \overline{\mathcal E^+_{\boldsymbol{\alpha}}(x_i)} \mathcal E^+_{\boldsymbol{\alpha}}(x_{n+i}).
\label{e:classical_mixture_correlation_functions}
\end{align*}

\begin{example}[Mixture states in phase space yield Cox processes]
Taking again $x_i=x_{n+i}=(\mathbf{r}, t_i)$ for some fixed position 
$\mathbf{r}\in\mathbb{R}^3$ in \eqref{e:coherent_correlation_functions}
and a broad-band detector, the detection times follow a point process with correlation functions
$$
\rho_n(t_1,\dots,t_n) = \mathbb{E}_{\boldsymbol{\alpha}\sim p_\text{classical}} 
	\vert \mathcal E^+_{\boldsymbol{\alpha}}(t_1) \cdots  \mathcal E^+_{\boldsymbol{\alpha}}(t_n) \vert^2.
$$
The detection times thus form a Cox point process; see \cref{s:point_processes}.
In particular, arrival times tend to exhibit bunching: they are more clustered
together than a Poisson process. 
Moreover, assuming that each $\alpha_k$ in \eqref{e:classical_field_in_a_box} 
is Gaussian, $\mathcal{E}_\alpha$ is a Gaussian process, and the detection
times then form a permanental point process. Permanental point processes
are the archetype of a point process exhibiting bunching; see 
\cref{s:permanental} and a few samples in \cref{f:three_examples:permanental_samples}.
The fact that we obtain a permanental point process validates \emph{a posteriori}
the semiclassical analysis carried out in \cref{s:point_processes}.
\end{example}

Additionally, note that while we mention ideally resolved detectors in time for simplicity, non-broad-band detectors also yield standard point processes when the detector's characteristic functions can be normalized.
Indeed, consider for simplicity the point process formed by the detection times at detectors that share the same characteristic function $\kappa$. The correlation functions then correspond to the point process formed by the marks in a marked point process with independent marks. 
In the vocabulary of \citep[Chapter 6]{DaVe03}, the ground process is formed by the ideal detection times, and for each such time $t_i$, an actual detection time is drawn from the probability density function proportional to $\kappa(\cdot - t_i)$, independently from all other ideal detection times and marks. 

Now we consider states that are not statistical mixtures of coherent states, the so-called \emph{non-classical} or \emph{genuinely quantum} states.
As we saw in \cref{s:coherent_states}, coherent states form an overcomplete family of states.
This suggests that even nonclassical states admit a decomposition as a linear combination of coherent states.
The decomposition of any mixed state as a linear combination of projectors onto coherent states is known as the \emph{Glauber-Sudarshan} decomposition. 
However, the linear decomposition is not a statistical mixture over a \emph{bona fide} probability density, and needs to be interpreted in a weak sense.
Informally, physicists write 
\begin{equation}
	\rho = \int P_\rho(\boldsymbol{\alpha}) \ketbra{\boldsymbol{\alpha}} \d\boldsymbol{\alpha},
\label{e:P-distribution}
\end{equation}
with $P_\rho(\boldsymbol{\alpha})$ a tempered distribution for each $\boldsymbol{\alpha}$; 
see \citep[Section 11.8]{MaWo95} and references therein.
For a general state, there is no reason for $ P_\rho$ to be a positive function, or even a function at all. 
We do not attempt to formalize the Glauber-Sudarshan decomposition here, 
but we study the simplest example of a nonclassical state: a 
Fock state. 

Consider a single-mode Fock state 
$$
\rho=\ketbra{k} \otimes I \otimes \dots.
$$ 
The associated Glauber-Sudarshan distribution, denoted for simplicity
$P_{\ket{k}}$, is the ``derivative of a Dirac delta", namely
\begin{align}
P_{\ket{k}}(\alpha) = 
\frac{\e^{\vert\alpha\vert^2}}{k!}
	\frac{\partial^{2k}}{\partial^k \alpha \partial^k\overline{\alpha}} 
		\delta(\alpha)
\label{e:P-fock}
\,;
\end{align}
see \citep[Section 11.8]{MaWo95}.
This representation of a Fock state is very singular and comes with no simple statistical interpretation. 
To get some intuition about \cref{e:P-fock}, let us fix two arbitrary states $\ket u$ and  $\ket v$.
We have in fact the matrix elements
\begin{align}
\matrixel{u}{\rho}{v} &= 
\int_{\mathbb{C}^2}
\left(
	\frac{\partial^{2k}}{\partial^k \alpha \partial^k\overline{\alpha}} 
		\frac{\e^{\vert\alpha\vert^2}}{k!} 
		\braket{u}{\alpha}\braket{\alpha}{v} 
\right)
\delta(\alpha)
\, \d \alpha\\
& = 
\left(
	\frac{\partial^{2k}}{\partial^k \alpha \partial^k\overline{\alpha}} 
		\frac{\e^{\vert\alpha\vert^2}}{k!} 
		\braket{u}{\alpha}\braket{\alpha}{v} 
\right)
\bigg\vert_{\alpha=\overline{\alpha}=0}
\,,
\label{e:P-fock-matrixelements}
\end{align}
as long as $\alpha\mapsto \braket{u}{\alpha}\braket{\alpha}{v}$ is 
smooth enough. This indeed corresponds to the informal 
\eqref{e:P-distribution}, when computing matrix elements and with 
$P_\rho$ a specific tempered distribution. 
In particular, if $\ket{m}$ and $\ket{n}$ are two single-mode Fock 
states, and remembering from \cref{s:coherent_states} that 
$\braket{n}{\alpha} = \e^{-\frac{\vert\alpha\vert^2}{2}} \alpha^n / \sqrt{n!}$, we verify that 
\eqref{e:P-fock-matrixelements} yields
$$
\matrixel{n}{\rho}{m} =  
\left(
	\frac{\partial^{2k}}{\partial^k \alpha \partial^k\overline{\alpha}} 
	\frac{\e^{\vert\alpha\vert^2}}{k!}
	\e^{-\vert \alpha\vert^2} \frac{\alpha^n \overline{\alpha}^m}{\sqrt{n!}\sqrt{m!}}
\right)
\bigg\vert_{\alpha=\overline{\alpha}=0}
= \delta_{kn} \, \delta_{km}
$$
which is $1$ if $n=m=k$ and otherwise $0$, as expected.

As for the coherence functions themselves, we can use this 
representation to give a form very similar to the statistical 
mixture we discussed above. Indeed, by linearity, we can
write
\begin{align}
G_\rho^{(n)}(x_1,\dots,x_{2n}) &= 
\int P_\rho(\boldsymbol{\alpha}) 
	\prod_{i=1}^n \overline{\mathcal E^+_{\boldsymbol{\alpha}}(x_i)} 
	\mathcal E^+_{\boldsymbol{\alpha}}(x_{n+i}) 
\, \d\boldsymbol{\alpha}
\,.
\label{e:P-coherence}
\end{align}
As an example, consider again a single-mode state in a Fock state
with $k$ photons. The coherence function can then be obtained
directly from \cref{e:P-fock,e:P-coherence}
\begin{align}
G_{\ket{k}}^{(n)}(x_1,\dots,x_{2n}) = 
\frac{\partial^{2k}}{\partial^k \alpha \partial^k\overline{\alpha}}
\left( 
	\frac{\e^{\vert\alpha\vert^2}}{k!} 
	\prod_{i=1}^n\overline{\mathcal E^+_\alpha(x_i)} \mathcal E^+_\alpha(x_{n+i})
\right)
\bigg\vert_{\alpha=\overline{\alpha}=0}
\,.
\end{align}

\begin{example}[A candidate point process corresponding to a Fock state.]
Taking again $x_i = x_{n+i}$ and a broad-band detector, we obtain 
candidate correlation functions to define a point process for a single-mode Fock state with $k$ photons
\begin{align}
\label{e:formula_derived_from_Mandel}
\rho_n(t_1,\dots,t_n) =  \frac{\partial^{2k}}{\partial^k \alpha \partial^k\overline{\alpha}} 
\left( 
	\frac{\e^{\vert\alpha\vert^2}}{k!}
	\vert \mathcal E^+_\alpha(t_1) \cdots  \mathcal E^+_\alpha(t_n) \vert^2
\right)
\bigg\vert_{\alpha=\overline{\alpha}=0}.
\end{align}
Explicitly, we have
\begin{equation}
	\mathcal{E}^+_\alpha(t) = \mathcal{N} \alpha \e^{-\i \omega t}.
	\label{e:monomode_analytic_field}
\end{equation}

Then \eqref{e:formula_derived_from_Mandel} implies 
$\rho_n(t_1,\dots,t_n) = 0$ for $k<n$. For $k\geq n$, we obtain 
\begin{align}
	\rho_n(t_1,\dots,t_n) &=  \cN^n 
		\frac{\partial^{2k}}{\partial^k \alpha \partial^k\overline{\alpha}} 
			\frac{\e^{\vert\alpha\vert^2}}{k!} \alpha^n \overline{\alpha}^n\bigg\vert_{\alpha=\overline{\alpha}=0}\\
	&= \frac{\cN^n}{k!} 
		\frac{\partial^{2k}}{\partial^k \alpha \partial^k\overline{\alpha}} 
			\frac{(\alpha\overline{\alpha})^k}{(n-k)!}\bigg\vert_{\alpha=\overline{\alpha}=0}\\
	&= \cN^n \frac{k!}{(k-n)!}. 
\label{e:correlation_functions_iid}
\end{align}
We find an expression reminiscent of the correlation functions of $k$ 
independent and uniformly distributed times \citep[Example 2.6]{Joh06}.
However, these tentative correlation functions are not integrable, a 
problem linked with the fact that the plane waves in 
\eqref{e:field_operator} or \eqref{e:classical_field_in_a_box}, while 
allowing simple formal derivations and a clear separation of positive
and negative frequencies, are not actually in $L^2$. 

In fact, it is possible to build a quantum field theory of radiation replacing plane waves by any orthonormal set of $L^2$ solutions $\phi_\ell(\mathbf{r},t)$ of the classical equation of motion \citep{Treps-2020}.
The positive part of the field can then be decomposed on these modes 
as
$\mathcal{E}^+_{\alpha}(\mathbf{r}, t) = \sum_\ell \alpha_\ell 
\phi^+_\ell(\mathbf{r}, t)$,
where the $+$ subscript still corresponds to the analytic signal \ref{e:analyticSignal} taken in the time variable.
The whole framework introduced in Section~\ref{s:qft} follows, with ladder operators and coherent states now relative to this new basis.
Similarly, \cref{e:formula_derived_from_Mandel} stills holds with 
$\boldsymbol{\alpha}$ corresponding to the coherent state parameters 
in the chosen modes. 
However, in the current example of a single-mode Fock state with $k$ photons, the electric field \eqref{e:monomode_analytic_field} becomes
$\mathcal{E}^+_\alpha(t) = \mathcal{N}\alpha\phi^+(t)$, where we have dropped the dependence on $\mathbf{r}$ as in \eqref{e:monomode_analytic_field} since the detector remains at a fixed position.
We thus obtain, in lieu of \eqref{e:correlation_functions_iid}, 
\begin{align}
\rho_n(t_1,\dots,t_n) = \cN^n \frac{k!}{(k-n)!} |\phi^+(t_1)|^2\dots |\phi^+(t_n)|^2
, \quad n\geq 1.
\label{e:correlation_functions_iid}
\end{align}
Since $\phi^+\in L^2$, these are valid correlation functions, namely of the point process formed by $k$ times $t_1, \dots, t_k$ drawn independepently with probability density function proportional to $t\mapsto |\phi^+(t)|^2$.
\end{example}



\subsection{Single-photon sources can lead to anti-bunching}
\label{s:single_photon_sources}

The fact that photons emitted by a Gaussian classical field yield bunched
detection times is known as the Hanbury-Brown \& Twiss (HBT) 
effect, and was first evidenced by \cite{HaTw58}.
While we shall come back to HBT-type experiments in Part II, 
we pause here to insist on the fact that bunching is a consequence 
of considering multiple independent sources. It is perfectly possible 
to obtain point processes that anti-bunch with photons, in the sense 
that the resulting second correlation function $\rho_2(x,y)$ of the point 
process is small close to the diagonal $\{(x,y): x=y\}$. 

To gain intuition without the quantum overhead, we first recast the HBT effect as a consequence of interference between classical sources.
Consider $S$ independent\footnote{For instance, we neglect
interatomic coupling through the radiation for atomic sources.} classical sources, represented by $S$ independent, zero-mean stochastic processes 
$\mathcal{E}_s(\mathbf{r}, t)$.
The total field is
$\mathcal{E}(\mathbf{r},t) = \sum_s \mathcal{E}_s(\mathbf{r},t)$.
%

As derived in Section~\ref{s:photodetection}, the first-order coherence function is
\begin{align}
G^{(1)}(\mathbf{r}' \,t', \mathbf{r} \, t) =
\mathbb{E}\,\left(\sum_s \mathcal{E}^-_s(\mathbf{r}',t')\right)
\left(\sum_s \mathcal{E}^+_s(\mathbf{r},t)\right)
\,.
\end{align}
By assumption, all cross-terms in the product have expectation zero, so that 
\begin{align}
G^{(1)}(\mathbf{r}' \,t', \mathbf{r} \, t)
= \sum_s \mathbb{E}\,\mathcal{E}^-_s(\mathbf{r}',t')\mathcal{E}^+_s(\mathbf{r},t)
:= \sum_s G^{(1)}_s (\mathbf{r}' \,t', \mathbf{r} \, t)
\,.
\label{eq/g1many}
\end{align}
The total first-order coherence is the sum of the first-order coherence
of each source. 
Quite generically, the sum will tend to
zero when $|\mathbf{r}-\mathbf{r}'|$ or $|t-t'|$ are large compared to the so-called
\emph{spatial coherence scale} and \emph{correlation time}
, respectively.

We can also compute physical quantities like the average field
intensity, which, in this simple model, is the sum of the average field intensities
of each source,
\begin{align}
\mathbb{E}\,I(\mathbf{r},t)
=\mathbb{E}\,\mathcal{E}^*(\mathbf{r},t) \mathcal{E}(\mathbf{r},t)
=G^{(1)}(\mathbf{r} \, t, \mathbf{r} \, t)
=  \sum_s \mathbb{E}\,(|\mathcal{E}_s(\mathbf{r},t)|^2)
= \sum_s \mathbb{E}\,I_s(\mathbf{r},t)
\,.
\label{eq/Imany}
\end{align}

To go further and be able to discuss bunching or anti-bunching, we need to consider the second-order coherence function, which characterizes the density of pairs of detection events; see \eqref{e:nth_coherence_function}.
It writes
\begin{align}
G^{(2)}& (\mathbf{r}_1' \,t_1', \mathbf{r}_2' \,t_2', \mathbf{r}_2 \, t_2, \mathbf{r}_1 \, t_1) \\
& =
\mathbb{E}\,
\left(\sum_s \mathcal{E}^-_s(\mathbf{r}'_1,t'_1)\right)
\left(\sum_s \mathcal{E}^-_s(\mathbf{r}'_2,t'_2)\right)
\left(\sum_s \mathcal{E}^+_s(\mathbf{r}_2,_2t)\right)
\left(\sum_s \mathcal{E}^+_s(\mathbf{r}_1,t_1)\right)
\,.
\label{eq/g2many}
\end{align}
Using our assumptions of independence and zero mean, this
coherence function is a sum of one-source and two-source terms.
We focus on two-source terms. 
Developing \eqref{eq/g2many}, the contribution of two-source terms to $\mathbb{E}I(\mathbf{r},t) I(\mathbf{r}',t')$ is
\begin{align}
\sum_{s \ne s'}
\mathbb{E}\,(
|\mathcal{E}^+_s(\mathbf{r},t)\mathcal{E}^+_{s'}(\mathbf{r}',t')
+
\mathcal{E}^+_s(\mathbf{r}',t')\mathcal{E}^+_{s'}(\mathbf{r},t)|^2)
\,.
\end{align}
At coincidence $(\mathbf{r},t) =(\mathbf{r}',t')$, both amplitudes are
the same and give a contribution
$4 |\mathcal{E}^+_s(\mathbf{r},t)\mathcal{E}^+_{s'}(\mathbf{r},t)|^2$.
However, when $(\mathbf{r},t)$ and $(\mathbf{r}',t')$ differ, a phase
appears between the two contributions and reduces the modulus of the sum.
We recover here the bunching effect of light as an interference effect
between pairs of complex amplitudes.
Note that $S=2$ sources are enough to create this interference.

If we want to observe the opposite effect, a \emph{reduction} in the rate at coincidence,
we need a single-photon source such as a single two-level atom described by a Hilbert space $\mathbb{C}^2$ excited by a pumping laser field.
A natural analogy is that of a gun (the atom) that needs to be reloaded (by the laser pump) before firing a new bullet (the photon). 
We shall see that the need to reload implies anti-bunching for the fired photons.

What we want to compute is once again the second-order coherence between two detections at the same photodetection location $r$, but delayed by a time $\tau$, and abusively denoted by $G^{(2)}_{\rho}(r,r,\tau)$.
The whole state space of the source, the field and the detector is then $\mathbb{C}^2\otimes \sH^\infty \otimes \mathbb{C}^2$.
An accurate model of the situation would require a proper description of the interaction between the field, the source and the detector but we only give some elements of the general computation; see \citep{Cohen-79-80} for details.
The new element here is the interaction of the field and the source and the main technical problem is to relate the field $E^+(t)$ emitted by the source to some of its physical properties. 
We ask the reader to admit that $E^+(t) \propto \sigma_-(t)$, where $\sigma_-(t) = U^\dagger(t) (\sigma_-(0) \otimes I \otimes I) U(t)$.
Physically, this expression means that when the source is excited and then relaxes to its ground state, it emits light.
%
%
The second-order coherence of the field can then be written as a function of the time evolution of a source operator as
\begin{align}
G^{(2)}_{\rho}(r,r,\tau)
&\propto
	\langle
		\sigma_+(0) \sigma_+(\tau) \sigma_-(\tau) \sigma_-(0)
	\rangle_{\rho}
\nonumber \\
&\propto
	\langle
		\sigma_+(0) U^\dagger(\tau) \sigma_+(0) U(\tau)
		U^\dagger(\tau) \sigma_-(0) U(\tau)\sigma_-(0)
	\rangle_{\rho}
\,.
\end{align}
%
%
Naturally, we assume that the initial state of the atom is excited and the
field is in the vacuum so that
$\rho = \ket{e}\!\!\bra{e} \otimes \ket{0}\!\!\bra{0}$.
With $\sigma_-(0) = \ket{g}\!\!\bra{e}$ we end up with the very simple
form:
\begin{align}
G^{(2)}_{\rho}(r,r,\tau) &\propto
	\tr\left(\big( \ketbra{g} \otimes \ketbra{0}\big) 
	U^\dagger(\tau)\ketbra{e}U(\tau)\right)
\nonumber \\
&\propto
\langle e |
	U(\tau) \big( \ketbra{g} \otimes \ketbra{0}\big) U^\dagger(\tau) 
| e\rangle
\,.
\label{eq/G2sourceatomic}
\end{align}
Two comments are in order. 
First, we could have also assumed a more general form for the atomic
density matrix which would amount to multiply the above expression
by the probability $p(e)$ for the atom to be initially excited. Its
appearance is natural since the atom has to be excited initially.
Second, the physical interpretation of the second-order coherence \eqref{eq/G2sourceatomic} is as follows. 
Starting from a factorized state in which the atom
has already emitted a photon and is thus in its ground state,
the system evolves until time $\tau$ where we compute the
probability for the atom to be excited again.
\Cref{eq/G2sourceatomic} is the conditional probability that,
knowing that the atom is in its ground state initially, it gets
excited at time $\tau$.

\begin{figure}
	\begin{center}
		\includegraphics[width=0.6\textwidth]{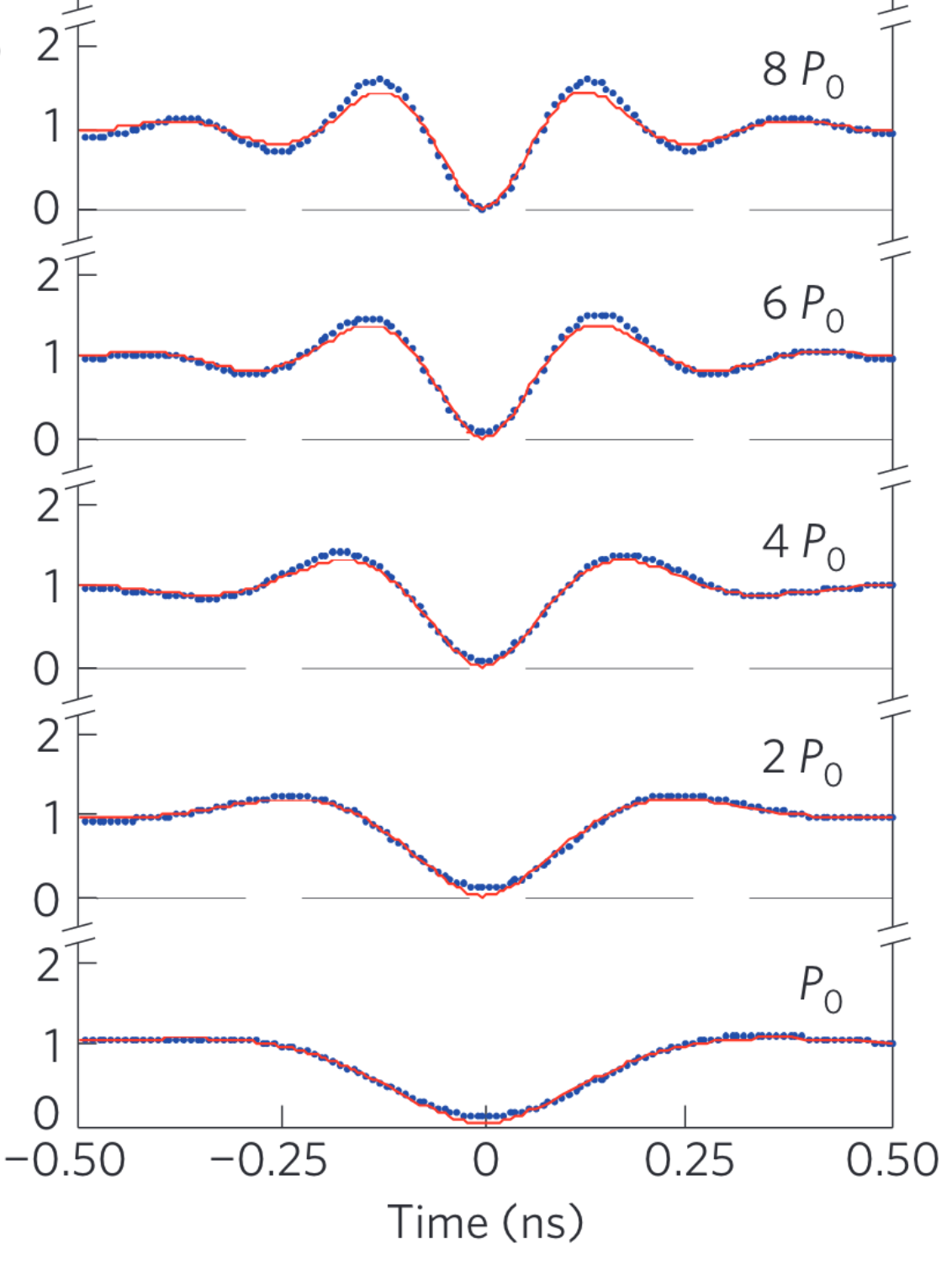}
	\end{center}
\caption{
Experimental measurement of the second-order correlation signal of light coming from a single quantum dot showing anti-bunching at coincidence (zero time delay). The traces show different pump laser powers, all of which are multiples of the lowest power $P_0$. The oscillations at larger delay correspond to so-called \emph{Rabi oscillations} between the two energy levels; as the pump power is increased these oscillations increase in frequency. 
The solid line is a fit to a model. 
Figures taken from \cite{Flagg-2009}. 
}
\label{fig/antibunch}
\end{figure}

From there, predicting photon anti-bunching at coincidence
is straightforward: near coincidence $\tau =0$, the evolution
operator is $U(0) = \id$. 
Since $\ket{e}$ and $\ket{g}$ are orthogonal, this implies
\begin{align}
G^{(2)}_{\rho}(r,r,0) = 0
\,.
\end{align}
In other words, at coincidence, the probability to have a second emission
is zero. 
This result is in fact physically quite intuitive if we think
at the single-source level. 
Indeed, after the atom emits one photon,
it is surely in its ground state and cannot emit a second photon
right away. 
It has to be re-excited first by the laser source before
emitting the second photon. 
A typical experimental signal is
shown in \eqref{fig/antibunch} where we clearly see the coherence signal
going to zero at coincidence. 
The oscillation pattern is also nicely
understood as Rabi oscillations between the two atomic levels~\citep{PIC}.
Nowadays, this anti-bunching is used as signature of a good single photon
source, along with a second major interference effect called
Hong-Ou Mandel; see e.g. \citep{PIC}.

%


%% file: electroncoherence.tex

While photonic optics experiments have reached a high level of sophistication
\citep{Haroche-2006,Aspect-2010}, it is only recently that electronic coherences have started to
be probed, thanks to experimental advances in the manipulation, control
and measurement of small quantum systems in condensed matter and atomic
physics. 
Formally, the main difference with bosons comes from the 
anti-commutation of the associated ladder operators, resulting in the
Pauli exclusion principle and in determinants appearing in Wick's 
Theorem~\ref{th:wick}. 
Moreover, while we limit ourselves to free (i.e., non-interacting) fermions in this section, in realistic physical models fermions are usually subject to the effects of interaction.
These two aspects drastically change the physics compared to photons. 
Photons are noninteracting objects with a truly empty reference state, the vacuum.  
Electrons, on the contrary, are subject to Coulomb interaction and 
their ground state, called the Fermi sea and representing a 
metal at equilibrium, is full of fermions: their vacuum is not empty. 

This Section follows the lines of \cref{s:correlation_photodetection}, 
explaining how similar computations are affected by the fermionic 
character of the particles. 
In \cref{s:modeling_electrodetection}, 
we remind what a fermionic field is and discuss how to model 
the detection of electrons. 
In \cref{s:correlation_electrodetection_events}, 
we discuss the coherence functions of arbitrary orders, and 
use the second-order one in \cref{s:electron_anti_bunching} to 
justify the default anti-bunching character of fermions. 
In 
\cref{s:recovering_classical_currents}, we finally discuss the 
difficulties in recovering classical currents in the way coherent 
states do for bosonic fields.

\subsection{Modeling electrodetection events}
\label{s:modeling_electrodetection}
We proceed as for photodetection by first modeling the field and
then the detection, so as to be able to write correlation functions.

\subsubsection{Modeling the fermionic field}

A \emph{fermionic quantum field} $\psi(\mathbf{r},t)$ is a collection 
of operators indexed by space and time and acting on a Fock space
$\cH^\infty_\text{Fermions}$ of fermions; see \cref{s:fock_spaces}.
The field models an indefinite number of fermionic particles, like 
electrons. Similarly to a bosonic field, the state
of a fermionic system is built by acting on a reference state with 
operators that create or annihilate fermions in a given state of a 
basis of the single-particle Hilbert space.
When it is clear that we speak of fermions,\footnote{
	$a^\dagger$ and $a$ are usually kept for photons, or general particles.
} these ladder operators are usually denoted as $c^\dagger$ 
and $c$, and indexed by the label of the single-particle state.  
Unlike for bosons, the creation and annihilation operators satisfy 
the canonical \emph{anti}commutation relations \eqref{e:CAR}.

In theory, the reference state can be taken to be the vacuum, meaning
the complete absence of any particle, as we did for bosons. But when 
modeling electronic experiments, physicists often have to take another 
reference state, labeled as $\ket{F}$, and called the \emph{Fermi sea}.
The Fermi sea is meant to represent, for instance, the metal used in experiments, 
which is itself full of electrons. 
In the simplest case, corresponding to 
temperature zero, the Fermi sea is a state with one fermion in every 
energy level of the considered system, up to some reference level 
called the \emph{Fermi level}. 
For instance, in a system built using 
a single-particle Hamiltonian with eigenpairs $(\phi_n, \epsilon_n)_{n\in\mathbb{N}}$, so that $c^\dagger_n$ creates a particle in mode $\phi_n$, the Fermi sea is 
$$
\ket{F} = \prod_{n:\epsilon_n< \varepsilon_F} c^\dagger_n\ket{0},
$$
with the Fermi level $\varepsilon_F$ defined by the experimental setting.
Like the vacuum, the Fermi sea can also be built as the limit of a Gaussian state when $\beta\rightarrow \infty$ and the chemical potential $\zeta$ is fixed as a function of $\varepsilon_F$; see the computations in Section~\ref{s:dppfermions}. 

\begin{remark}
	Depending on the situation to model, more physical parameters
	can be introduced to describe the field in addition to the spacetime
	coordinates. 
	For instance, the spin of the fermionic excitation could
	contribute, or more fermionic fields can be used to describe different
	channels, as in the description of the quantum Hall effect \citep{Eza08}.
\end{remark}

\subsubsection{First-order electronic coherence}

As we did for photons, we consider a two-level detector with Hilbert space $\mathbb{C}^2$
probing a fermionic field $\psi(\mathbf{r})$ living in the Hilbert space $\mathcal{H}^{\infty}_{\text{Fermions}}$.\footnote{
	Actual experimental settings are more intricate to describe than for photons exciting a two-level atom; Part II will contain examples.
}
The dynamics of the detector is described here by a fermionic operator $d$ that satisfies the anticommutation rule $\lbrace d,d^\dagger \rbrace =0$.
Their interaction, again assumed to be weak, is modeled by Hamiltonian (in the interaction picture)
\begin{align}
	H_I(t) = d^\dagger(t) \psi(\mathbf{r},t) + d(t) \psi^\dagger(\mathbf{r},t)
\,,
\label{eq/electrodetechamil}
\end{align}
Note that we place ourselves in the interaction picture, so that the time dependence in \eqref{eq/electrodetechamil} results from the evolution of the free part of the Hamiltonian. 
The first term of the interaction Hamiltonian describes the absorption 
process of an electron that excites the detector while the second term
describes an electron being emitted by the detector. 
With the same assumptions as in the photodetection problem, we 
can compute the excitation probability of the detector as a function 
of an electronic correlation function, naturally called in this context 
the \emph{first-order electronic coherence function}, and defined as
\begin{align}
	G^{(1e)}_{\rho}(\mathbf{r}' \, t', \mathbf{r} \, t)
	=
	\langle \psi^\dagger(\mathbf{r}',t') \psi(\mathbf{r},t) \rangle_\rho
\,.
\end{align}

\begin{remark}
In concrete condensed matter systems, we argued above that the natural
ground state is in fact a Fermi sea $\ket{F}$, contrary to the photonic
case where the natural ground state is the true vacuum $\ket{0}$. In
such a context, it is then also interesting to consider the
de-excitation probability of the detector, sending back the
electron into the system, or equivalently creating a \emph{hole} in the Fermi sea.
This transition probability is controlled
by the \emph{first-order hole coherence function} 
\begin{align}
	G^{(1h)}_{\rho}(\mathbf{r}' \, t', \mathbf{r} \, t)
	=
	\langle \psi(\mathbf{r}',t') \psi^\dagger(\mathbf{r},t) \rangle_\rho
\,.
\end{align}
Note however that the two types of coherence functions are not
independent. Indeed, thanks to the canonical anticommutation relations,
they satisfy the equal time relation
$$
G^{(1e)}_{\rho}(\mathbf{r}' \, t, \mathbf{r} \, t) + 
G^{(1h)}_{\rho}(\mathbf{r}' \, t, \mathbf{r} \, t) =
\delta(\mathbf{r}'-\mathbf{r}).
$$
In the following, we work with one of the two coherence functions, 
and we pick $G^{(1e)}_{\rho}$ by convention.
\end{remark}

\begin{example}[Free fermions at non-zero temperature $T$]
	\label{ex:free_fermions}
	The simplest example we can consider is a set of free fermions.
	In particular, we consider the free field Hamiltonian $\Hff = \sum_{\mathbf{k}} \nu_{\mathbf{k}} c_{\mathbf{k}}^\dagger c_{\mathbf{k}}$, and further assume that the energy level $\nu_\mathbf{k} = \nu_k$ only depends on $k = \Vert\mathbf{k}\Vert$.
	We prepare the state in the grand-canonical ensemble \eqref{e:grandcanonical}, which we denote here as $\rho=\rho_{\beta,\zeta}$.
	Note that $\rho$ commutes with $\Hff$, so that Schrödinger's equation yields that $\rho$ does not change in time: it is an equilibrium state.
	In particular, the coherence function $G^{(1e)}_{\rho}(\mathbf{r}' t', \mathbf{r} t)$ depends only on $t-t'$. 
	We arbitrarily set $t=t'$, and write $G^{(1e)}_{\rho}(\mathbf{r}', \mathbf{r})$.
 	Introducing the Fourier transform of the annihilation operator 
	$$
	\psi(\mathbf{r}) = \int_{\mathbb{R}^3} c_{\mathbf{k}} \, \e^{\i (\mathbf{k}\cdot\mathbf{r})} \, \d\mathbf{k}/(2\pi)^3,
	$$ 
	we can write the first electronic coherence function as
\begin{align}
G^{(1e)}_{\rho}(\mathbf{r}', \mathbf{r})
	=
	\langle \psi^\dagger(\mathbf{r}') \psi(\mathbf{r}) \rangle_\rho
	=
	\int_{\mathbb{R}^6}
	\langle c^\dagger_{\mathbf{k}'} c_{\mathbf{k}} \rangle_\rho \, 
	\e^{\i (\mathbf{r}\cdot \mathbf{k}- \mathbf{r}'\cdot\mathbf{k}')}
	\, \frac{\d \mathbf{k}}{(2\pi)^3}\frac{\d \mathbf{k}'}{(2\pi)^3}
\,.
\end{align}
Now note that\footnote{
	This computation is standard in physics. 
	At this stage, we take it for granted; but see \cref{e:mean_number} for an explicit derivation.
}
$$
\langle c^\dagger_{\mathbf{k}'} c_{\mathbf{k}} \rangle_{\rho} 
= f_{\beta,\zeta}(k) \delta(\mathbf{k}-\mathbf{k}')
$$ 
with 
$k =  \| \mathbf{k} \|$ where $f_{\beta,\zeta}$ 
is the Fermi-Dirac distribution
\begin{align}
f_{\beta,\zeta}(k) = \frac{1}{\e^{\beta(\nu_k - \zeta)} +1}
\,.
\end{align}
We obtain
\begin{align}
G^{(1e)}_{\rho}(\mathbf{r}', \mathbf{r}) =
	\int_{\mathbb{R}^3}
	f_{\beta,\zeta}(k) \,
	\e^{\i (\mathbf{r}-\mathbf{r}')\cdot \mathbf{k}}
	\, \frac{\d \mathbf{k}}{(2\pi)^3}
\,.
\end{align}
A practically relevant case is the zero-temperature
limit, where the grand-canonical ensemble state is replaced by the Fermi 
sea itself $\ketbra{F}$. 
The Fermi-Dirac distribution then simplifies to the indicator $\mathbbm{1}_{B(k_F)}$ of the centered ball $B(k_F)$ of radius $k_F$ known as the Fermi momentum, which is an input of the model\footnote{
	Equivalently, we could have specified the chemical potential $\zeta$ to a certain value $\zeta_F$, called the Fermi chemical potential. 
	Moreover, $k_F$ is related to the Fermi velocity $v_F$ or the Fermi energy $\nu_F$ by the free Hamiltonian so these too can be used as inputs of the model.}. 
The first-order electronic coherence function is 
then
\begin{align}
G^{(1e)}_{\ket{F}}(\mathbf{r}', \mathbf{r})
	&=
\frac{1}{\pi^2 \|\mathbf{r}-\mathbf{r}'\|} \int_0^{k_F} k \sin(k \|\mathbf{r}-\mathbf{r}'\|) \, \d k
\nonumber \\
&= \frac{1}{\pi^2 \|\mathbf{r}-\mathbf{r}'\|} 
	\left( \frac{\sin(k_F \|\mathbf{r}-\mathbf{r}'\|)}{\|\mathbf{r}-\mathbf{r}'\|^2} 
	- \frac{k_F}{\|\mathbf{r}-\mathbf{r}'\|}\cos(k_F \|\mathbf{r}-\mathbf{r}'\|) \right) 
\,.
\label{e:firstorderfreefermion}
\end{align}

\end{example}

\begin{example}[Chiral free fermions at non-zero temperature $T$]
\label{ex/G1wire}
A second example appears naturally when discussing electronic coherences. 
This time, we consider a simple model of a \emph{chiral quantum wire}, that is, electrons moving in one dimension and only in a given direction (e.g. to the right).
The Hilbert space is similar to Example~\ref{ex:free_fermions}, $\mathbb{C}^2 \times \cH_{\text{Fermions}}^\infty$, with $\cH = L^2(\mathbb{R})$ as a single-particle Hilbert space.
Rather than describing the Hamiltonian, a typical shortcut to describe the model is to directly discuss the solutions to Schrödinger's equation, or, as we do here for the field operators, how operators evolve in the Heisenberg picture.

Compared to Example~\ref{ex:free_fermions}, we are not at equilibrium, and thus need to keep track of the time dependence
of the field $\psi(x,t)$.
The Fourier variable conjugated to the time $t$ is denoted by $\omega$, just like the momentum
$\mathbf{k}$ is conjugated to the position $\mathbf{r}$. 
Being in one dimension in this example, the vectors $\mathbf{r}$ and $\mathbf{k}$
are in fact scalar quantities that we will denote $x$ and $k$ respectively.
To further constrain the model, we posit an equation of motion for the field, 
i.e., a relation between its time- and space derivatives\footnote{
	Again, this is a shortcut to avoid describing the Hamiltonian and Schrödinger's equation.
}.
The non-zero solutions of this equation are described by a relation between all the
Fourier variables, called \emph{a dispersion relation}. 
In our quantum wire model, the dispersion relation is assumed to be linear, meaning that the eigenvalues of the free field Hamiltonian are given by $\nu_k = v_F  |k|$, with $v_F$ the so-called Fermi velocity of the
electron in the quantum wire. 
Finally, we again assume that the electrons
coming into the wire come from a reservoir of particles, whose state 
is the grand-canonical ensemble $\rho = \rho_{\beta, \zeta}$ at temperature 
$\beta = 1/k_B T$ and chemical potential $\zeta$. 

\begin{figure}
	\begin{center}
		\includegraphics[width=0.8\textwidth]{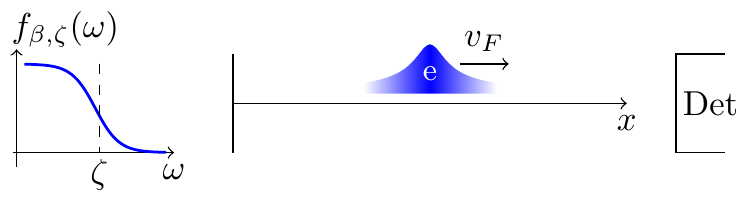}
	\end{center}
\caption{Representation of a simple model of a quantum wire where
fermions are sent from a reservoir at equilibrium (grand-canonical ensemble)
into a one-dimensional wire where they can travel and are probed by an
electro-detection device. Detection probabilities are given by electronic
coherence functions.
\rb{For a v2, can we avoid transparency?}}
\label{fig/1dchiral}
\end{figure}

Now that we have all the additional elements for this model, we can proceed to compute the first-order coherence function. 
Assuming that the field lives in a large box, from the Fourier decomposition $\psi(x,t) = \sum_{k} c_{k} \, \e^{\i (kx-\nu_k t )}$, the first-order coherence expands as
\begin{align}
G^{(1e)}_{\rho}(x' \, t', x \, t)
	=
	\langle \psi^\dagger(x',t') \psi(x,t) \rangle_\rho
	=
	\sum_{k,k'} 
	\langle c^\dagger_{k'} c_k \rangle_\rho \, 
	\e^{\i (kx- k'x' - (\nu_k t - \nu_{k'} t'))}
\,.
\end{align}
The average value of $c^\dagger_{k'} c_k$ on the grand canonical state $\rho$ can be shown  to be $\langle c^\dagger_{k'} c_k \rangle = \delta(k-k') f_{\beta,\zeta}(k)$.
In short, the first-order electronic coherence is simply a Fourier
transform of the Fermi-Dirac distribution. 
Doing the change
of variable to $\omega$, we have
\begin{align}
G^{(1e)}_{\rho}(x' \, t', x \, t)
	=\int_{\mathbb{R}}
	\frac{1}{\e^{\beta(\hbar \omega - \zeta)} +1}\, 
	\e^{\i (x-v_Ft - (x'-v_Ft'))\frac{\omega}{v_F}}
	\, \frac{\d \omega}{2\pi v_F}
\,.
\end{align}
As it is traditionally done in this field, we will use a rescaled
time variable $t - x/v_F$ as $t$. This simplification between time
and space is very special to the one-dimensional case thanks to
the ballistic propagation of the electron. We then write the simpler
expression
\begin{align}
G^{(1e)}_{\rho}(t', t)
	=\int_{\mathbb{R}}
	\frac{1}{\e^{\beta(\hbar \omega - \zeta)} +1}\, 
	\e^{\i (t'-t)\omega}
	\, \frac{\d \omega}{2\pi v_F}
\,.
\end{align}
As such, the integral is divergent when $t=t'$ and has to
be regularized. 
This divergence has a physical origin: the linear dispersion relation posits a Fermi sea with an infinite number of fermions with negative energy. 
In practice however, the system has a finite size and the Fermi sea has a
finite bandwidth. 
The strategy is then to introduce a small parameter
$\epsilon >0$, which corresponds to the inverse of that bandwidth,
and compute the integral with the substitution$t-t' \to t-t' + \i\epsilon$.
With these modifications, the residue theorem gives
\begin{align}
G^{(1e)}_F (t', t) = 
	\frac{\i}{2\pi v_F \tau_{\text{th}}} 
	\frac{\e^{-\i\frac{\zeta}{\hbar} (t-t')}}{\sinh\left( \frac{t-t' + \i \epsilon}{\tau_{\text{th}}}\right)},
\end{align}
where $\tau_{\text{th}} = \hbar \beta /\pi$ is called the \emph{thermal
coherence time}. 
A particular case to have in mind is the 
zero-temperature limit of the coherence function. 
It can in fact be computed directly as the
Fourier transform (in the sense of distributions) of the zero-temperature
Fermi-Dirac energy density distribution 
$f_{0,\zeta}(\omega) = \Theta(\zeta- \omega)$,
where $\Theta$ is the Heavyside function. 
It reads\footnote{The $+\i \epsilon$ notation is a shorthand notation
for the more rigorous distribution
\begin{align}
\lim\limits_{\epsilon \to 0} \frac{1}{x + \i \epsilon} =
	\text{P}\left(\frac{1}{x}\right) -\i \pi \delta (x)
\,.
\end{align}}
\begin{align}
G^{(1e)}_F (t', t) = 
	\frac{\i}{2\pi v_F} 
	\frac{\e^{-\i\frac{\zeta}{\hbar} (t-t')}}{t-t' + \i \epsilon}\,.
\end{align}
\end{example}

\subsection{Correlation between electrodetection events}
\label{s:correlation_electrodetection_events}

The first-order coherence gives information about the system at
the one-particle level. If we want to probe two-particle or higher
levels of information like the correlations between electrodetection
events, we have to study the coherence functions of higher order.
They are defined similarly as the photonic coherences of Section~\ref{s:photoncoherence}. 
For instance, the second-order electronic and hole coherence functions are 
defined as
\begin{subequations}
\begin{align}
	G^{(2e)}_{\rho}(\mathbf{r}'_1 \, t'_1, \mathbf{r}'_2 \, t_2', \mathbf{r}_2 \, t_2, \mathbf{r}_1 \, t_1)
	&=
	\langle \psi^\dagger(\mathbf{r}'_1,t'_1) \psi^\dagger(\mathbf{r}'_2,t'_2) 
	\psi(\mathbf{r}_2,t_2) \psi(\mathbf{r}_1,t_1) \rangle_\rho 
	\\
	G^{(2h)}_{\rho}(\mathbf{r}'_1 \, t'_1, \mathbf{r}'_2 \, t_2', \mathbf{r}_2 \, t_2, \mathbf{r}_1 \, t_1)
	&=
	\langle \psi(\mathbf{r}'_1,t'_1) \psi(\mathbf{r}'_2,t'_2) 
	\psi^\dagger(\mathbf{r}_2,t_2) \psi^\dagger(\mathbf{r}_1,t_1) \rangle_\rho
\,.
\label{eq/defg2gn}
\end{align}
\end{subequations}
This straightforwardly generalizes to higher orders. 
Using the shorthand notation
$x =(\mathbf{r}, t)$ for the spatio-temporal coordinates, the
$n-$th-order electronic and hole coherences are
\begin{subequations}
\begin{align}
	G^{(ne)}_{\rho}(x_1', \cdots, x_n',x_n,\cdots,x_1)
	&=
	\langle \psi^\dagger(x_1') \cdots \psi^\dagger(x_n')\psi(x_n) \cdots \psi(x_1) \rangle_\rho \\
	G^{(nh)}_{\rho}(x_1', \cdots, x_n',x_n,\cdots,x_1)
	&=
	\langle \psi(x_1') \cdots \psi(x_n')\psi^\dagger(x_n) \cdots \psi^\dagger(x_1) \rangle_\rho
\,.
\end{align}
\end{subequations}

\begin{example}[Continuation of \cref{ex:free_fermions}]
The grand canonical ensemble $\rho=\rho_{\beta,\zeta}$ for free fermions is Gaussian, so that
\cref{th:wick} (Wick's) applies and we obtain a $2\times 2$ determinant
\begin{subequations}
\begin{align}
G^{(2e)}_{\rho}(\mathbf{r}'_1, \mathbf{r}'_2, \mathbf{r}_2, \mathbf{r}_2)
	&=
	\langle \psi^\dagger(\mathbf{r}'_1) \psi^\dagger(\mathbf{r}'_2)
	\psi(\mathbf{r}_2) \psi(\mathbf{r}_1)\rangle_{\rho} \\
&\underset{\text{Wick}}{=}
G^{(1e)}_{\rho}(\mathbf{r}'_1, \mathbf{r}_1) G^{(1e)}_{\rho}(\mathbf{r}'_1, \mathbf{r}_1)
-
G^{(1e)}_{\rho}(\mathbf{r}'_1, \mathbf{r}_2)G^{(1e)}_{\rho}(\mathbf{r}'_2, \mathbf{r}_1)
\,.
\end{align}
\end{subequations}
The minus sign is of course a consequence of the fermionic 
character of the electronic field. In fact, we can already
see at this level the repulsion between different fermions.
Indeed, if we look for instance at the diagonal part $\mathbf{r}'_1 = \mathbf{r}_1$ and
$\mathbf{r}'_2 = \mathbf{r}_2$, in the zero temperature limit,
we have
\begin{align}
G^{(2e)}_{\ket{F}}(\mathbf{r}_1, \mathbf{r}_2, \mathbf{r}_2, \mathbf{r}_1)
\propto
\left(1 - \left| g^{(1e)}_{\ket{F}}(\|\mathbf{r}_1 - \mathbf{r}_2 \|) \right|^2 \right)
\,,
\end{align}
where $g^{(1e)}_{\ket{F}}(\|\mathbf{r}_1 - \mathbf{r}_2 \|)$ is the 
first-order electronic coherence function \eqref{e:firstorderfreefermion} 
normalized to unity. 
We clearly see that when $\mathbf{r}_1 = \mathbf{r}_2$, 
the second-order coherence function is exactly zero: two fermions cannot 
be at the same place at the same time. 
This is another avatar of the general anti-bunching effect of fermionic systems.
\end{example}

\begin{example}[Second-order coherence of $N$ excitations]
\label{ex:G2_for_fermions}
To gain some intuitive understanding of the information content
of the second-order coherence function, let's continue \cref{ex/G1wire}
of a quantum wire. Remember that in this simple one-dimensional model, 
time and space coordinates are identified thanks to the linear
dispersion relation and that we are looking for the
coherence functions at the fixed position of the detector (position that we omit
in the arguments to simplify the notations). Consider a state containing
two electrons above the true vacuum $\ket{0}$ (not the Fermi sea) in
two orthogonal wavefunctions
$\ket{\phi_1, \phi_2} =
\psi^\dagger[\phi_1] \psi^\dagger[\phi_2] \ket{0}$,
where $\psi[\phi] = \int \psi(t)\phi(t)\d t$. 
Note again that wavefunctions, strictly speaking functions of $x$ and $t$, are here functions of $x-v_F t$ only because of the linear dispersion relation. 
Since we further consider a fixed position here (that of the detector), we simply write $t\mapsto\psi(t)$ and $t\mapsto \phi(t)$.

Since the true vacuum is a Gaussian state, 
we can apply Wick's \cref{th:wick} to the second-order coherence
function
\begin{align}
G^{(2e)}_{\ket{\phi_1, \phi_2}} (t'_1, t'_2, t_2,t_1) &= 
	\int_{\mathbb{R}^4}
		\langle 0 | 
			\psi(y'_1)\psi(y'_2)
			\psi^\dagger(t'_1) \psi^\dagger(t'_2) \psi(t_2)  \psi(t_1) 
			\psi^\dagger(y_2)\psi^\dagger(y_1)
		| 0 \rangle \,
\nonumber\\		&\phi_1 ^*(y'_1) \phi_2 ^*(y'_2)  \phi_2 (y_2) \phi_1 (y_1) 
	\,\d y_1 \d y'_1 \d y_2 \d y'_2
	\nonumber \\
&= \phi^*_1(t_1') \phi^*_2(t_2') \phi_1(t_1)\phi_2(t_2)
+ \phi^*_1(t_2') \phi^*_2(t_1') \phi_1(t_2)\phi_2(t_1) 	
\nonumber \\
&- \phi^*_1(t_2') \phi^*_2(t_1') \phi_1(t_1)\phi_2(t_2)
- \phi^*_1(t_1') \phi^*_2(t_2') \phi_1(t_2)\phi_2(t_1) 
\,.
\end{align}
We see that the form of this coherence contains the expected
anti-symmetries coming with fermionic statistics. 
It can
be written even more explicitly by introducing the antisymmmetrized
wave-function
\begin{align}
\Phi_{12}(t_1,t_2) = \det
\left[\begin{matrix}
\phi_1(t_1) & \phi_2(t_1)\\
\phi_1(t_2) & \phi_2(t_2)
\end{matrix}\right]
\,.
\end{align}
The second-order electronic coherence function then reads
\begin{align}
G^{(2e)}_{\ket{\phi_1, \phi_2}} (t'_1, t'_2, t_2,t_1) = 
\Phi^*_{12}(t'_1,t'_2) \Phi_{12}(t_1,t_2)
\,.
\end{align}
The second-order coherence function is thus essentially the full 
many-body wavefunction of a two-particle state.

When more than two fermions are present, the second-order 
coherence function only keeps track of two-particle terms. 
To see this, consider a state containing $N$ electrons above 
the vacuum in mutually orthogonal wavepackets $\ket{\psi_N} =
\psi^\dagger[\phi_1] \dots \psi^\dagger[\phi_N] \ket{0}$.
A straightforward, but cumbersome, application of Wick's theorem
gives
\begin{align}
G^{(2e)}_{\ket{\psi_N}} (t'_1, t'_2, t_2,t_1) = 
\sum_{k<l}
\Phi^*_{kl}(t'_1,t'_2) \Phi_{kl}(t_1,t_2)
\,.
\label{e:second_coherence}
\end{align}
Informally, the second-order coherence function contains only 
the ``two-particle physics".

\end{example}

\begin{example}[$k$th-order coherence of $N$ excitations]
	\label{ex:Gk_for_fermions}
Continuing the quantum wire example, we can understand its
$k-$particle physics by computing the $k$th-order coherence function.
Consider again $N$ fermions prepared in mutually orthogonal wavepackets
$\ket{\psi_N} = \psi^\dagger[\phi_1] \dots \psi^\dagger[\phi_N] \ket{0}$.
By Wick's theorem, the $k$th-order coherence function is
\begin{align}
G^{(ke)}_{\ket{\psi_N}} (t'_1,\cdots, t'_k,t_1,\cdots, t_k) = 
\sum_{i_1<\cdots<i_k}
\Phi^*_{i_1,\dots,i_k}(t'_1,\cdots, t'_k) \Phi_{i_1,\dots,i_k}(t_1,\cdots, t_k)
\,,
\end{align}
with
\begin{align}
\Phi_{i_1,\dots,i_k}(t_1,\cdots, t_k)= \det
\left[\begin{matrix}
\phi_{i_1}(t_1) & \cdots & \phi_{i_k}(t_1)\\
\vdots & \ddots & \vdots \\
\phi_{i_1}(t_k) & \cdots & \phi_{i_k}(t_k)
\end{matrix}\right]
\,.
\end{align}
In particular, the $N$-th coherence function is
\begin{align}
\label{e:Nth_coherence_fermions}
G^{(Ne)}_{\ket{\psi_N}} (t'_1,\cdots, t'_N,t_1,\cdots, t_N) = 
\Phi^*_{1,\dots, N}(t'_1,\cdots, t'_N) \Phi_{1,\dots, N}(t_1,\cdots, t_N)
\,.
\end{align}
As expected, the $N$-th coherence function is essentially the
full many-body wavefunction of a quantum state with $N$
excitations in mutually orthogonal wavefunctions.
\end{example}

\begin{figure}
\begin{center}
\begin{tabular}{| c | c c c c c |}
\hline
 & $\phi_{e_1}$  & $\cdots$ & $\phi_{e_k}$ & $\cdots$ & $\phi_{e_N}$ \\ \hline
$G^{(1e)}$ & $\phi^*_{e_1}\phi_{e_1}$ & $\cdots$ & 
	$\sum_{p=1}^k \phi^*_{e_p}\phi_{e_p}$ & $\cdots$  & 
		$\sum_{p=1}^N \phi^*_{e_p}\phi_{e_p}$ \\
$\vdots$ & $\vdots$ & $\ddots$ & &  & $\vdots$ \\
$G^{(ke)}$ & 0 & & $\Phi^*_{1,\dots, k}\Phi_{1,\dots, k}$ &  &
		$\sum_{i_1<\cdots<i_k} \Phi^*_{i_1,\dots,i_k} \Phi_{i_1,\dots,i_k}$ \\
$\vdots$ & $\vdots$ & & & $\ddots$  & $\vdots$ \\
$G^{(Ne)}$ & 0 & $\cdots$ & 0 & $\cdots$ & $\Phi^*_{1,\dots, N}\Phi_{1,\dots, N}$  \\ \hline
\end{tabular}
\end{center}
\caption{Table showing the structure of the electronic coherence functions of
order up to $N$ in the presence of one up to $N$ fermions in some
given wavepackets.}
\end{figure}

To see the correlation functions of determinantal point processes appear, one can use an ideal broad-band detector as in \cref{s:electroncoherence}. 
In particular, imagine that we could prepare $N$ electrons in mutually 
orthogonal wavepackets $\phi_1, \dots, \phi_N$, and set up an ideal 
detector as we did for photons in \cref{s:photoncoherence}, resulting 
in a point process with correlation functions given by the diagonal of 
$G^{(ne)}_{\ket{\psi_N}}$, 
\begin{align}
	\rho_n(t_1,\dots,t_n) &= G^{(ne)}_{\ket{\psi_N}} (t_1,\cdots, t_n,t_1,\cdots, t_n), \\
	&= \left \vert \Phi_{1,\dots, N}(t_1,\cdots, t_N)\right\vert^2\\
	&= \det \left[K(t_i,t_j)\right]_{i,j=1}^n,
\end{align}
where 
\begin{equation}
	\label{e:kernel_N_fermions}
	K(t,s) = \sum_{i=1}^N \phi_{e_i}(t)\phi_{e_i}^*(s).
\end{equation} 
We recognize a DPP with (projection) kernel $K$, as introduced in 
\cref{s:determinantal}. 

\begin{remark}
To illustrate what reference computations different scientists may 
have in mind, note that, following a physics viewpoint, we computed 
in Examples~\ref{ex:G2_for_fermions} and \ref{ex:Gk_for_fermions} 
coherence functions applying Wick's theorem, one after the other, 
starting from the lower ones. Probabilists and statisticians are more
used to derive correlation functions starting from the definition of the
point process. Had we started by defining the DPP with kernel $K$ in 
\eqref{e:kernel_N_fermions}, and asked for the diagonal of, say, the 
second coherence function in~\eqref{ex:G2_for_fermions}, it would 
have been enough to notice, from the definition \eqref{rhoDet} of 
the correlation functions of a DPP, that
\begin{align}
	\rho_2(t_1,t_2) &= \det
	\left[\begin{matrix}
		K(t_1,t_1) & K(t_1,t_2)\\
		K(t_2,t_1) & K(t_2,t_2)
	\end{matrix}\right].
\end{align}
By definition \eqref{e:kernel_N_fermions} of $K$,
\begin{align}
	\rho_2(t_1,t_2) &= \det
	\left[\begin{matrix}
		\phi_1(t_1) &  \dots & \phi_N(t_1)\\
		\phi_1(t_2) &  \dots & \phi_N(t_N)
	\end{matrix}\right]
	\left[\begin{matrix}
		\phi_1^*(t_1) & \phi_1^*(t_2) \\
		\vdots & \vdots \\
		\phi_N^*(t_1) & \phi_N^*(t_2)
	\end{matrix}\right]
	\,.
\end{align}
The diagonal version of \eqref{e:second_coherence} then results 
from the Cauchy-Binet formula, which allows rewriting $\rho_2(t_1,t_2)$ 
as a sum of $2\times 2$ determinants; see e.g. \citep{KuTa12}.
\end{remark}

\subsection{Electron anti-bunching}
\label{s:electron_anti_bunching}

For photons, we saw that HBT-type experiments can
reveal both a bunching effect (for classical light beams) and an
anti-bunching effect (for quantum beams), depending on the source.

A similar experimental setup as the one pictured for photons in \cref{s:photoncoherence} can be imagined for fermionic excitations. 
We defer the discussions on how to properly do this experimentally with fermionic atoms or single electronic excitations to Part II, but we can already foresee from either DPP constructions like the one with kernel \eqref{e:kernel_N_fermions}, or directly from the coherence functions of \cref{s:correlation_electrodetection_events} that fermions will exhibit anti-bunching detection times.
Unlike bosons, this is a direct consequence of the anti-symmetric statistic of fermions, and does not qualitatively change when introducing source models.

If the arrival times follow a DPP with an Hermitian kernel like \eqref{e:kernel_N_fermions}, we saw indeed in \cref{s:determinantal} that samples tend to spread regularly, and form less clusters than a Poisson point process.
Alternately, this statistical anti-bunching effect can be directly seen from the coherence funtions. 
Indeed, as for photons, the HBT signal is controlled by the second-order coherence function at the position $\mathbf{r}$ of the detector, 
\begin{align}
	G^{(2e)}_{\rho}(\mathbf{r}, t_2, t_1)
	=
	\langle \psi^\dagger(\mathbf{r},t_1) \psi^\dagger(\mathbf{r},t_2) 
	\psi(\mathbf{r},t_2) \psi(\mathbf{r},t_1) \rangle_\rho
\,.
\end{align}
At coincidence $\tau = t_2-t_1=0$, because the CAR \eqref{e:CAR} impose $\psi(\mathbf{r},t_1)^2 = 0$, we obtain
\begin{align}
	G^{(2e)}_{\rho}(\mathbf{r}, t, t) = 0
\,.
\end{align}
If $G^{(2e)}_\rho$ is smooth, then we expect few coincidences at small time differences.

As a final remark, we insist that bunching or anti-bunching of
quantum excitations is not a signature of the statistics of the
elementary excitations. 
While it is a consequence of the
statistics for fermions, bosons also anti-bunch when we add to the model a single-excitation source. 
Hence, it is better to think of
anti-bunching as a signature of the quantum nature of the
fundamental excitation.

\subsection{Recovering a classical current is not as easy as for bosons}
\label{s:recovering_classical_currents}

For photons, coherent states allowed for relating the quantum formalism to classical fields; see Section~\ref{s:photoncoherence}. 
Describing a classical electric current or, more generally,
a classical theory of fermionic fields, is more difficult. 
In this section, we examine reasons why it is not possible to build a set of states having all the properties that make bosonic coherent states handy.
First, even considering a single mode, bosonic coherent states are superpositions of arbitrary numbers of bosons, which allows for having very small fluctuations around the average boson number.
Indeed the number of photons in the mode is a Poisson variable with arbitrarily large parameter, and thus arbitrarily small relative variance.
Such small fluctuations are not achievable with fermions, because it is impossible to have more than one fermion per mode.

A second difference is that the eigenvalue of a tentative fermionic coherent state of the
fermionic annihilation operator cannot be a complex number. 
Indeed, were $\ket{\alpha}$ an eigenstate of the fermionic annihilation operator $a$, then the CAR \eqref{e:CAR} would imply that its eigenvalue
$\alpha$ satifies $\alpha^2 = 0$. 
Moreover, in the many-mode case, a two-mode coherent state $\ket{\alpha_1,\alpha_2}$  would satisfy the anti-commutation 
relation  $a_1a_2\ket{\alpha_1,\alpha_2} =\alpha_1\alpha_2\ket{\alpha_1,\alpha_2}
=-a_2 a_1\ket{\alpha_1,\alpha_2} =-\alpha_2\alpha_1\ket{\alpha_1,\alpha_2} $.\footnote{This result is independent on whether one assumes that $\alpha_i$ commutes or anti-commutes with $a_j$.}
The eigenvalues of a fermionic coherent state thus cannot be complex numbers.\footnote{  
It is however possible to define fermionic
coherent states in terms of anti-commuting variables (elements of an
exterior algebra) by the action of the analogue of a displacement operator \citep[Chapter 1]{Berezin1966}.
While they are very useful objects to define and use (especially to write
path integrals), they do not possess the nice set of properties of
bosonic coherent states that allow for deriving coherence functions. 
}

A more subtle difference is that fermionic fields cannot have an
observable average amplitude,\footnote{
	Just like $a$ for photons, $\psi$ is not Hermitian, and thus not an observable, \emph{stricto sensu}. Yet, for $a$, the average value corresponds to the intensity of the field, and is thus experimentally accessible as a statistical average. 
	It thus makes sense to wonder whether the average value of $\psi$ can be non-zero.
}
unlike bosonic fields prepared in a coherent state.\footnote{
	This property can actually be used as a definition of bosonic coherent states \Citep{CohenBookI}.
} 
This impossibility is fundamental and is related to \emph{superselection rules}~\citep{Wigner-1952}. 
For concreteness, consider a general fermionic state 
$\ket{\phi} = \alpha \ket{0} + \beta\ket{1}$ with  $\ket{1} = \psi^\dagger\ket{0}$, where $\psi^\dagger$ is any fermionic creation operator.
Now, physicists usually impose symmetries on their models, which implies that some states should yield the same measurements. 
In particular, one would like to model the fact that rotating the physical system around any axis by an angle of $2\pi$ should not modify the law of measurements. 
For fermions, the so-called \emph{spin-statistics theorem} of relativistic quantum field theory tells us that this kind of rotations should be modelled by the action of the half-integer representations of the group $SU(2)$. 
For instance, in the representation of order $1/2$, a rotation of angle $2\pi$ corresponds to multiplying the state by $\e^{\i\pi N}$, where $N$ is the number operator. 
Applied to $\ket{\phi}= \alpha \ket{0} + \beta\ket{1}$, this yields $ \ket{\phi'} := \e^{\i\pi N}\ket{\phi}= \alpha \ket{0} - \beta\ket{1}$, where the minus sign comes from the fact that $1$ is an odd number.
However, since such a rotation should not correspond to any change of the system, we impose that $\ket{\phi'} =\ket{\phi}$ up to a phase, and consequently either  $\alpha$ or $\beta$ is zero. Thus $\ket{\phi}$ is in fact not a superposition, implying that $\expval{\psi}_{\ket{\phi}}=0$.
More generally, the \emph{parity superselection rule} imposes that a given state can only consist of superpositions of states with the same parity of the number of fermions.
Therefore, a single fermionic annihilation operator can only have a vanishing average.\footnote{
	This generalises beyond physically realistic fermions of half-integer spins, and applies to non-relativistic quantum mechanics, see for instance  \citep{Johansson-2016,Szalay-2021}.
}

For all these reasons, defining a classical
regime with fermionic fields is an open question.
This actually raises deep conceptual questions when
fermions are involved. 
For instance, it is not clear what is meant by a \emph{classical
electronic current} if we start from a pure quantum description of
the electronic current.

%% file: dppfermions.tex


To conclude this Part I, we first show how to use the formalism of Sections~\ref{s:point_processes} to \ref{s:electroncoherence} to write a generic permanental or determinantal point process as the detection process of a system of free bosons or fermions, respectively. 
Second, we propose a list of questions, from simple ones to open problems, that are raised by the connections between point processes and physical measurements in quantum field theory. 
Third, we announce the sections that are to appear in Part II of this document.

\subsection{From a point process to free particles}
\label{s:generic_construction}

In Sections~\ref{s:photoncoherence} and \ref{s:electroncoherence}, we obtained some point processes, including permanental and determinantal point processes, by modeling the detection of physical particles. 
A natural question is whether all permanental and determinantal point processes arise in this form. 
We now show that, at least formally and under weak assumptions on the kernel, the answer is yes.
More precisely, let $\mu$ be a Borel measure on a complete metric space $\X$, and $(\phi_i)$ an orthonormal family in $L^2(\mu)$.
Consider the kernel
\begin{equation}
  \label{e:desired_kernel}
  K(x,y) := \sum_{i} \lambda_i \overline{\phi_i(x)} \phi_i(y),
\end{equation}
where, for all $i$, $\lambda_i\in[0,1]$ if $\eta=-1$, and $\lambda_i\in\mathbb{R}_+$ if $\eta=+1$.
We now build a Fock space, a quantum state, and a measurement model that together lead to a permanental or determinantal point process with $K$ as its kernel. 
Let $\eta=\pm1$ depending on whether one wants a permanental or a determinantal point process.

\paragraph{Building a Fock space.} 
Build the Fock space like in Section~\ref{s:fock_spaces}, using the basis $(\phi_i)$ as single-particle basis and the symmetrization property dictated by $\eta$.
Consider the operator
$$
H = \sum_i \nu_i a_i^\dagger a_i,
$$
where $(\nu_i)$ is left as a free parameter for now, and will later be chosen in relation to the spectrum of $K$.
We think of $H$ as a Hamiltonian.

\paragraph{A Gaussian density matrix.}
Let $\zeta>0$, and define the so-called grand canonical ensemble as in \eqref{e:grandcanonical} by the Gaussian density matrix
$$
\rho = \frac{\e^{-\beta\left ({H}-\zeta \sum_i a_i^\dagger a_i \right )}}{Z_\eta} = \frac{1}{Z_\eta} \e^{-\beta \sum_i  (\nu_i-\zeta) a_i^\dagger a_i}.
$$
We explicitly write the dependence of $Z$ to $\eta$, as bosons and fermions lead to different normalization constants. 
More precisely, since the Fock states $\ket{\mathbf n} = \ket{n_1, n_2, \dots}$ are eigenvectors of all the number operators $a_i^\dagger a_i$ in the exponential, the normalization constant is 
\begin{equation}
  Z_\eta = \tr  \e^{-\beta \sum_i  (\nu_i-\zeta) a_i^\dagger a_i} 
= \sum_{\mathbf n} \expval{\e^{-\beta \sum_i  (\nu_i-\zeta) a_i^\dagger a_i}}{\mathbf n} 
= \sum_{\mathbf n} \e^{-\beta \sum_i  (\nu_i-\zeta)n_i}.
  \label{e:normalization}
\end{equation}

For bosons, we sum over all sequences of integers with only $p$ non-zero components, and this for all $p\geq 1$. 
In particular,
\begin{equation}
  \label{e:normalization_bosons}
  Z_1 = \prod_{p\in\mathbb{N}} \sum_{n\in\mathbb{N}} e^{-\beta (\nu_p-\zeta)n} 
= \prod_{p\in\mathbb{N}} \frac{1}{1-\e^{-\beta(\nu_p-\zeta)}},
\end{equation}
where we implicitly assumed $\nu_p-\zeta>0$ for all $p$, for the geometric sums in \eqref{e:normalization_bosons} to converge.\footnote{
  Note that for a given $p$, the limit $\nu_p\to\zeta$ corresponds to the onset of Bose-Einstein condensation, where a macroscopic number of bosons start to occupy the state $p$.
}
For $Z_1$ to be finite, we further need a condition on the $(\nu_p)$, such as 
\begin{equation}
  \label{e:condition_bosons}
  \sum_p \log \left ( \frac{1}{\e^{\beta(\nu_p-\zeta)}-1} \right ) < \infty.
\end{equation}

  For fermions, we only sum in \eqref{e:normalization} over all sequences in $\{0,1\}^\mathbb{N}$ that have $p$ non-zeros components, for all sizes $p\geq 1$, so that
\begin{equation}
  \label{e:normalization_fermions}
  Z_{-1} = \prod_{p\in\mathbb{N}} \sum_{n\in\{0,1\}} \e^{-\beta (\nu_p-\zeta)n} 
	= \prod_{p\in\mathbb{N}} \left(1+\e^{-\beta (\nu_p-\zeta)}\right).  
\end{equation}
Guaranteeing convergence is easier than for bosons, e.g. assuming that
$\sum_p (\nu_p-\zeta)$ converges.

Now that we have simple expressions for $Z_\eta$, $\eta=\pm1$, we can use the partition function trick\footnote{Statisticians might say the \emph{score function} trick.} to compute the expected number of particles in mode $i$, 
\begin{subequations}
\begin{align}
  \expval{a_i^\dagger a_i}_\rho &= \frac{1}{Z_\eta} \tr a_i^\dagger a_i \rho\\
  &= \frac{\sum_n n_i  \e^{-\beta \sum_j(\nu_j-\zeta)n_j}}{\sum_n \e^{-\beta \sum_j(\nu_j-\zeta)n_j}}\\
  &= -\frac1\beta \frac{\partial}{\partial\nu_i} \log Z_\eta(\nu).
\end{align}
\end{subequations}
Using \eqref{e:normalization_bosons} and \eqref{e:normalization_fermions} in turn, this yields 
\begin{equation}
  \expval{a_i^\dagger a_i}_\rho = \frac{1}{\e^{\beta(\nu_i-\zeta)}-\eta},
  \label{e:mean_number}
\end{equation}
which is the so-called \emph{Fermi-Dirac} distribution\footnote{The word \emph{distribution} is used here in a loose sense.} for $\eta=-1$ and the \emph{Bose-Einstein} distribution for $\eta=1$.
Note that along the lines of Lemma~\ref{lemma:wick}, one can show that $\langle a_i^\dagger a_j\rangle_\rho = 0$ if $i\neq j$.

\paragraph{A measurement basis.}
Now, consider the alternative basis $\ket{x}$, $x\in \X$, with annihilation operator
$$
\psi(x) = \sum_i \langle x, \phi_i\rangle a_i;
$$
see \eqref{e:change_of_basis}.
We think of $\ket{x}$ as the position basis, so that we write $\langle x, \phi_i\rangle = \phi_i(x)$, and we actually mean a generalized basis if necessary, e.g., if $X=\mathbb{R}^d$; see Section~\ref{s:quantum_framework}.
In the basis $\ket{x}$, the state $\rho$ rewrites
$$
\rho = \frac{1}{Z} \exp \left( -\beta \sum_i (\nu_i-\zeta) \int \phi_i(x) \overline{\phi_i(y)} \psi^\dagger(x) \psi(y) \d\mu(x)\d\mu(y) \right).
$$

\paragraph{Wick's theorem applies.}
By \cref{th:wick}, see also Example~\ref{ex:wick_for_correlation_functions}, for all $n\geq 1$, we obtain the coherence functions
$$
\expval{\psi^\dagger(x_1)\psi(x_1) \dots \psi^\dagger(x_n)\psi(x_n)} = f_\eta(\mathbf{K}), 
$$
where $$
\mathbf{K} = \left( \expval{\psi(x_i)^\dagger \psi(x_j)}_\rho \right)_{1\leq i,j\leq n}
$$ 
and $f_\eta$ is the determinant if $\eta=-1$, and the permanent if $\eta=1$.
If the framework is consistent, an ideal detection experiment that measures all particles in $\rho$ in the basis $\ket{x}$ should thus correspond to a permanental/determinantal point process on $\X$, with kernel 
\begin{align}
  \expval{\psi^\dagger(x) \psi(y)}_\rho
   &= \sum_{i,j} \overline{\phi_i(x)} \phi_j(y) \expval{a_i^\dagger a_j}_\rho.
  \label{e:obtained_kernel}
\end{align}
This is precisely $K$ in \eqref{e:desired_kernel}, provided that we choose $\nu_i$ so that
\begin{equation}
  \lambda_i = (\e^{\beta(\nu_i-\zeta)}-\eta)^{-1}.
  \label{e:lambda_as_a_function_of_nu}
\end{equation}
This is achieved by setting
$$
\beta(\nu_i-\zeta)= \log(\frac{1+\eta\lambda_i}{\lambda_i}).
$$ 
Note that when $\eta=1$, one can check that the corresponding permanental point process exists by an adaptation of Macchi's Cox process construction of Section~\ref{s:permanental}.
When $\eta=-1$, the determinantal point process exists by the Macchi-Soshikov theorem; see Section~\ref{s:determinantal}.

\paragraph{Obtaining any kernel.}
We conclude with several comments. 
First, inverting \eqref{e:mean_number} allows us to obtain any spectrum $(\lambda_i)$ with $\lambda_i\in (0,1)$ for fermions, and $\lambda_i>0$ for bosons. 
With the right assumptions, projection kernels for DPPs can also be obtained, but as a limit as $\beta\rightarrow +\infty$, i.e. when the temperature $1/\beta$ goes to zero.
The corresponding projection operator is onto the span of the $\phi_i$s for which $\nu_i<\zeta$. 
In other words, the chemical potential controls the rank of the limiting projection operator.
Note that for DPPs, uniform convergence of the kernel on compact subsets of $\mathbb{X}$ implies convergence of the point process in a natural sense; see e.g. \citep[Section 4.2.8]{AnGuZe10}.
Another way to formally obtain projection DPPs is as a ground state; see Section~\ref{s:example_N_particle_states}.

Second, the construction in this section, with a state $\rho$ and measuring correlators, is as close as one can hope for a Cox process-like decomposition of determinantal point processes. 
Loosely speaking, there is a Gaussian object above determinantal point processes with self-adjoint kernels, but it is a Gaussian density matrix, i.e., an operator, not a functional process.\footnote{
Alternatively, the formulation in terms of operators can be replaced by a ``Gaussian functional integral'' over complex commuting (anti-commuting) fields for permanents (determinants). 
The connection with Gaussian measures is then more direct, at the price of subtle mathematical difficulties. 
Furthermore, in the case of fermions, this implies generalising the notion of integration to variables that anti-commute, see for instance \citep[Chapter 1]{Berezin1966}.
}
Third, while we restricted here to Hermitian kernels, thus corresponding to an observable quantity, the Hermitian assumption does not play a major role in the mathematical framework. 
Furthermore, non-Hermitian free fermions may not be so removed from physics, see e.g. \citep{Ashida2020} for a recent survey and \citep{Guo2021} for a formal example.

\paragraph{On physical realizability.}
Finally, we have seen in Section~\ref{s:electroncoherence} examples of experimentally realizable DPPs, in the sense that the corresponding idealized experimental setups can, in principle, be built in a physics laboratory. 
We will see in Part II concrete examples of DPP samples obtained from lab experiments.
Yet, the generic construction given in Section~\ref{s:generic_construction} remains rather theoretical, as it is not clear how to build an experimental setup leading to \emph{any} given DPP, that is, any choice of underlying space $(\mathbb{X},\mu)$ and any valid kernel $K$. 
One might be able to treat $\X=\mathbb{R}^d$ for $d$ up to $3$ by associating mathematical coordinates to space-coordinates. 
Going slightly beyond $3$ might be doable using either time or additional physical degrees of freedom (``pseudo-dimensions") as the next dimensions, but it is unclear to physicists where fermions in arbitrary dimension would be needed as a model. 
Outside the problem of dimensionality, fixing the spectrum of the Hamiltonian to an arbitrary sequence can also be experimentally difficult.

\subsection{Going further: from simple to open questions}
Many natural questions come to mind when one looks back at the correspondence between permanental point processes and free bosons, or between DPPs and free fermions. 

\subsubsection{A quantum state is more than a point process}
A point process is the result of observing a state $\rho$ with a detector, but the state itself contains a lot more information. 
For instance, the correlation functions that we identified in Sections~\ref{s:photoncoherence} and \ref{s:electroncoherence} are only the diagonals of the corresponding coherence functions. 
Relatedly, depending on the measurement basis, a single state $\rho$ can yield many point processes.

As a concrete example, consider for simplicity fermions living in $\X = \{1,\dots,N\}$, $\mu$ the counting measure on $\X$, and the Hamiltonian
$$ 
H = \sum_{i=1}^N \nu_i a^\dagger(\phi_i) a(\phi_i)
$$
acting on the corresponding Fock space.
By \eqref{e:obtained_kernel}, measuring in an orthonormal basis $(v_i)$ of $\mathbb{C}^N$, with $\mathbf{V}=(\braket{\phi_i}{v_j})_{i,j}$, we obtain the DPP with kernel matrix 
$$
\mathbf{K} = \left(\sum_{k} \lambda_k \braket{\phi_k}{v_i} \braket{v_j}{\phi_k} \right)_{1\leq i,j\leq N}  = \mathbf{V}\text{diag}(\lambda_k) \mathbf{V}^\dagger,
$$
where $(\lambda_i)$ is given by \eqref{e:lambda_as_a_function_of_nu}. 
All Hermitian kernels with spectrum $(\lambda_k)$ can thus be obtained from the same Hamiltonian, just changing the measurement basis.

At one extreme, measuring in the original basis used to build the Hamiltonian, i.e. taking $\mathbf{V} = \mathbf{I}$, leads to a diagonal kernel matrix.
This means that the corresponding point process is a set of independent Bernoulli samples for fermions, and independent geometric samples for bosons \citep{HKPV06}.
Correlation appears when the observation basis $(v_k)$ and the creation basis $(\phi_k)$ do not match. 
For instance, let $v_k = \phi_k$ for $k=1,\dots,N-2$, but apply a special unitary transformation to the last two vectors. 
That is, for $\alpha, \beta \in \mathbb{C}$ such that $\vert\alpha\vert^2+\vert\beta\vert^2 = 1$, complete the observation basis with 
$$
v_{N-1} = \alpha \phi_{N-1} - \overline\beta \phi_N \quad \text{ and } \quad v_N = \beta \phi_{N-1} + \overline\alpha \phi_N.
$$ 
The resulting permanental or determinantal point process, call it $\gamma^{\alpha, \beta}$, has a block diagonal kernel 
$$
\mathbf{K}^{\alpha, \beta} = 
  \begin{pmatrix} 
    \lambda_1 & & & & \\
    & \ddots & & & \\
    & & \lambda_{N-2} & & \\
    & & & \lambda_{N-1}\vert\alpha\vert^2 + \lambda_N\vert\beta\vert^2 & (\lambda_{N-1} - \lambda_N) \overline\alpha \beta \\
    & & & (\lambda_{N-1} - \lambda_N) \alpha\overline \beta& \lambda_{N-1}\vert\beta\vert^2 + \lambda_N\vert\alpha\vert^2
  \end{pmatrix}.
$$
Note that since the diagonal of $\mathbf{K}^{\alpha, \beta}$ varies with $(\alpha, \beta)$, the first correlation function of the point process is also altered.
For instance, the marginal probability that item $N-1$ belongs to the point process is  $\lambda_{N-1}\vert\alpha\vert^2 + \lambda_N\vert\beta\vert^2$.
If $\lambda_{N-1}\geq \lambda_N$, then as soon as $\beta\neq 0$, the marginal probability of item $N-1$ occurring decreases.
Note that for $\eta=-1$, though, the average number of points in $\gamma^{\alpha, \beta}$ remains constant, as the trace of $\mathbf{K}^{\alpha, \beta}$ does not depend on $(\alpha, \beta)$.
Meanwhile, the second correlation function $\rho_{2,\theta}(N-1,N)$ is the permanent/determinant of the trailing block of $\mathbf{K}^{\alpha,\beta}$, namely
$$
\rho_{2}^{\alpha, \beta}(N-1,N) = \vert\alpha\vert^2\vert\beta\vert^2 \left (\lambda_{N-1}^2 + \lambda_N^2 + \eta(\lambda_{N-1}- \lambda_N)^2 \right ) + \left (\vert\alpha\vert^4 + \vert\beta\vert^4 \right ) \lambda_{N-1}\lambda_N. 
$$
For the sake of illustration, we focus now on $\eta=-1$, in which case
$$
\rho_{2}^{\alpha, \beta}(N-1,N) = \lambda_{N-1}\lambda_N \left (\vert\alpha\vert^2 + \vert\beta\vert^2 \right )^2 =  \lambda_{N-1}\lambda_N.
$$  
The probability of co-occurrence of items $N-1$ and $N$ thus does not depend on $\alpha$ and $\beta$, while the product of their marginal probabilities of occurrence does change: we have thus introduced correlation between the events $\{N-1\in \gamma^{\alpha, \beta}\}$ and $\{N\in \gamma^{\alpha, \beta}\}$.
Because $\gamma^{\alpha,\beta}$ is a DPP with Hermitian kernel, this correlation is nonpositive.

To go further, an interesting question would be to relate the different ways to quantify repulsiveness in a DPP \citep{BiLa16, MoOR21} to properties of the bases used in the construction of free fermions.

\subsubsection{A DPP from bosons in a non-Gaussian state}
Another example that illustrates the subtleties of associating correlation functions to states is that (non-Gaussian) density matrices corresponding to interacting (i.e., non-free) bosons can give rise to a DPP for some observables.

One example is the Tonks-Girardeau gas \citep{Girardeau1960}, which consists of impenetrable (statisticians would say \emph{hardcore}) bosons in $\X=\mathbb R$. 
The very strong interactions prevent two bosons to be at the same position, which is reminiscent of the Pauli exclusion principle for fermions. 
In particular, one considers the usual bosonic field operators $a(x), a^\dagger(y)$, satisfying the canonical commutation relations as long as $x\neq y$, but the hardcore constraint is enforced by requiring that $a(x)^2=a^\dag(x)^2=0$ and $[a(x),a^\dag(x)]_+=1$. 
These additional constraints are the results of adding a large penalty term to the Hamiltonian at coincidence.

Now, there actually is a transformation, called the \emph{Jordan-Wigner transformation}, that maps these hardcore bosons onto free fermions \citep{Lieb1961}.
Concretely, one can write $a(x) = S(x) \psi(x)$ and $a^\dag(x) =  \psi^\dag(x)S^\dag(x)$, where $\psi(x), \psi^{\dag}(x)$ obey the standard canonical anti-commutation relations, and $S(x)$ is a unitary operator called a \emph{string operator}, which allows for preserving the commutation relations of the bosonic operators.\footnote{
  In particular, it satisfies $[S(x),S(y)]=0$ and $S(x)\psi(y)+{\rm sign}(x-y)\psi(y)S(x)=0$.
}

By construction, all observables that are built from number operators\footnote{Usually rather called \emph{density operators} when we use a generalized basis like here.} are the same as for free fermions, since $a^\dag(x)a(x)=\psi^\dag(x)\psi(x)$, and the correlation functions of an ideal detection experiment are those of a DPP with kernel $\langle \psi^\dag(x)\psi(y)\rangle$. 
However, the coherence function $\langle a^\dag(x)a(y)\rangle = \langle \psi^\dag(x)S^\dag(x)S(y)\psi(y)\rangle$ is very different from that of the free fermions. 
For instance, the largest eigenvalue of the operator with kernel $\langle a^\dag(x)a(y)\rangle$ evaluated in a pure state typically scales like the square root of the number of particles; see e.g.  \cite{Forrester2003} for investigations on the ground state properties of a Tonks gas in a harmonic trap.

\subsubsection{Interacting field theories and point processes}

The relationship between, on one side, permanental and determinantal point processes, and, on the other side,
bosonic and fermionic quantum field theory, is based on free models, i.e., with a quadratic Hamiltonian in the fields. 
However, in physical models, Hamiltonians are rarely quadratic, but take into account interactions between the particles. 
For a given Hamiltonian, success is achieved when it is possible to approximate the coherence functions of the field, which correspond to concrete experimental measurements. 
The main issue is that Wick's theorem does not apply.

There are interacting Hamiltonians that still yield closed-form correlation functions for the underlying ideal detection experiments. 
Such systems are usually called \emph{integrable} by physicists.
For instance, the Calogero-Sutherland model is a one-dimensional system of interacting fermions, for which ideal detection leads to well-known point processes, called $\beta$-ensembles in random matrix theory; see e.g. \citep[Chapter 11]{For10} as well as \citep{stephan2019free,smith2021full}.

Yet, most interacting systems are not integrable. 
In Sections~\ref{s:photoncoherence} and \ref{s:electroncoherence}, we have used simple perturbation-theoretic arguments like Assumption~\ref{a:weak_coupling} to work with the interaction Hamiltonian. 
There is now a significant and sophisticated toolbox to approximate coherence functions in the presence of interaction. 
Key tools include mean-field methods~\citep[Chapter 3]{Goldenfeld-2018}, Feynman diagrams \citep[Chapter 6]{Fol08} and renormalisation \citep[Chapter 7]{Fol08}. 
At a high level, it would be interesting to investigate what these methods say about point processes.
In words, one of the lessons of quantum field theory is that it is possible, even in the presence of interaction,
to get ``close" to a free situation in some regimes. 
Does this imply approximation results for the sophisticated point processes behind models with interaction? 
Conversely, can we use the mathematical technology of DPPs to perform non-trivial calculations for many-body problems?

Moreover, a central concept in quantum field theory is that of universality, i.e.,  the fact that many different interacting models behave, in a certain regime usually qualified
as the \emph{low-energy regime}, essentially in the same way. 
Put differently, the detailed form of the interactions are irrelevant
to understand this low-energy regime, only the dimension of the  space and the symmetries of the model usually matter. 
The archetypal phenomenon illustrating this is phase transitions~\citep{Goldenfeld-2018}.
Does this universality connect with similar results for point processes, such as those obtained for the eigenvalues of random matrices \citep{AnGuZe10}?

\subsubsection{Constructive arguments for point processes}

The fermionic system ``above" a DPP can help address fundamental questions on DPPs, about their invariance or their construction. 
For instance, a single DPP corresponds to many kernels. 
In particular, for any $f$, the kernel
$$
x, y \mapsto \frac{f(x)}{f(y)}K(x,y)
$$ 
yields the same correlation functions \eqref{rhoDet} as $K$. 
Without strong assumptions on the kernel, it has proven difficult to find all the transformations of a kernel that leaves a DPP invariant. 
It would be interesting if the fermionic framework helped us to understand these invariances; see e.g. recent partial results by \cite{Ols20}.


As another example, generalizing permanental and determinantal point processes, one can define $\alpha$-DPPs as having correlation functions 
\begin{align}
  \label{rhoDet}
  \rho_k(x_1,\ldots,x_k)={\det}_\alpha \Big[K(x_i,x_j)\Big]_{i,j=1}^k
  \,, \quad k\geq 1,
\end{align}
where, for an $n\times n$ matrix $\mathbf A = ((a_{ij}))$, 
$$
{\det}_\alpha(\mathbf A) := \sum_{\sigma\in\frak{S}_n} \alpha^{n-\nu(\sigma)} \prod_{i=1}^n a_{i\sigma(i)}.
$$ 
\cite{ShTa03} have studied the existence of $\alpha$-DPPs for $\alpha\in[-1,1]$, where $\alpha=-1$ corresponds to a DPP, and $+1$ to a permanental point process. 
Some $\alpha$-DPPs appear naturally when marginalizing a projection DPP with a separable kernel over coordinates \citep{MaCoAm20}.
It is a natural question whether $\alpha$-DPPs correspond to any physical system of particles. 
As a partial affirmative answer, \cite{CuMaOC19} give a limit procedure to construct certain $\alpha$-DPPs out of fermionic processes. 

In the same vein, can we build point processes from particles with more exotic commutation rules for their ladder operators? A natural physical example is anyons \citep[Chapter 8]{Eza08}, for which a phase factor appears for each transposition.

The links of fermions to Pfaffian point processes, another generalization of DPPs are also a promising research direction \citep{Kos21}, as well as the point processes behind quasi-free states \citep{BaLiSo94,Lyt02,LyMe07,Ols20}, namely (possibly non-Gaussian) states to which Wick's theorem still applies.

\subsection{A teaser for Part II}
This manuscript is intended to become Part I of a monograph.
In Part II, we shall present selected topics at the intersection of point processes and quantum optics, using the vocabulary of Part I. 
We will describe landmark experimental measurements of HBT signals, with both photons and (bosonic and fermionic) atoms.
We will discuss the application of determinantal point processes to the study of non-interacting trapped fermions in statistical physics.
We will show how the formalism of quantum field theory can help to prove fundamental results on point processes appearing in combinatorics.
Finally, we will discuss electronic quantum optics and its interactions with signal processing.

\subsection*{Acknowledgments}
We thank all participants to the Lille and Lyon workshops, and all participants to the Lille workgroup on point processes and applications; the discussions held in these circles helped a lot in the construction of this document.
We hope that the document, in return, participates to developing this cross-disciplinary research field.  
In particular, RB thanks all physicists with whom he has interacted over the last years for their enthusiasm and their patience, explaining basic physical concepts over and over until we converged to a description that spoke to all of us.

We acknowledge support from ERC grant \textsc{Blackjack}
(ERC-2019-STG-851866), ANR AI chair \textsc{Baccarat}
(ANR-20-CHIA-0002), ANR grant cl\textsc{1shot reloaded} (ANR-14-CE32-0017), ANR grant \textsc{QuSig4QuSense} (ANR-21-CE47-0012), the Joint Research Project \textsc{SEQUOIA} (17FUN04) within the European Metrology Programme for Innovation and Research (EMPIR), ANR grants \textsc{QRITiC} I-SITE ULNE and ANR-16-IDEX-0004 ULNE, ANR grant \textsc{COSQUA} (ANR-20-CE47-0001-01), ANR grant \textsc{MENTA} (ANR-22-QUA2-0008-01),  ANR Grant \textsc{RaMaTraF} (ANR-17-CE30-0027-01), ANR grant \textsc{Dimers} (ANR-18-CE40-0033), ANR grant \textsc{Combin\'e} (ANR-19-CE48-0011), Labex \textsc{PALM} (ANR-10-LABX-0039-PALM), and Labex \textsc{MILYON}.
\jb{y a-t-il un ordre particulier sur les grants?}
